%% file: ms.tex
\documentclass{article}
\usepackage[utf8]{inputenc}
\usepackage[margin=0.75in]{geometry}

\title{Geometric Uncertainty in Patient-Specific Cardiovascular Modeling with Convolutional Dropout Networks}
\author{Gabriel~D.~Maher$^{1}$, Casey~M.~Fleeter$^{1}$, Daniele~E.~Schiavazzi$^{2}$, Alison~L.~Marsden$^{3}$\footnote{corresponding author: amarsden@stanford.edu}}
\date{\small\it $^{1}$Institute for Computational and Mathematical Engineering, Stanford University, Stanford, CA, USA\\
$^{2}$Department of Applied and Computational Mathematics and Statistics, University of Notre Dame, Notre Dame, IN, USA\\
$^{3}$Departments of Pediatrics and Bioengineering, Stanford University, Stanford, CA, USA
}

\usepackage[english]{babel}
\usepackage{bm} 
\usepackage{multirow}
\usepackage{multicol} 
\usepackage[neverdecrease]{paralist} 
\usepackage{graphicx} 
\usepackage{hyperref} 
\usepackage[Gray,squaren,thinqspace,thinspace]{SIunits} 
\usepackage{eurosym} 
\usepackage{subcaption}
\usepackage{morefloats} 
\usepackage{pdfpages} 
\usepackage{fancyhdr} 
\usepackage{algpseudocode} 
\usepackage{algorithm}
\usepackage{booktabs}
\usepackage{wrapfig}

\input{header.tex}

\begin{document}

\maketitle

\input{abstract}

\input{introduction}

\input{method}

\input{experiment}

\section{Results}\label{sec:results}

\input{res_dropout}

\input{res_Bifurcation}

\input{res_AAA}

\input{res_LCA}

\input{discussion}

\input{conclusion}

\section*{acknowledgements}
This research was funded by the National Science Foundation (NSF) Software Infrastructure for Sustained Innovation (SSI) grants 1663671 and 1562450, Computational and Data-Enabled Science and Engineering (CDSE) grant 1508794, NSF CAREER award 1942662,  National Institute of Health grants R01HL123689 and R01EB018302 and the American Heart Association Precision Medicine Platform.

\bibliographystyle{plain}
\bibliography{suthesis}

\end{document}

%% file: header.tex
\usepackage{amsmath} 
\usepackage{amssymb}	
\usepackage{graphicx} 
\usepackage{multicol} 

\newcommand{\intt}[4]{\int_{#1}^{#2}\! #3 \, \mathrm{d}#4}
 
\let\baraccent=\= 
\renewcommand{\=}[1]{\stackrel{#1}{=}} 


%% file: abstract.tex
\begin{abstract}
\noindent We propose a novel approach to generate samples from the conditional distribution of patient-specific cardiovascular models given a clinically aquired image volume.
A convolutional neural network architecture with dropout layers is first trained for vessel lumen segmentation using a regression approach, to enable Bayesian estimation of vessel lumen surfaces. This network is then integrated into a path-planning patient-specific modeling pipeline to generate families of cardiovascular models. 
We demonstrate our approach by quantifying the effect of geometric uncertainty on the hemodynamics for three patient-specific anatomies, an aorto-iliac bifurcation, an abdominal aortic aneurysm and a sub-model of the left coronary arteries. 
 A key innovation introduced in the proposed approach is the ability to learn geometric uncertainty directly from training data.
The results show how geometric uncertainty produces coefficients of variation comparable to or larger than other sources of uncertainty for wall shear stress and velocity magnitude, but has limited impact on pressure.
Specifically, this is true for anatomies characterized by small vessel sizes, and for local vessel lesions seen infrequently during network training.
\end{abstract}

%% file: introduction.tex
\section{Introduction}
Results from cardiovascular models are affected by a number of uncertainty sources, including material properties, image-data resolution, and boundary condition selection to match clinical target data.
A rigorous determination of simulation uncertainty and the development of numerical approaches to efficiently quantify its effects on patient-specific models is necessary to increase clinical adoption of simulation tools and improve their effectiveness for early treatment planning and non-invasive diagnostics.

Prior studies have investigated a range of methods for characterizing the influence of specific sources of uncertainty in cardiovascular modeling~\cite{eck15,steinman18}. To reduce the computational burden with respect to Monte Carlo sampling, a stochastic collocation approach was proposed in~\cite{Sankaran2011}, while a multi-fidelity approach was proposed by~\cite{biehler15} in the context of mechanical stress analysis of abdominal aortic aneurysms. 
Additionally, a generalized polynomial chaos expansion is presented in~\cite{quicken16} and applied to two pathological anatomies, i.e., an abdominal aortic aneurysm and an arteriovenous fistula, respectively. A generalized multiresolution expansion for uncertainty quantification was developed in~\cite{Schiavazzi2017} to better handle uncertainty in the presence of non-smooth stochastic responses, while mitigating the exponential complexity of multi-dimensional multi-wavelet refinement.
Combined uncertainty in vessel wall material properties and hemodynamics are investigated in~\cite{Tran2019} for several patient-specific models of coronary artery bypass grafting, leveraging a novel submodeling approach to focus the analysis only on venous and arterial bypass grafts. This and other studies focusing on the coronary circulation, (see, e.g.,~\cite{Seo2019}) contributed to show a loose coupling between hemodynamics and wall mechanics for such anatomies.
One dimensional models have been used to better understand main pulmonary artery pressure uncertainty in mice due to material property and image segmentation uncertainty \cite{colebank19,colebank19a}.

Finally, multifidelity simulations based on approximate control variate variance reduction in Monte Carlo sampling, were thorougly analyzed in the context of deformable cardiovascular models in~\cite{Fleeter2019}.

While the above contributions focus on the propagation of uncertainty from model inputs to outputs, an end-to-end (or clinical data to simulation results) uncertainty analysis pipeline is proposed in~\cite{schiavazzi2016uncertainty} in the context of virtual stage II single ventricle palliation surgery.
Additionally, the solution of inverse problems is discussed in~\cite{Tran2017}, where  automated Bayesian estimation is applied to tune close-loop boundary condition parameters for patient-specific multi-scale models of the coronary circulation, in order to match a number of non-invasive clinical measurements.

The vast majority of studies in the literature focus on uncertainty in the boundary conditions and mechanical properties of the vascular walls. A third major source is geometric uncertainty which results from errors and operator subjectivity in vessel segmentation from image data, which constitutes a fundamental step in the generation of cardiovascular models. 
Acquisition of medical image volumes is inherently noisy, has limitations related to the achievable resolution as well as artifact, motion, and aliasing errors. Construction of patient-specific model geometries from image volumes is therefore affected by image uncertainty.
In the literature, analysis of the effects of geometrical uncertainty on the results of high-fidelity cardiovascular models has remained elusive due to the complexity of assembling end-to-end pipelines for automatic model generation and analysis.
As discretization approaches invariably require the geometry to be represented through a discrete surface mesh, a popular technique within the biomechanics community is mesh morphing~\cite{sigal08}. Other methods have focused on modeling the variation of geometry via segmentation approaches. For example, for a given input image the STAPLE algorithm generates a distribution of possible segmentations, but requires a set of ground-truth segmentations as input.
%
Gaussian processes have also been used to model preexisting segmentation variation in~\cite{le16}.
Segmentation priors and multivariate sensitivity analysis proved useful for segmentation variability estimation in~\cite{joskowicz18}, however it is unclear how to extend the method to multiple simultaneous images.

Only a few studies consider the effect of geometrical uncertainty for cardiovascular models.
Sensitivity of hemodynamics to geometry variation in patient-specific cerebral aneurysms was investigated in~\cite{gambaruto11}, which considered two model samples generated using heuristic smoothing techniques.
Manual segmentation uncertainty was shown to have varying influence on FFR-CT calculations in~\cite{venugopal19}, where uncertainty depended on the mean FFR-CT value.
In an aortic flow simulation, geometric uncertainty was shown to be a dominant factor when compared to computational fluid dynamics (CFD) model parameter uncertainty and boundary condition uncertainty \cite{xu20}.
Geometric uncertainty was also investigated in~\cite{sankaran15} for coronary artery simulations, and obtained through local perturbations of an idealized stenosis model. Effects on entire cardiovascular models were investigated in~\cite{sankaran14,sankaran15b,sankaran16} by perturbing the area and surface points of selected vessel segments using a spatial Gaussian function with uniform parameterization. The variation in geometry was found to produce sensitivities of up to 10\% in simulated FFR-CT measurements. 
We would like to point out that geometric uncertainty is assumed \emph{a-priori} in the above studies, instead of being directly \emph{learned} from the image data.

More recently, Bayesian Neural Networks, neural networks that are able to learn uncertainty from data, have been increasingly adopted in applications where it is crucial to quantify confidence in predictions~\cite{gal16,jena19}.
In particular, \cite{gal16} showed that augmenting neural networks with dropout layers enables them to learn uncertainty from the training dataset.
In the medical imaging field dropout networks have been used to model segmentation uncertainty for MRI volumes~\cite{nair20}.
In particular, the network's prediction uncertainty was found to be a useful marker for detecting human expert prediction error.

In this work, we use Bayesian deep learning to develop a cardiovascular model generation technique that learns the geometry distribution from a dataset of existing geometries and images.
We then use this network along with Monte Carlo sampling and numerical blood flow simulation to characterize the change in model outputs due to geometric uncertainty. 

In section \ref{section:paper3:method} we discuss our dropout network architecture and path-planning cardiovascular model generation process.
Sections \ref{section:paper3:experiments} and \ref{sec:results} provide an overview of the anatomical benchmarks we selected and the results we obtained.
Finally, Section~\ref{section:paper3:discussion} contains a discussion and \ref{section:paper3:conclusion} presents our conclusions.

%% file: method.tex
\section{Methods}\label{section:paper3:method}

\begin{figure*}[!ht]
\centering
\includegraphics[scale=0.5]{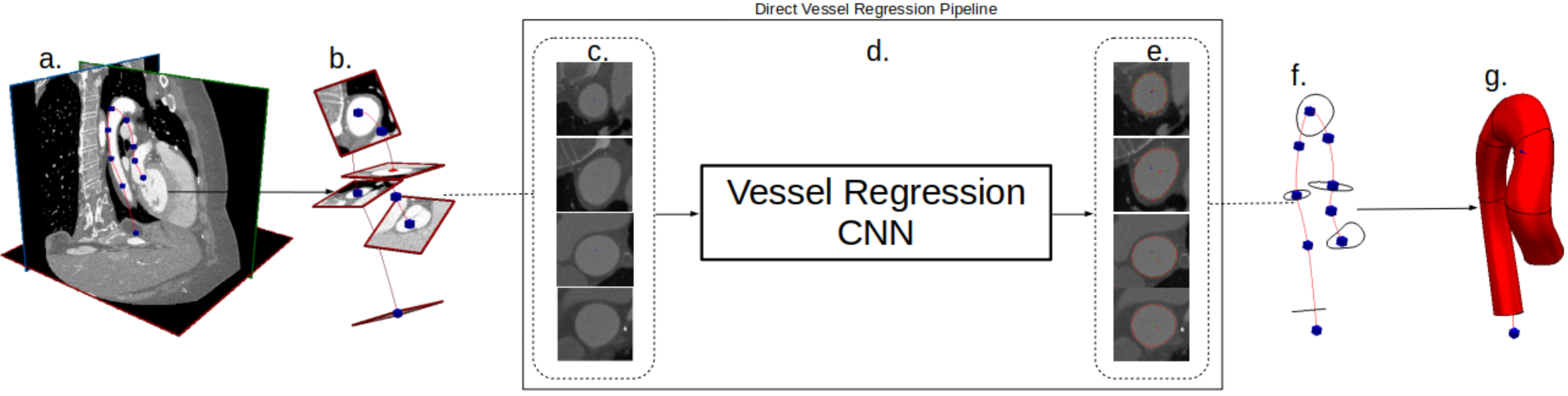}
\caption{Proposed model building pipeline. 
\textbf{(a)} Image data and vessel pathline are supplied by the user.
\textbf{(b)} Path information is used to extract local 2D cross-sectional images in the plane orthogonal to the vessel path.
\textbf{(c)} Two-dimensional images are extracted along vessel pathlines and fed to the CNN as inputs.
\textbf{(d) and (e)} The proposed CNN processes the cross-sectional images and directly outputs an array of point coordinates, characterizing a two-dimensional lumen segmentation.
\textbf{(f)} The collection of two-dimensional points is transformed back to three-dimensional coordinates on the image volume.
\textbf{(g)} The cross-sectional segmentations are lofted along the pathline to form the final lumen surface.}
\label{fig:method:modeling}
\end{figure*}

\noindent Given a medical image volume $\mathbf{X}$ and a set of vessel pathlines $\mathbf{V}$, our method produces samples from the distribution of patient-specific cardiovascular models, $\mathbf{Y}\sim P(\mathbf{Y}|\mathbf{X},\mathbf{V},\mathbf{z})$, compatible with both the image data, pathline and a collection of latent random variables $\mathbf{z}$.
To better explain how this is accomplished, we summarize in Figure~\ref{fig:method:modeling} a typical \emph{two-dimensional segmentation} or \emph{path-planning} approach, i.e., a widely used method to generate anatomical surfaces developed in a prior work \cite{maher_20} and based on the cardiovascular model format developed for SimVascular \cite{updegrove2016} and the Vascular Model Repository (VMR) \cite{wilson13}.
This requires to first define a vessel pathline (we will use the term \emph{centerline} interchangeably) by connecting user-specified point locations inside the lumen of the vessel of interest (e.g., the aortic arch in Figure~\ref{fig:method:modeling}). 
The tangent vector to this centerline is then used to generate a continuous collection of local 2D images slices.
Two-dimensional vessel lumen segmentation on this slice is accomplished through a parametric estimator, trained using a large collection of 2D cross-sectional images and corresponding ground truth lumen boundary.
The resulting two-dimensional segments, representing the intersection between the lumen wall surface and the cross-section plane, are then \emph{lofted} into a three-dimensional lumen surface and the final model generated by boolean union of multiple vessels.

\begin{figure}[!ht]
	\centering
	\includegraphics[scale=0.5]{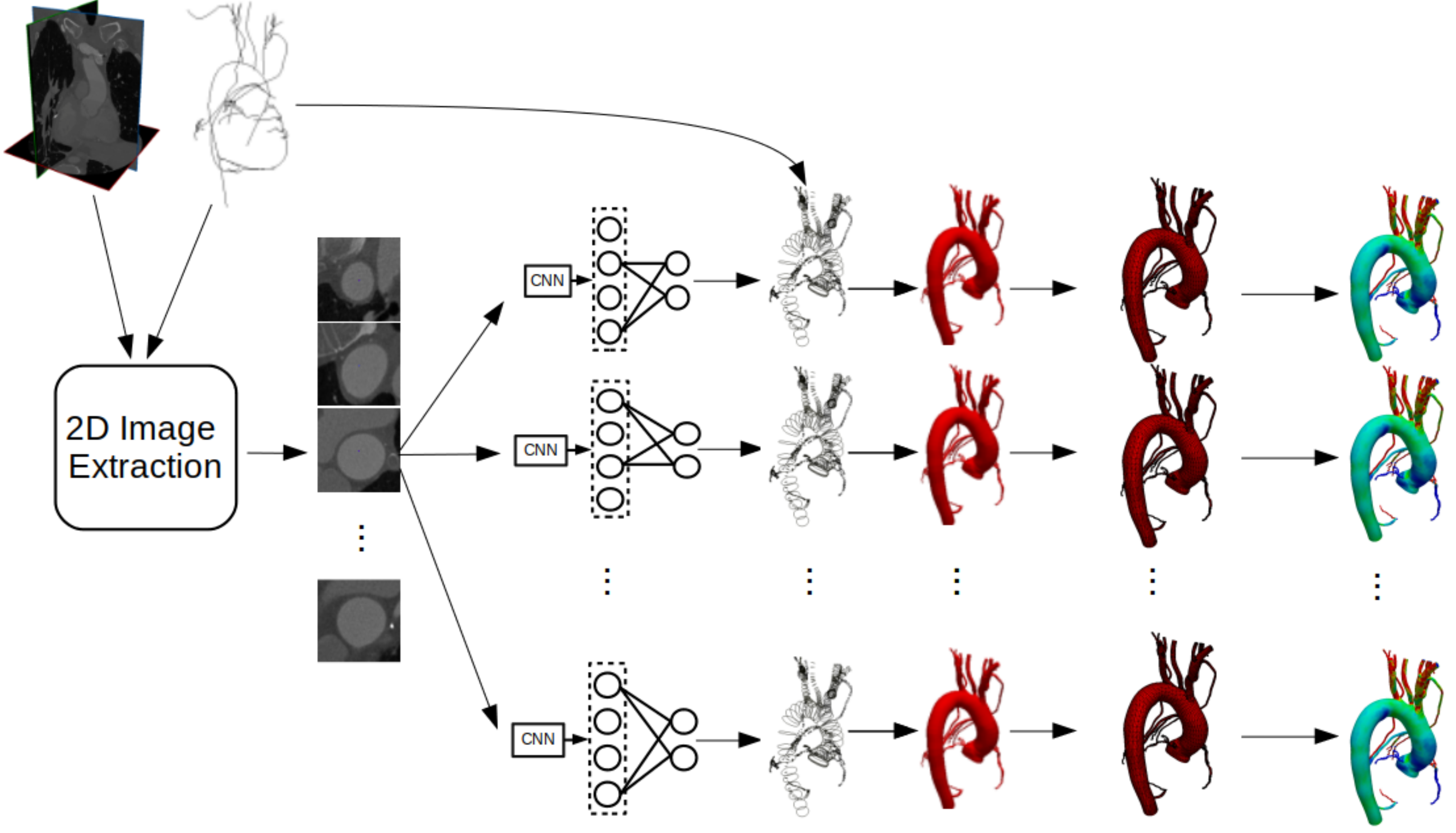}
	\caption{Generation of geometrically uncertain cardiovascular model solutions following Algorithm \ref{alg:method:montecarlo}}
	\label{fig:method:montecarlo}
\end{figure}

Note how our vessel lumen estimator also depends on a collection of latent variables $\mathbf{z}$, where two different realizations $\mathbf{z}^{(1)},\mathbf{z}^{(2)}\sim p(\mathbf{z})$ will produce two distinct anatomical surfaces. This way, we can naturally generate families of cardiovascular models $\mathcal{Y}:=\{\mathbf{Y}^{(1)},\dots,\mathbf{Y}^{(n)}\}$, where $\mathbf{Y}^{i}\sim P(\mathbf{Y}|\mathbf{X},\mathbf{V},\mathbf{z}),\,i=1,\dots,n$. 
Given this ability to \emph{sample} from distributions of cardiovascular anatomies, we adopt Monte Carlo sampling and a computational fluid dynamics (CFD) solver to estimate the changes in hemodynamics induced by geometrical uncertainty in the segmented anatomy.
Specifically, each sample $\{\mathbf{Y}^{(1)},\dots,\mathbf{Y}^{(n)}\}$ provides a computational domain where we numerically solve the Navier-Stokes equations using the finite element method, as further discussed in Section \ref{section:ns}.
This Monte Carlo process is described in Algorithm~\ref{alg:method:montecarlo} and outlined in Figure \ref{fig:method:montecarlo}.

\begin{algorithm}
\caption{Monte Carlo Sampling of Geometrically Uncertain Hemodynamic Solutions}
\begin{algorithmic}
\State medical image volume $\mathbf{X}$
\State vessel pathlines $\mathbf{V}$
\State element size $h$
\State parametric estimator $m_{\boldsymbol{\theta}}$
\State $\mathbf{x}:=\{\mathbf{x}_1,...,\mathbf{x}_n\}\gets extract(\mathbf{X},\mathbf{V})$
\State $\mathcal{U}:=\{\}$
\For{$i=1,...,K$}
\State $\mathbf{z}\sim P(\mathbf{z})$
\State $\mathcal{Y}:=\{\hat{\mathbf{y}}_1,...,\hat{\mathbf{y}}_n\}\gets\{m_{\boldsymbol{\theta}}(\mathbf{x}_1;\mathbf{z}),...,m_{\boldsymbol{\theta}}(\mathbf{x}_n;\mathbf{z})\}$
\State $\mathbf{Y}_i \gets model(\mathcal{Y},\mathbf{V})$	
\State $\mathbf{Y}_i^h \gets mesh(\mathbf{Y}_i,h)$
\State $\mathbf{U}_i^h \gets simulate(\mathbf{Y}_i^h)$
\State $\mathcal{U}\gets \mathcal{U}\cup \mathbf{U}_i^h$
\EndFor
\State return $\mathcal{U}$
\end{algorithmic}\label{alg:method:montecarlo}

\vspace{3pt}

\footnotesize{
$extract(\mathbf{X},\mathbf{V})$ creates a collection of cross-section images by slicing the image $\mathbf{X}$ orthogonal to the vessel pathlines $\mathbf{V}$.\\
$model()$ generates a cardiovascular model by lofting the two-dimensional lumen segments along the vessel pathline, and merging multiple vessels together by boolean union.\\
$mesh(\mathbf{Y}_{i},h)$ generates a tetrahedral mesh of the domain $\mathbf{Y}_{i}$, using an element size $h$ (see details in~\cite{updegrove16a,updegrove2016}).\\
$simulate(\mathbf{Y}^h)$ computes the solution of the Navier-Stokes equations on $\mathbf{Y}^h$, with appropriately chosen initial and boundary conditions.
}

\vspace{0pt}

\end{algorithm}	

\subsection{Cardiovascular Model Construction using Path-Planning}

\begin{figure}[!ht]
	\centering
	\begin{subfigure}[b]{0.23\textwidth}
		\centering\includegraphics[width=\textwidth]{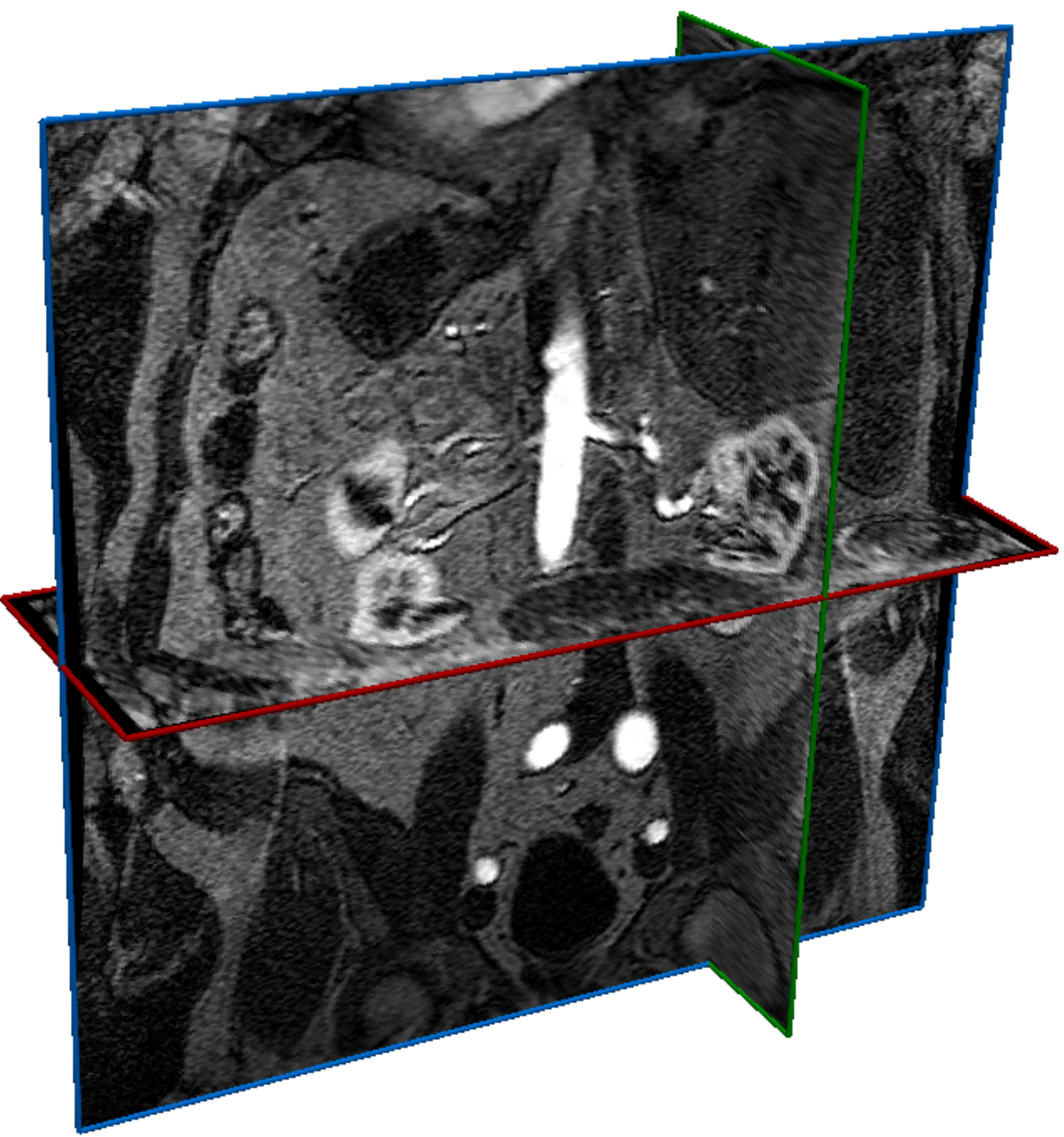}
		\caption{\label{fig:modelconstruction-img}}
	\end{subfigure}
	\begin{subfigure}[b]{0.12\textwidth}
		\centering\includegraphics[width=\textwidth]{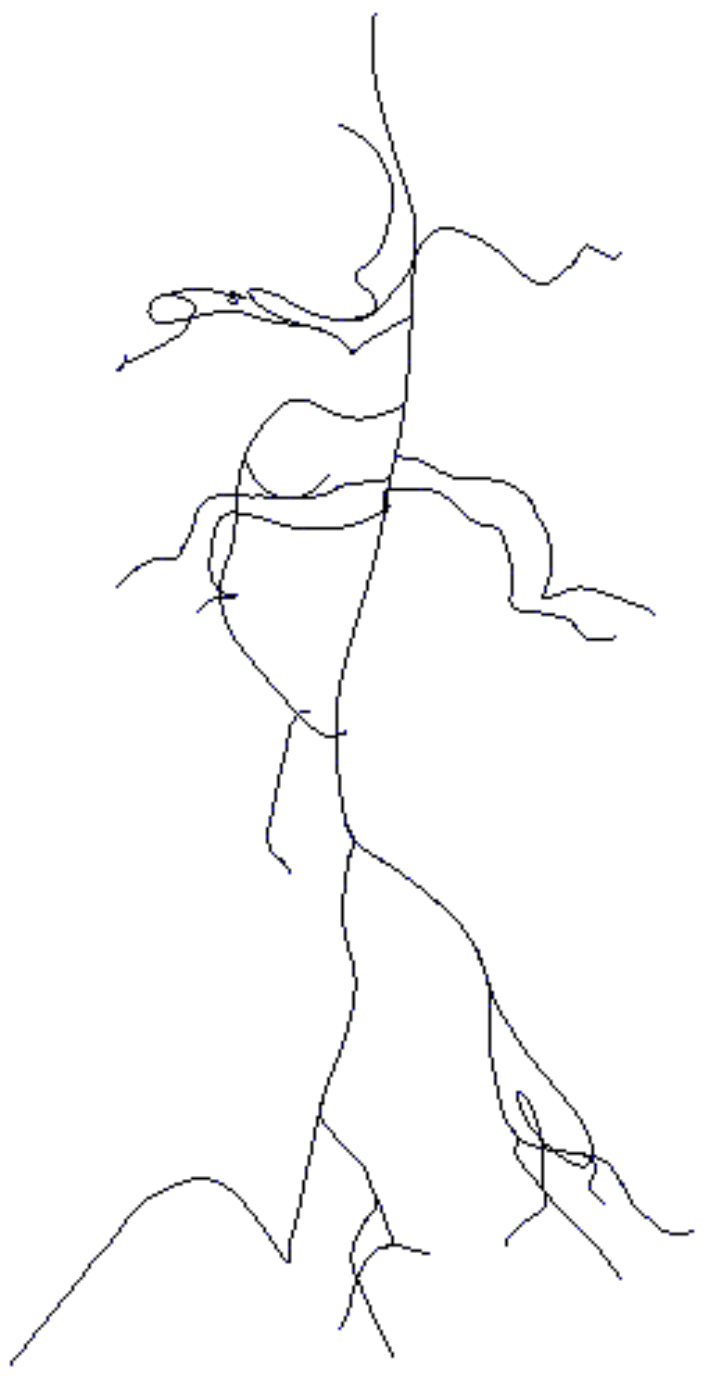}
		\caption{\label{fig:modelconstruction-path}}
	\end{subfigure}
	\begin{subfigure}[b]{0.12\textwidth}
		\centering\includegraphics[width=\textwidth]{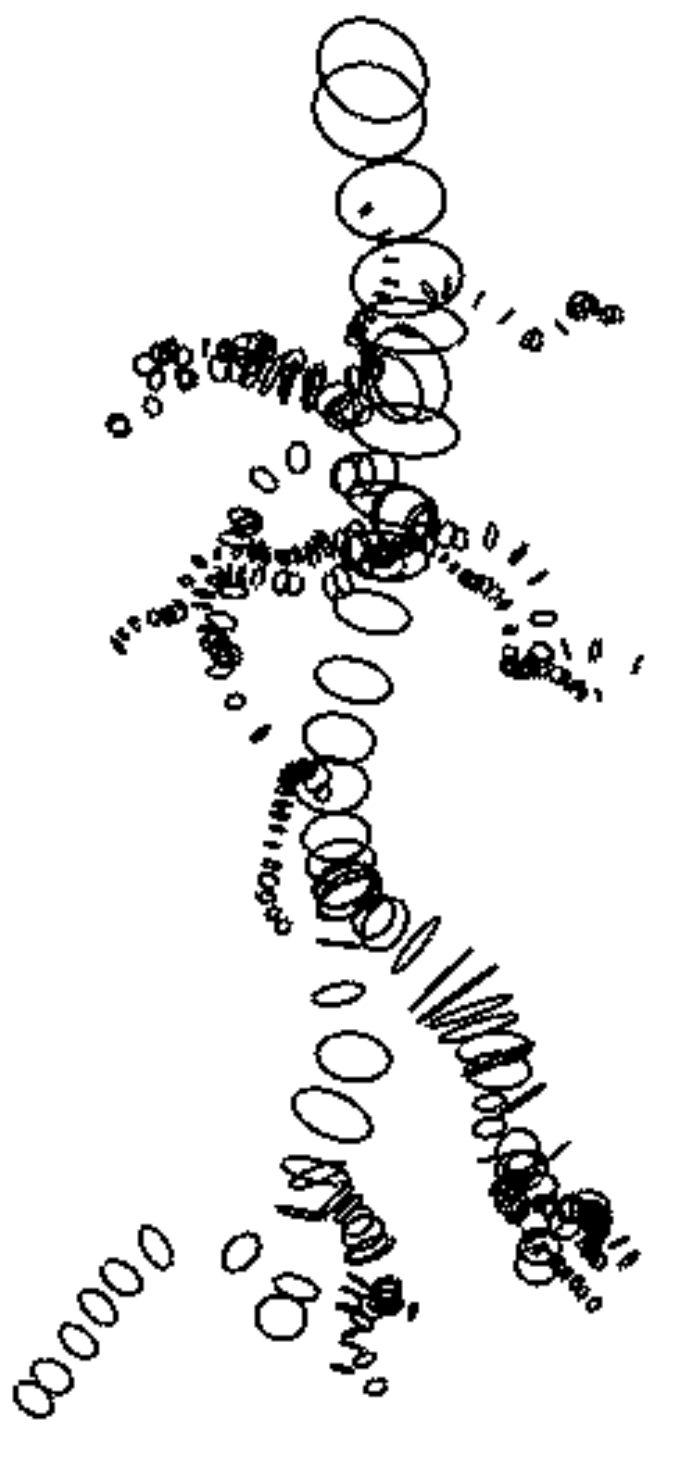}
		\caption{\label{fig:modelconstruction-loft}}
	\end{subfigure}
	\begin{subfigure}[b]{0.12\textwidth}
		\centering\includegraphics[width=\textwidth]{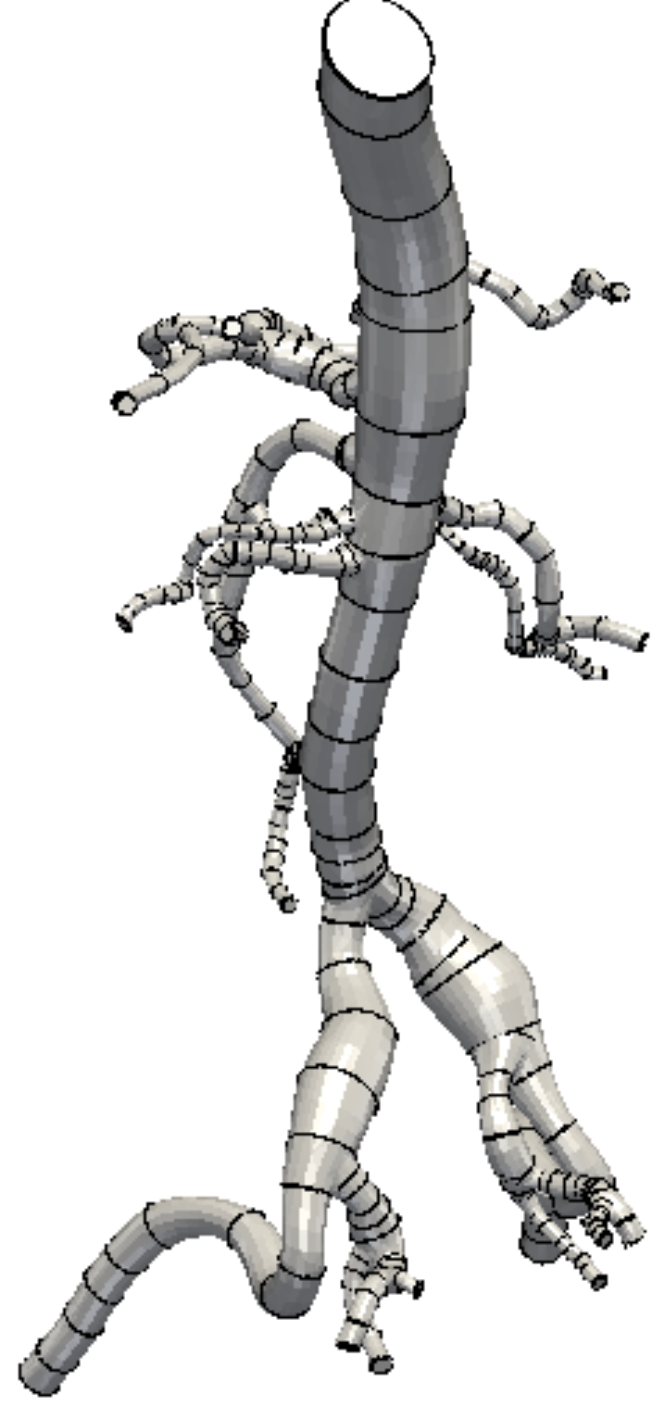}
		\caption{\label{fig:modelconstruction-model}}
	\end{subfigure}
	\begin{subfigure}[b]{0.12\textwidth}
		\centering\includegraphics[width=\textwidth]{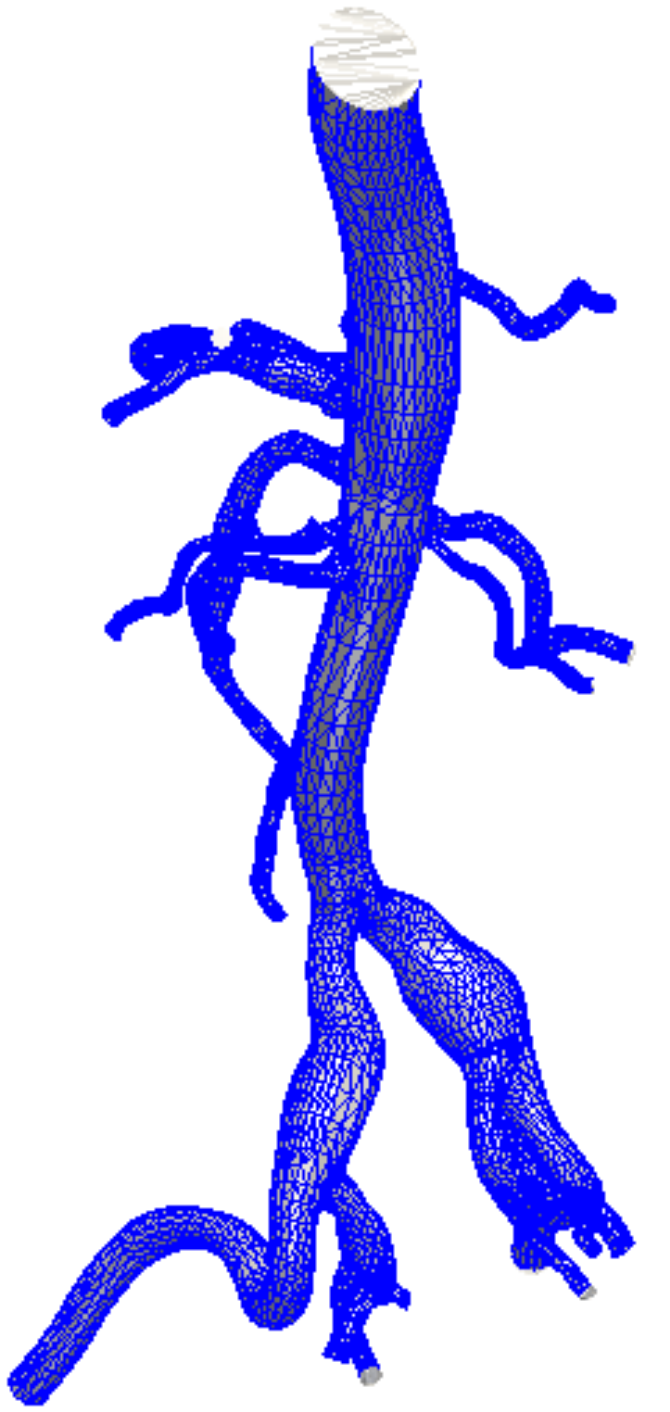}
		\caption{\label{fig:modelconstruction-mesh}}
	\end{subfigure}
	\caption{Cardiovascular model construction workflow used in SimVascular~\cite{updegrove2016}.
		Starting from \textbf{(a)} Image data,
		\textbf{(b)} pathlines are manually generated by the users,
		\textbf{(c)} two-dimensional lumen segmentations are generated at each cross section $\mathbf{x}_{i},\,i=1,\dots,N_{s}$,
		\textbf{(d)} the entire vessel lumen surface is reconstructed by lofting,
		\textbf{(e)} a Boolean union of multiple vessels is meshed to generate a 3D cardiovascular model.}
	\label{fig:modelconstruction}
\end{figure}

\noindent In this section we provide a more formal description of the path-planning process (Figure~\ref{fig:modelconstruction}).
The first input consists of a gray-scale medical image volume with $H,W$ and $D$ voxels in the axial, sagittal and coronal direction, respectively, i.e. $\mathbf{X} \in \mathbb{R}^{H\times W \times D}$.
The second input is a single pathline which consists of a collection of spline segments $\mathbf{V}:=\{v_1(s),...,v_{N_v}(s)\}$. Each segment, $v_i(s):[0,1]\to \mathbb{R}^3,\,i=1,\dots,N_{v}$ is a function that maps the relative length along the segment to three-dimensional locations within the image volume, and is obtained through spline interpolation from a collection of user-specified points locations~\cite{updegrove2016}.
A collection of $N_{s}$ local two-dimensional cross-sectional images of the vessel lumen are then extracted for each pathline at the discrete image space locations $\mathcal{S}:=\{v_{i_1}(s_{1}),...,v_{i_{N_s}}(s_{N_s})\}$ and the local tangent and normal vectors to the pathline at point $v_{i_{j}}(s_k)$ are used to construct a planar grid where the voxel intensities from $\mathbf{X}$ are interpolated, to create a gray-scale cross-sectional image of the vessel lumen $\mathbf{x}_{i},\,i=1,\dots,N_{s}$.
Repeating this process for all selected $N_{s}$ locations along the pathline produces a set of two-dimensional images $\mathbf{x}:=\{\mathbf{x}_1,...,\mathbf{x}_{N_s}\}$.

Realizations from a Bernoulli random vector $\mathbf{z}$ and the images $\mathbf{x}_i,\,i=1,\dots,N_{s}$ constitute the inputs to an artificial neural network designed to produce two-dimensional segmentations of the form
\begin{equation}
\mathbf{y}:=\{\mathbf{y}_1,...,\mathbf{y}_{N_s}\}=\{m_{\boldsymbol{\theta}}(\mathbf{x}_1;\mathbf{z}),...,m_{\boldsymbol{\theta}}(\mathbf{x}_{N_s};\mathbf{z})\},
\end{equation}
at all the $N_{s}$ locations along the pathline. It is important to note how the \emph{same} realization from the Bernoulli vector $\mathbf{z}$ is used to set the dropout layer in the network across all the $N_{s}$ segmentation instances for each anatomical surface realization.
In contrast to generating the segmentations with independent dropout vectors, our process ensures that the same network weights are used to segment all the $\mathbf{x}_{i},\,i=1,\dots,N_{s}$ images, leading to a consistent bias across the whole cardiovascular model geometry.
Finally, the normal and tangent vectors to the pathline are used to re-orient the 2D lumen segments back to image space (Figure~\ref{fig:modelconstruction-loft}), where they are interpolated and joined together~\cite{updegrove16a} to form a triangular surface mesh of the full cardiovascular model~\cite{updegrove2016}. A volumetric finite element Delaunay triangulation is then generated using TetGen~\cite{si15} for finite element analysis.

\subsection{Convolutional Dropout Networks For Lumen Segmentation}

\noindent For the parametric vessel lumen estimator, $m_{\boldsymbol{\theta}}(\mathbf{x};\mathbf{z})$, we use a convolutional neural network.
In particular, $m_{\boldsymbol{\theta}}$ maps the input 2D gray-scale image slice, $\mathbf{x}\in\mathbb{R}^{H\times H}$, to a vector of $K$ normalized radii $\mathbf{y}\in [0,1]^K\subset\mathbb{R}^{K}$.
The radii correspond to the distance of the vessel lumen from the center of the image along rays oriented according to angular intervals $\boldsymbol{\phi}:=\{\phi_1,...,\phi_K\}$. 
This allows the radius $y_{i}^{j}\in\mathbf{y}_{i},\,i=1,\dots,N_{s},\,j=1,\dots,K$ to be converted to a single location in the cross-sectional slice $\mathbf{x}_{i}$ using the expression
\begin{equation}
\mathbf{p}_{i}^{j} = \left( y_{i}^{j}\,H\,\cos\phi_j, \hat{y}_{i}^{j}\,H\,\sin\phi_j \right),\,i=1,\dots,N_{s},\,j=1,\dots,K,
\end{equation}
and the full lumen segmentation from image slice $\mathbf{x}_{i}$ in the set of points $\mathbf{p}_{i}:=\{\mathbf{p}^{1}_{i},...,\mathbf{p}^{K}_{i}\}$.
Even though the literature has witnessed an explosion in new layouts and arrangements in recent years \cite{lecun_deep_2015, schmidhuber_deep_2015}, a CNN generally consist of a collection of layers, each applying a mathematical operation, such as a linear transformation or convolution, followed by an elementwise nonlinear activation. 
In our case, the transformation in layer $l$ is expressed as
\begin{equation}
\mathbf{o}^{(l)} = m^{(l)}(\mathbf{a}^{(l-1)};\boldsymbol{\Theta}^{(l)}),\,\,
\mathbf{a}^{(l)} = g^{(l)}(\mathbf{o}^{(l)}),
\end{equation}
where $m^{(l)}(\cdot)$ is the specific mathematical transformation occurring through layer $l$, $\mathbf{a}$ represents the generic input vector from the previous layer and output to the next one, and $g^{(l)}(\cdot)$ the selected non linear activation. 
The learnable parameters for the $l$-th layer are denoted by $\boldsymbol{\Theta}^{(l)}\subset\boldsymbol{\theta}$.
In this study, we employ a CNN combining dense and convolutional layers. 
Dense layers operate on vector inputs and outputs through the linear transformation
\begin{equation}
\mathbf{o}^{(l)} = \boldsymbol{\Theta}^{(l)}\,\mathbf{a}^{(l-1)} + \mathbf{b}^{(l)},
\end{equation}
where $\boldsymbol{\Theta}^{(l)}$ and $\mathbf{b}^{(l)}$ are a weight matrix and bias term, respectively. The convolutional layers instead transform a third order tensor input using
\begin{equation}
\mathbf{o}^{(l)}_{ijk} = \sum_o \sum_p \sum_q \boldsymbol{\Theta}_{opqk}^{(l)}\,\mathbf{a}_{i+o,j+p,q}^{(l-1)} + \mathbf{b}^{(l)},
\end{equation}
where $\boldsymbol{\Theta}^{(l)}$ is a fourth order tensor of trainable weights. 
Note how the output from convolutional layers are flattened into one-dimensional vectors before being fed to dense layers.

The activation functions $g^{(l)}(\cdot)$ allow the neural network to learn nonlinear relationships in the data~\cite{hornik91} and determine the types of output it can produce. Since the output in our case is a vector of radii in $[0,1]$ we use the elementwise sigmoid activation function
\begin{equation}
g^{(l)}(x) = \frac{1}{1+e^{-x}}.
\end{equation}
For the intermediate layers we instead use Leaky Rectified Linear Units (Leaky-RELU) because they avoid the problem with \emph{vanishing gradients} when optimizing the network weights using gradient-descent \cite{maas13}
\begin{equation}
g^{(l)}(x) = \begin{cases}
x, & x > 0 \\
\alpha\cdot x, & x\leq 0.
\end{cases}
\end{equation}
We augment our convolutional network to sample from the distribution of vessel lumens for a given image by adding dropout layers to the network, which sets the outputs of the previous layer to zero through Hadamard (elementwise) products by a vector of Bernoulli random variables $\mathbf{z}^{(l)}$. In practice, this is implemented as
\begin{align}
\mathbf{a}^{(l)} &= m^{(l)}_\theta(\mathbf{o}^{(l-1)}) \\
\mathbf{o}^{(l)} &= \mathbf{a}^{(l)} \odot \mathbf{z}^{(l)},\ \mathbf{z}^{(l)}\sim \mathbf{B}(1-p) \\
\mathbf{a}^{(l+1)} &= m^{(l+1)}_\theta(\mathbf{o}^{(l)}), 
\end{align}
where $l$ denotes the layer number after which a dropout layer is applied, $\mathbf{B}$ is a multivariate distribution with indepenent Bernoulli components, and $p$ the selected dropout probability. 
The inclusion of dropout layers induce stochasticity to the vessel lumen segmentation process, resulting in random collections of points $\mathbf{y}$ obtained as
\begin{equation}
\mathbf{y} = m_{\boldsymbol{\theta}}(\mathbf{x};\mathbf{z}),\ \mathbf{z}\sim \mathbf{B}(1-p),
\end{equation}
where $\mathbf{z}$ is a vector containing all Bernoulli dropout variables throughout the network.
As discussed above, these variables are kept the same for every two-dimensional segmentation in a single cardiovascular model instance. 

\begin{figure}[!ht]
	\centering
	\includegraphics[scale=0.4]{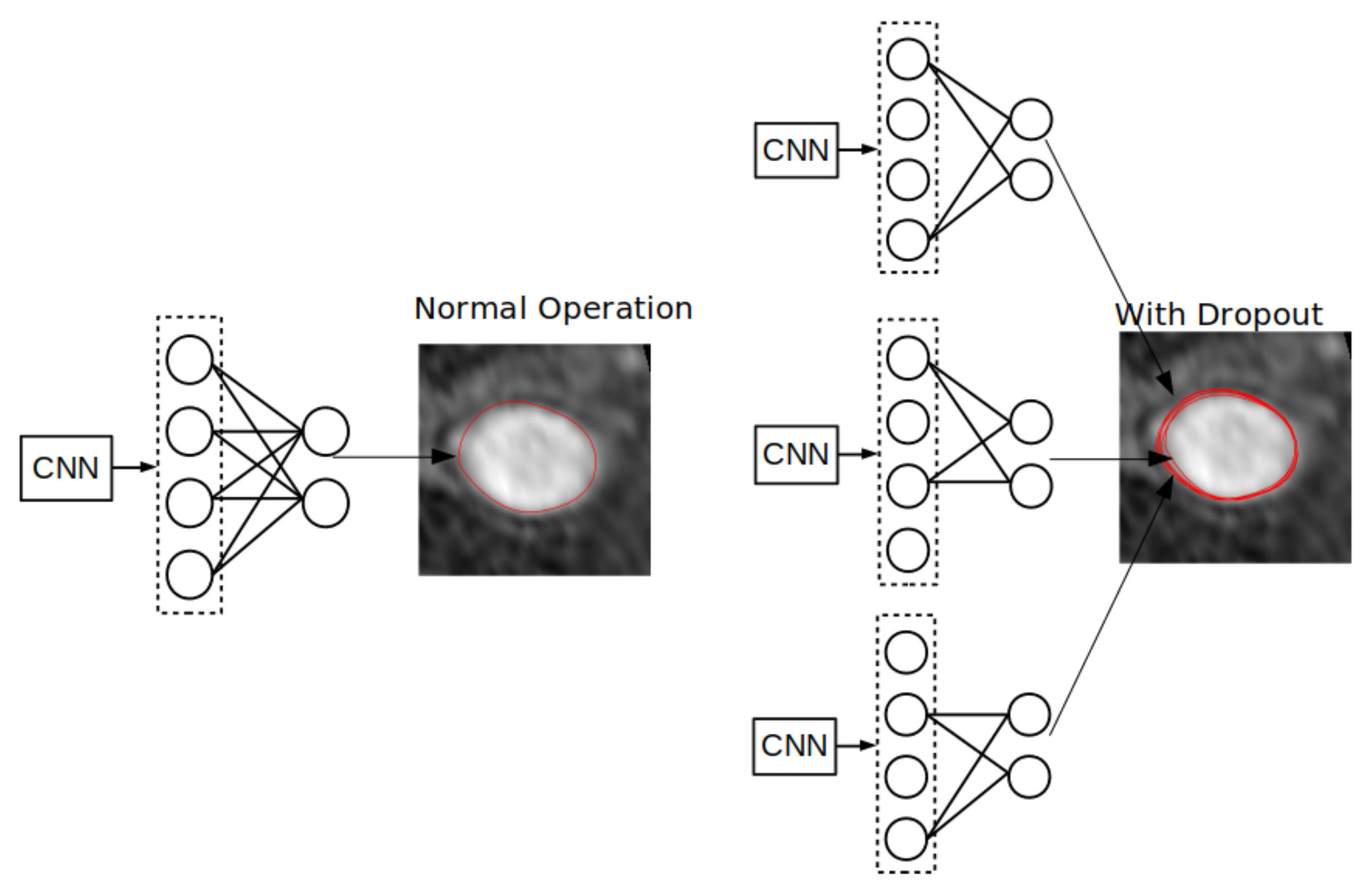}
	\caption{CNN based vessel lumen regression with and without dropout sampling.}
	\label{fig:method:dropout}
\end{figure}

For the network architecture, we build on our previous work \cite{maher_20} and use the GoogleNet architecture~\cite{szegedy_15} appropriately modified for vessel lumen regression, which consists of a CNN encoder followed by fully-connected layers to transform the encoded vector into the vessel lumen space. A dropout layer is applied to the output of the penultimate layer in the network in order to inject stochasticity (see Figure~\ref{fig:method:dropout}). 
A GoogleNet architecture was selected for computational efficiency, as the proposed algorithm is distributed as a SimVascular plug-in, targeting users without specialized hardware.
The GoogleNet architecture is computationally efficient due to the use of convolutional and pooling layers with different dimensions to compress the input image while still retaining necessary input information.
For more details the reader is referred to \cite{szegedy_15}.
While more recent networks have been developed, earlier studies we conducted \cite{maher_20} showed that the GoogleNet network achieved accuracy comparable to human experts on a 2D vessel lumen segmentation task and so is sufficient for the purposes of this work.
%
%

\subsubsection{Network Training}\label{sec:training}

\noindent During training, the network's weights are initialized according to the variance-scaling approach discussed in~\cite{he_15} and optimized using stochastic gradient-descent ADAM algorithm \cite{adam}.
We apply the angular distance transform to each ground-truth lumen surface $\mathbf{p}^{1}_i,\dots,\mathbf{p}^{K}_i$ to transform it to a ground-truth vector $\mathbf{y}_i$ and create a training dataset of size $N_{d}$ consisting of the following collection of image-radii pairs
\begin{equation}
\mathcal{D}=\{(\mathbf{x}_1,\mathbf{y}_1),...,(\mathbf{x}_{N_d},\mathbf{y}_{N_d})\}.
\end{equation}
Additionally, we employ a $l_2$ loss of the form 
\begin{equation}
l(\mathbf{y}_{i},\widehat{\mathbf{y}_{i}}) = \sqrt{\sum_{j=1}^{K} (y^{j}_{i}-\widehat{y}^{j}_{i})^2},\,\,L(\mathbf{x},\mathbf{y};\boldsymbol{\theta},\mathbf{z},N_{b}) = \frac{1}{N_b}\,\sum_{i=1}^{N_b}\,l(\mathbf{y}_i,\widehat{\mathbf{y}}_i) = \frac{1}{N_b}\,\sum_{i=1}^{N_b}\,l[\mathbf{y}_i,m_{\boldsymbol{\theta}}(\mathbf{x}_i;\mathbf{z})],
\end{equation}
where $\mathbf{y}_i$ and $\widehat{\mathbf{y}}_i = m_{\boldsymbol{\theta}}(\mathbf{x}_i;\mathbf{z})$ represent the $i$-th ground-truth collection of normalized lumen radii and neural network prediction, respectively, while $L(\mathbf{x},\mathbf{y};\boldsymbol{\theta},\mathbf{z},N_{b})$ represents the expected loss over a given \emph{batch} of training examples.
We also pre-process each input gray-scale image $\mathbf{x}_{i}\in\mathbb{R}^{H\times W}$, by computing a normalized image $\widetilde{\mathbf{x}}_{i}$ having zero mean and unit variance pixel intensities expressed as
\begin{equation}
\widetilde{{x}}^{jk}_{i} = \frac{x^{jk}_{i}-\mu_x}{\sigma_x},\,j=1,\dots,H,\,k=1,\dots,W,\,\,\mu_x = \frac{1}{H^2}\,\sum_{j=1}^H\sum_{k=1}^W x^{jk}_{i},\,\,\sigma_x = \left(\frac{1}{H^2}\,\sum_{j=1}^H\sum_{k=1}^W(x^{jk}_{i}-\mu_x)^2\right)^{1/2},
\end{equation}
where $\mu_x$ and $\sigma_x$ are the mean and standard deviation pixel intensities, respectively.
Finally, the training dataset $\mathcal{D}$ is augmented by randomly rotating and cropping each pair of image slice and 2D vessel lumen segmentation.

\subsubsection{Dataset}

\noindent Our dataset consists of 50 CT and 54 MR contrast-enhanced 3D medical image volumes, all publicly available from the Vascular Model Repository (VMR) \footnote{\url{http://www.vascularmodel.com}}~\cite{wilson13}.
For each image volume, the VMR contains vessel pathlines, segmentations, 3D patient-specific models and hemodynamic simulation results (see Figure~\ref{fig:modelconstruction}) created in SimVascular by expert users, in many cases with supervision from a radiologist.
To avoid anisotropic voxel spacing, all image volumes were re-sampled keeping an isotropic voxel spacing of 0.029 cm, which ensures the largest vessel diameter to be around 100 pixels, a relatively small window size which reduces the network computation and memory requirements.
Specifically, we used a window size $H\times W$ of $160\times160$ pixels to allow the full range of vessel sizes to be represented with sufficient resolution by each two-dimensional slice.
Finally, we split the data into training, validation and testing sets, of 86, 4 and 14 volumes, respectively. This resulted in 16004, 239 and 6317 cross-sectional images and vessel lumen surface point labels for the training, validation and testing sets, respectively.

\subsection{Patient-Specific Hemodynamics Simulations}\label{section:ns}

\noindent The cardiovascular model generation process discussed above results in three-dimensional tetrahedral meshes which provide a domain $\Omega\subset\mathbb{R}^{3}$ where the incompressible Navier-Stokes equations are solved.
These are the equations describing the evolution in time of a Newtonian fluid of constant density $\rho$, in a domain $\Omega$ with boundary $\partial \Omega =\Gamma _{D}\cup \Gamma _{N}$, partitioned according to the application of Dirichlet and Neumann boundary conditions, respectively
\begin{equation}\label{equ:NS}
\begin{cases}
\rho\,\dfrac{\partial{\boldsymbol{u}}}{\partial t} + \rho ({\boldsymbol {u}}\cdot \nabla ){\boldsymbol {u}}-\nabla \cdot {\boldsymbol {\tau }}={\boldsymbol {f}}&{\text{ in }}\Omega \times [0,T]\\\nabla \cdot {\boldsymbol {u}}=0&{\text{ in }}\Omega \times [0,T]\\{\boldsymbol {u}}={\boldsymbol {g}}&{\text{ on }}\Gamma _{D}\times [0,T]\\\tau\cdot{\boldsymbol {\widehat {n}}}={\boldsymbol {h}}&{\text{ on }}\Gamma _{N}\times [0,T]\\{\boldsymbol {u}}(0)={\boldsymbol {u}}_{0}&{\text{ in }}\Omega \times \{0\},
\end{cases}
\end{equation}
in which $\boldsymbol {u}$ is  fluid velocity, $p$ is the fluid pressure, $\boldsymbol {f}$ is a given forcing term, $\boldsymbol{\widehat{n}}$ is the outward directed unit normal vector to $\Gamma _{N}$, and $\boldsymbol{\tau}$ the viscous stress tensor defined as $\boldsymbol{\sigma}=-p\,\boldsymbol{I} + 2\,\mu\,\boldsymbol{\epsilon}(\boldsymbol {u})$.
Let $\mu$ be the dynamic viscosity of the fluid, $\boldsymbol {I}$ the second order identity tensor and $\boldsymbol{\epsilon}(\boldsymbol{u})$ the strain-rate tensor defined as $\boldsymbol{\epsilon}(\boldsymbol{u})=\frac {1}{2}(\nabla{\boldsymbol{u}}+\nabla{\boldsymbol{u}}^{T})$.
The functions $\boldsymbol{g}$ and $\boldsymbol{h}$ are given Dirichlet and Neumann boundary data, while $\boldsymbol{u}_{0}$ is the initial condition.

We numerically solve the system~\eqref{equ:NS} using a Streamline Upwind Petrov Galerkin (SUPG) finite element method implemented in the SimVascular flow solver (svSolver)~\cite{brooks82}, which contains specialized routines for cardiovascular CFD such as backflow stabilization~\cite{moghadam11}, algebraic system solvers and preconditioners~\cite{seo19} and a large collection of physiologic boundary conditions (see, e.g.,~\cite{moghadam11,moghadam13,moghadam15,seo19}). The numerical solution is integrated in time using a second-order generalized-$\alpha$ method~\cite{jansen00}.
We also apply RCR boundary conditions for generic outlets and  a coronary lumped parameter boundary condition for coronary artery outlets, respectively (see, e.g.,~\cite{moghadam13,clementel13,clementel10}).
Finally, we restrict our attention to simulations with rigid walls.

\subsection{Monte Carlo Sampling of Cardiovascular Flow Solutions}

Our model generation procedure generates a set of discrete meshes $\mathbf{Y}^h:=\{\mathbf{Y}_1^h,...,\mathbf{Y}_{N_y}^h\}$ (here the superscript $h$ is used to indicate the size of the discrete mesh). Numerical solution of the Navier-Stokes equations on each mesh subsequently produces a set of velocity fields $\mathbf{U}^h:=\{\mathbf{U}_1^h,...,\mathbf{U}_{N_y}^h\}$ and pressure fields $\mathbf{P}^{h}:=\{\mathbf{P}_1^h,...,\mathbf{P}_{N_y}^h\}$ with $\mathbf{U}_i^h : \mathbf{Y}_i\times[0,T] \to \mathbb{R}^3$ and $\mathbf{P}_i^h: \mathbf{Y}_i\times[0,T] \to \mathbb{R}$.

Our objective is to calculate relevant Monte Carlo (i.e., sample) statistics using the ensembles $\mathbf{U}^{h}$ and $\mathbf{P}^{h}$. However, the mesh geometry and hence the solution domain $\Omega$ is not constant for different realizations of the flow and pressure fields. This precludes us from considering quantities of interest defined at specific point locations in $\Omega$, and we focus instead on output quantities that do hold meaning in the context of varying geometry. 
Consider a generic model result $\mathbf{r}_{i}(\boldsymbol{\omega},t)$ for the $i$-th geometry realization, and the cross-sectional area $A^{j}_i$ corresponding to the $j$-th slice location. We define the quantity
\begin{equation}\label{eq:experiment:crossavg}
q^{j}_i(t) = \frac{1}{|A^{j}_i|}\int_{A^{j}_i}\,r_i(\boldsymbol{\omega},t)\,d\Gamma,\,\,\text{or the quantity}\,\,q^{j}_i(t) = \frac{1}{|\partial A^{j}_i|}\int_{\partial A_i(s)}\,r_i(\boldsymbol{\omega},t)\,d\Gamma,
\end{equation}
for situations where model outputs are only defined on the lumen surface $\partial A^{j}_i$ (e.g., wall shear stress). The second type of quantities are time-average versions of the $q^{j}_i(t)$, such as,
\begin{equation}
q^{j}_i = \frac{1}{(T_2-T_1)}\int_{T_1}^{T_2}\,q^{j}_i(t)\,dt.
\label{eq:experiment:timeavg}
\end{equation}
In this study, we focus on quantities $\mathbf{r}_{i}(\boldsymbol{\omega},t)$ such as the pressure as well as the wall shear stress and velocity magnitudes.
Finally, for our Monte Carlo trials we choose to report a relative measure of variability, i.e., the coefficient of variation, defined as
\begin{equation}
\text{CoV} = \frac{\sigma}{\mu},
\end{equation}
where $\sigma$ and $\mu$ are the sample mean and the standard deviation of the quantities of interest $q^{j}_i(t)$ or $q^{j}_i$, respectively.

We further report confidence intervals for our Monte Carlo estimates.
By the Central Limit Theorem the sample mean tends to a Gaussian random variable with standard deviation
\begin{equation}
    \tilde{\sigma} = \frac{\sigma}{\sqrt{N}},
\end{equation}
where $N$ is the sample size.
In many cases the absolute value of the sample mean is small, therefore, for the sake of clarity, we report the confidence interval as a percentage of the sample mean, that is
\begin{equation}
    CI = \pm \alpha\frac{\tilde{\sigma}}{\mu},
\end{equation}
where $\alpha$ is a constant depending on the confidence level (e.g. $\alpha=2$ for $95\%$ confidence).
The CoV is thus an estimate of the variability of a particular QOI and $CI$ is a measure of the accuracy of our sample mean estimates.
\subsection{Shape Variability Assessment Through Principal Component Analysis}

\noindent To better understand the shape variation in our cardiovascular model samples we use the Principal Component Analysis (PCA) algorithm to compute a low rank factorization of the matrix $X\in \mathbb{R}^{M \times N}$ constructed from the samples. 
In particular we compute
\begin{equation}
X-\bar{X} = U_K\Sigma_K V_K^T,
\end{equation}
where $\bar{X}$ is the mean of the model sample, $\Sigma_K$ is a diagonal matrix containing the singular values on its diagonal and the columns of $U\in \mathbb{R}^{M \times K}$ represent a reduced-order basis for the deviation of the cardiovascular models from the mean model. 
%
%
We use $U$ to study the modes of variation in our cardiovascular model samples. 

By the properties of the PCA factorization, the columns of $U_K$ are ordered starting with the modes that capture the most variance of $X$. 
The first column thus represents the most significant mode, the second column the second most significant mode etc. revealing the dominant modes in which segmentation uncertainty is causing the cardiovascular model geometry to vary.

%% file: experiment.tex
\section{Demonstration in Selected Cardiovascular Anatomies}\label{section:paper3:experiments}

\subsection{Aorto-iliac Bufircation Model}

\begin{figure}[h!]
\centering
\begin{subfigure}[T]{0.4\textwidth}
\begin{subfigure}[T]{1.0\textwidth}
\centering
\begin{tabular}{lccc}
\hline
Vessel & $R_1$ & $C$ & $R_2$ \\
\hline
aorta & 237.0 & 0.00115 & 2370 \\
left iliac & 376.0 & 0.00085 & 3760 \\
\hline
\end{tabular}
\caption{RCR boundary condition parameters. Resistance in $dyne\cdot s/cm^5$, capacitance in $cm^5/dyne$.}\label{tab:experiment:0110rcr}
\end{subfigure}
\begin{subfigure}[T]{1.0\textwidth}
\centering

\vspace{10pt}

\includegraphics[width=0.9\textwidth]{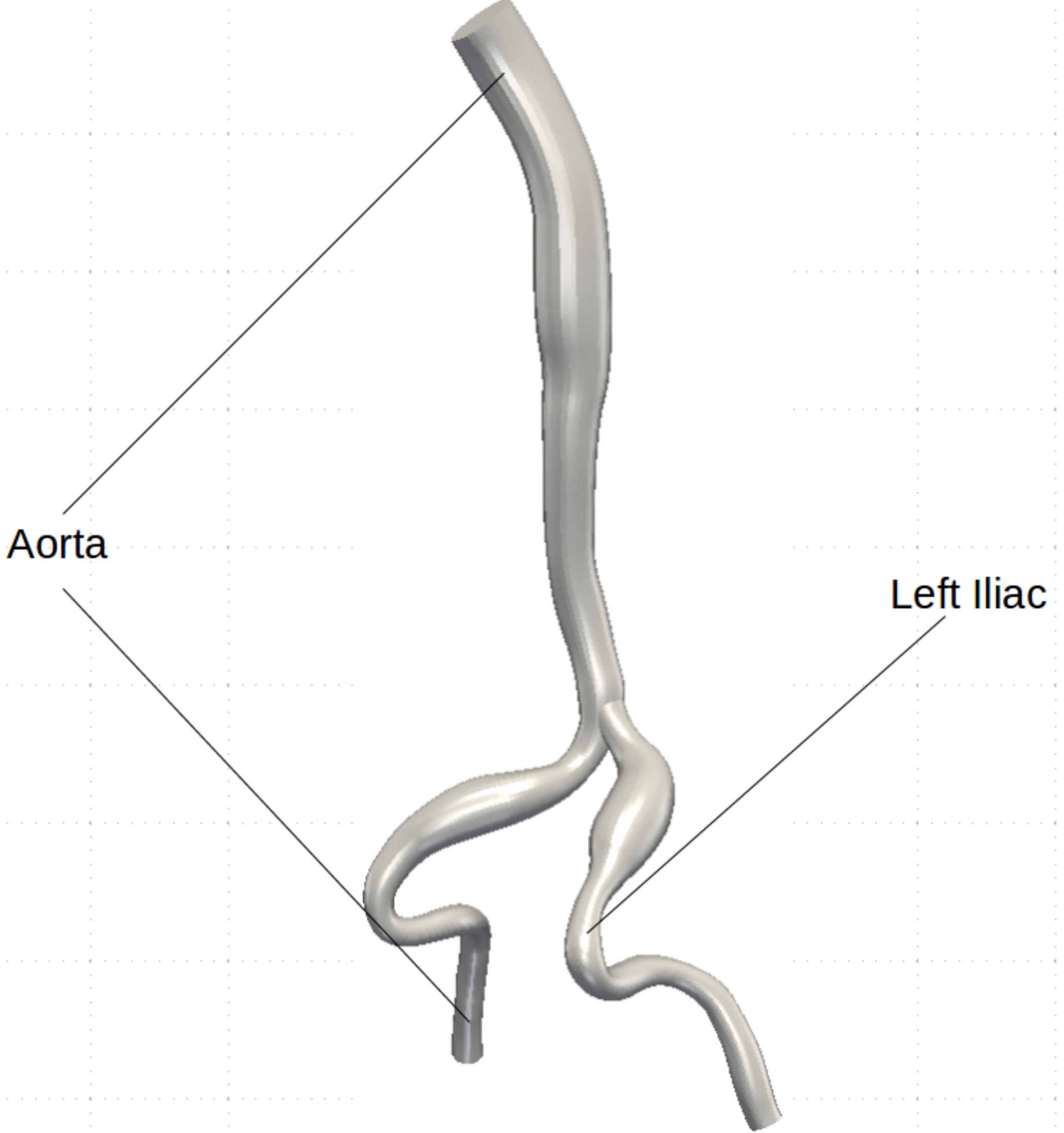}
\caption{Patient-specific model surface.}\label{fig:exp:aortailiac}
\end{subfigure}
\end{subfigure}
\begin{subfigure}[T]{0.55\textwidth}
\begin{subfigure}[T]{0.45\textwidth}
\centering
\includegraphics[width=\textwidth]{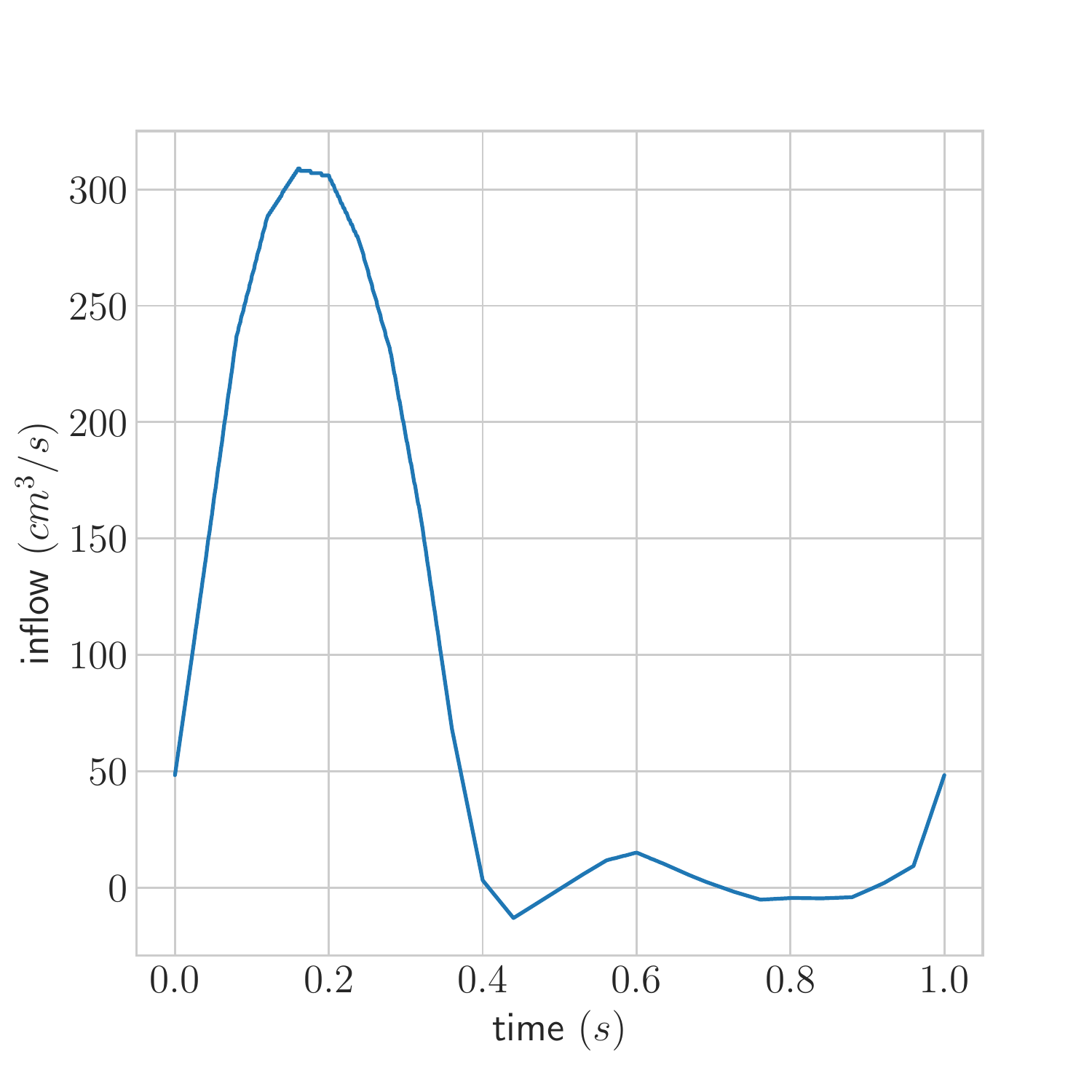}
\caption{Aortic inlet waveform}
\label{fig:exp:aortainflow}
\end{subfigure}
\begin{subfigure}[T]{0.45\textwidth}
\centering
\includegraphics[width=\textwidth]{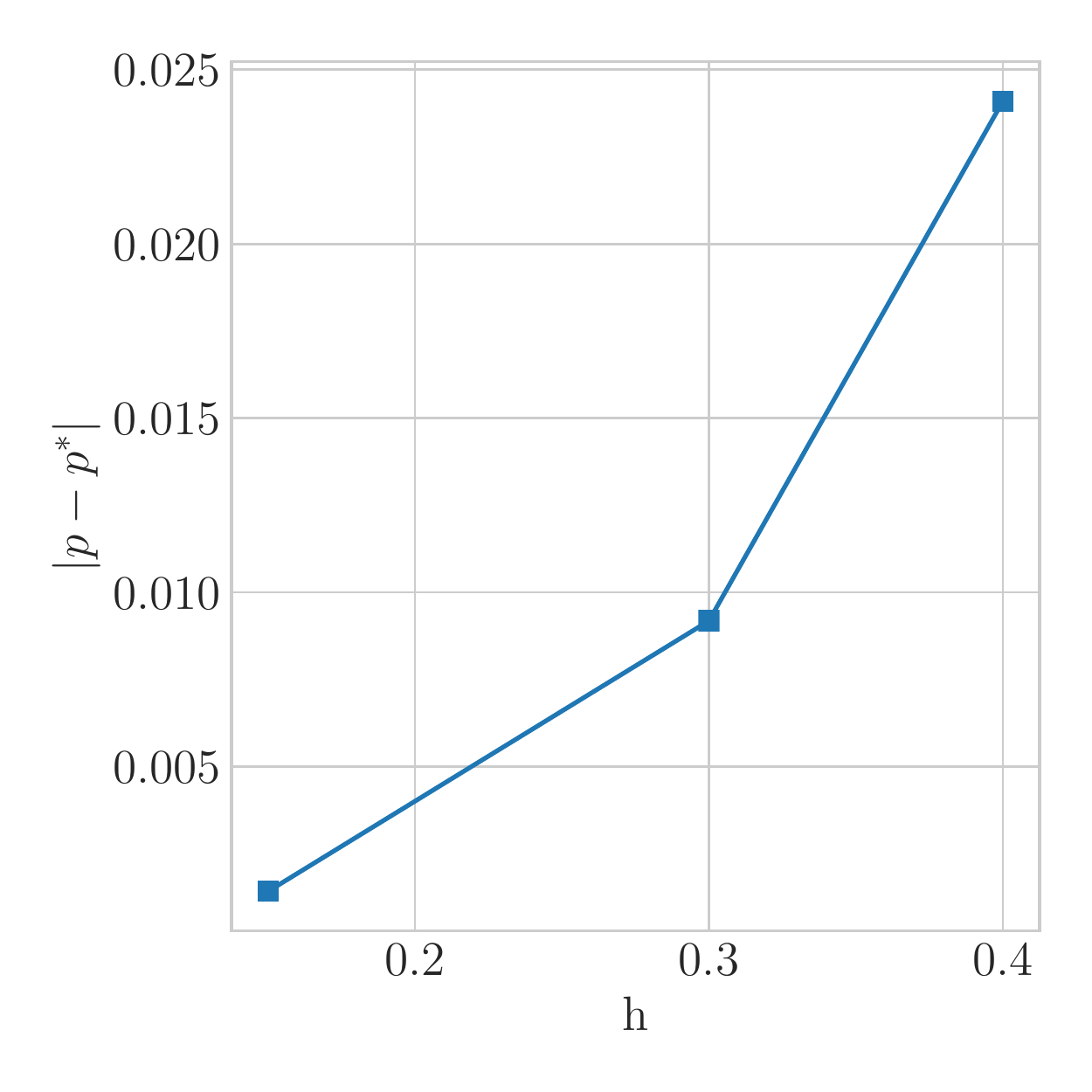}
\caption{Pressure convergence study.}
\label{fig:experiment:0110pressure}
\end{subfigure}

\begin{subfigure}[T]{0.45\textwidth}
\centering
\includegraphics[width=\textwidth]{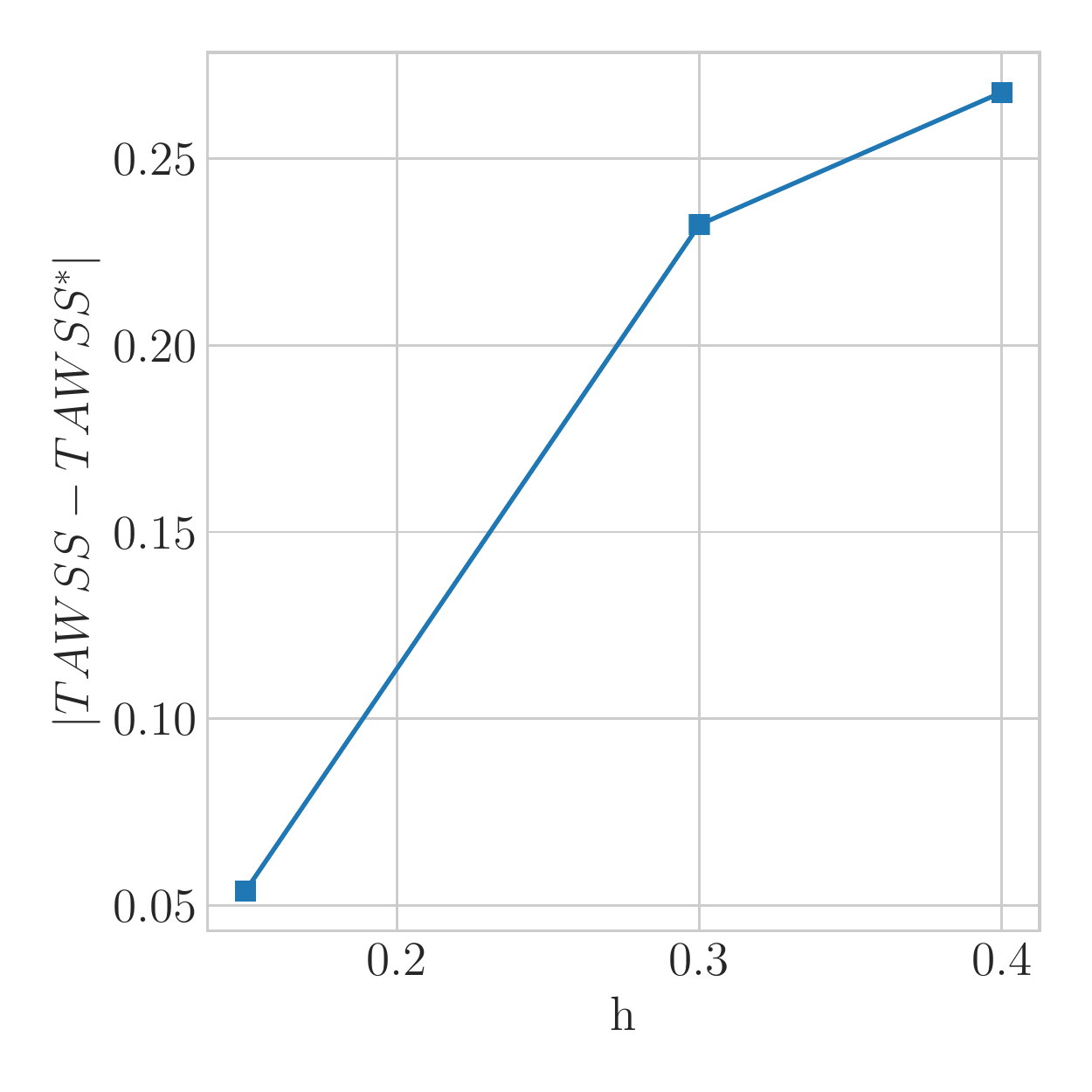}
\caption{WSS convergence study.}
\label{fig:experiment:0110wss}
\end{subfigure}
\begin{subfigure}[T]{0.45\textwidth}
\centering
\includegraphics[width=\textwidth]{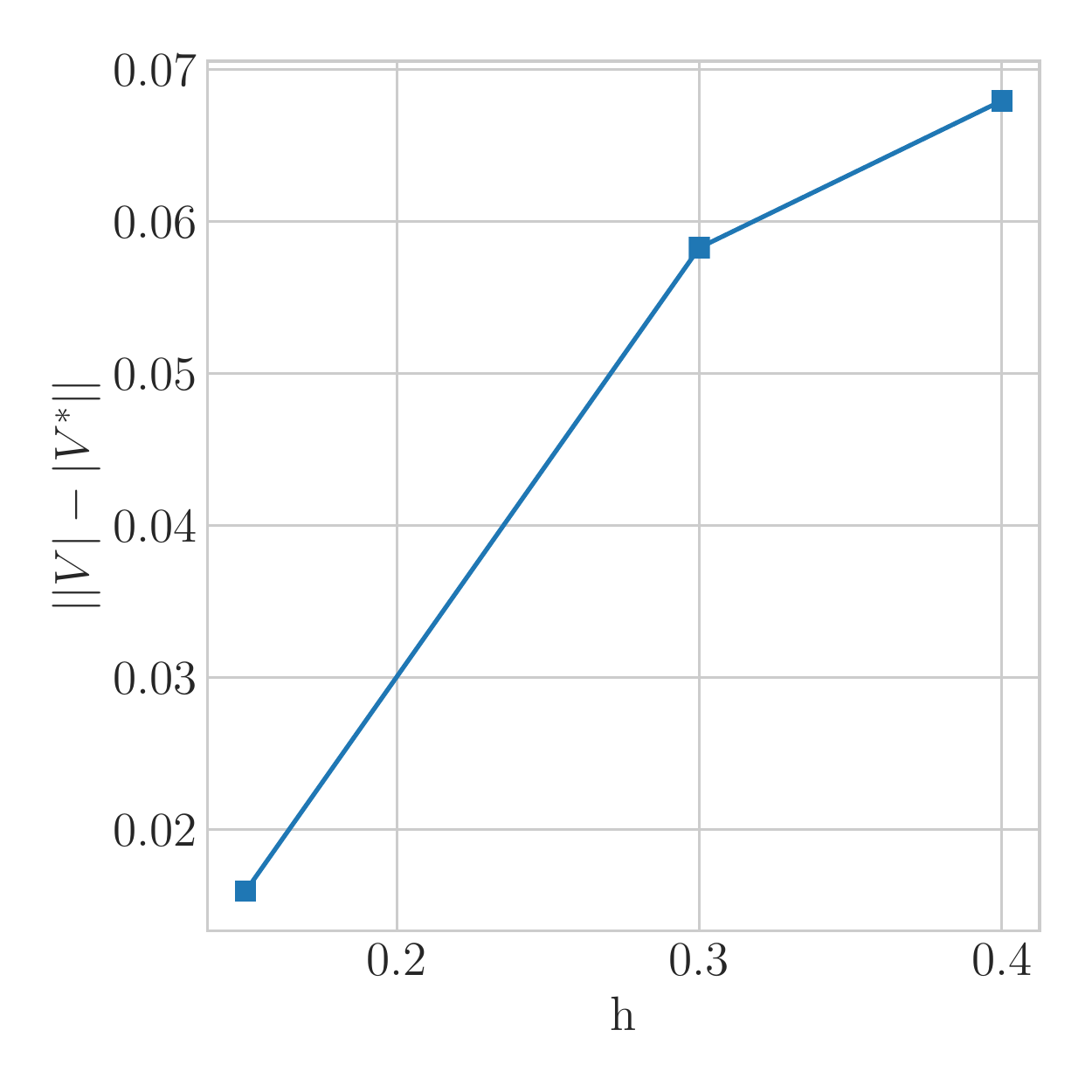}
\caption{Velocity convergence study.}
\label{fig:experiment:0110velocity}
\end{subfigure}
\end{subfigure}
\caption{Aorto-iliac bifurcation model with boundary conditions (a,c), lumen surface (b) and mesh convergence analysis (d,e,f).}\label{fig:twodmatern}
\end{figure}

\noindent The first anatomy we consider consists of the bifurcation of the abdominal aorta in the two iliac arteries, presented through a model having one inlet and two outlets (see Figure~\ref{fig:exp:aortailiac}).
The inlet boundary condition is chosen to be a typical physiological waveform, corresponding to an average inflow of 6 L/min (see Figure~\ref{fig:exp:aortainflow}), while outflow RCR boundary conditions are applied, where the resistance and compliance parameters were preliminarily tuned to produce a realistic outlet pressure range of 80-120 mmHg (see Table~\ref{tab:experiment:0110rcr}).
Initially, we conducted a mesh convergence study with meshes comprising $100,000$, $250,000$ and $1,500,000$ tetrahedral elements and boundary layer mesh with 5 layers, and compared these to a reference mesh with $3,500,000$ elements.
Mesh convergence was assessed by first time-averaging the QOI and then using the mean absolute error
\begin{equation}\label{eq:mae}
\epsilon^h = \frac{1}{|\Omega|}\intt{\Omega}{}{|\bar{f}^h(\boldsymbol{\xi})-\bar{f}^*(\boldsymbol{\xi})|}{\Omega},
\end{equation}
where $\bar{f}^h$ and $\bar{f}^*$ are the time-averaged QOI on the investigated and reference mesh respectively.
The $1,500,000$ element mesh showed a less than $0.2\%$, $6\%$ and $1.7\%$ error for the pressure, WSS magnitude and velocity magnitude respectively (see Figures~\ref{fig:experiment:0110pressure}, \ref{fig:experiment:0110wss} and \ref{fig:experiment:0110velocity}) and is employed in all additional experiments.
The size of the Monte Carlo ensemble is finally selected equal to $150$ for this anatomy. 

\subsection{Abdominal Aortic Aneurysm model}

\begin{figure}[!ht]
\centering
\begin{subfigure}[T]{1.0\textwidth}
\centering
\begin{subfigure}[c]{0.38\textwidth}
\centering
\includegraphics[width=\textwidth]{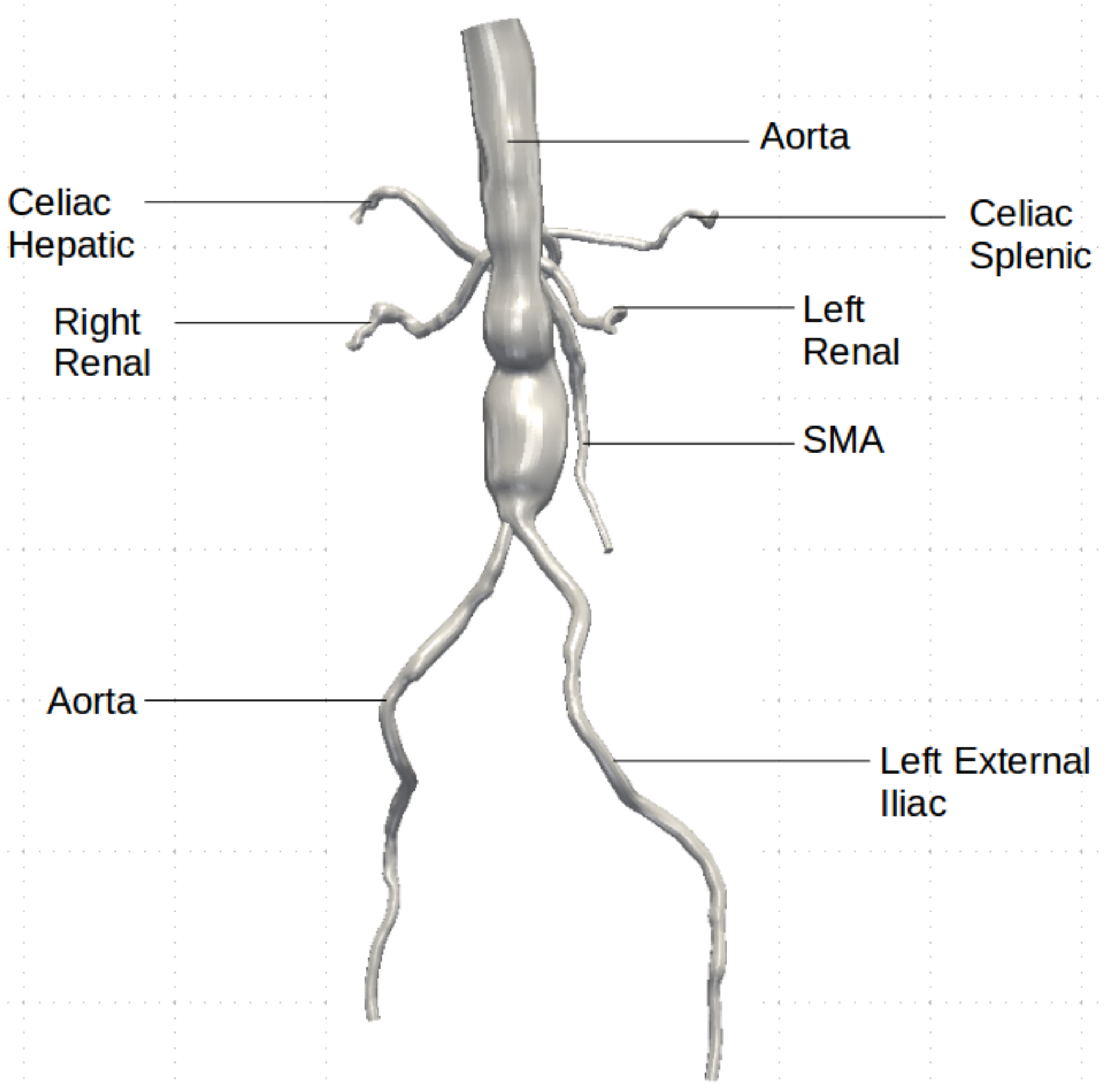}
\caption{Patient-specific AAA model.}\label{fig:experiment:0144}
\end{subfigure}
$\quad\quad$
\begin{subfigure}[c]{0.38\textwidth}
\centering
\begin{tabular}{lccc}
\hline
Vessel & $R_1$ & $C$ & $R_2$ \\
\hline
Aorta & 660 &  0.0004 & 6600 \\
Celiac Hepatic & 1600 & 0.00017 & 16000  \\
Celiac Splenic & 910 & 0.0003 & 9100\\
Ext Iliac Left & 1155 & 0.00024 & 11550 \\
Renal Left & 691 & 0.00039 & 6910\\
Renal Right & 1220 & 0.00022 & 12200\\
SMA & 1040 & 0.00026 & 10400\\
\hline
\end{tabular}
\caption{RCR boundary conditions for AAA model, resistance in $dyne\cdot s/cm^5$, capacitance in $cm^5/dyne$.}\label{tab:experiment:tab0144rcr}
\end{subfigure}
\end{subfigure}

\begin{subfigure}[T]{1.0\textwidth}
\centering
\begin{subfigure}[T]{0.3\textwidth}
\centering
\includegraphics[width=\textwidth]{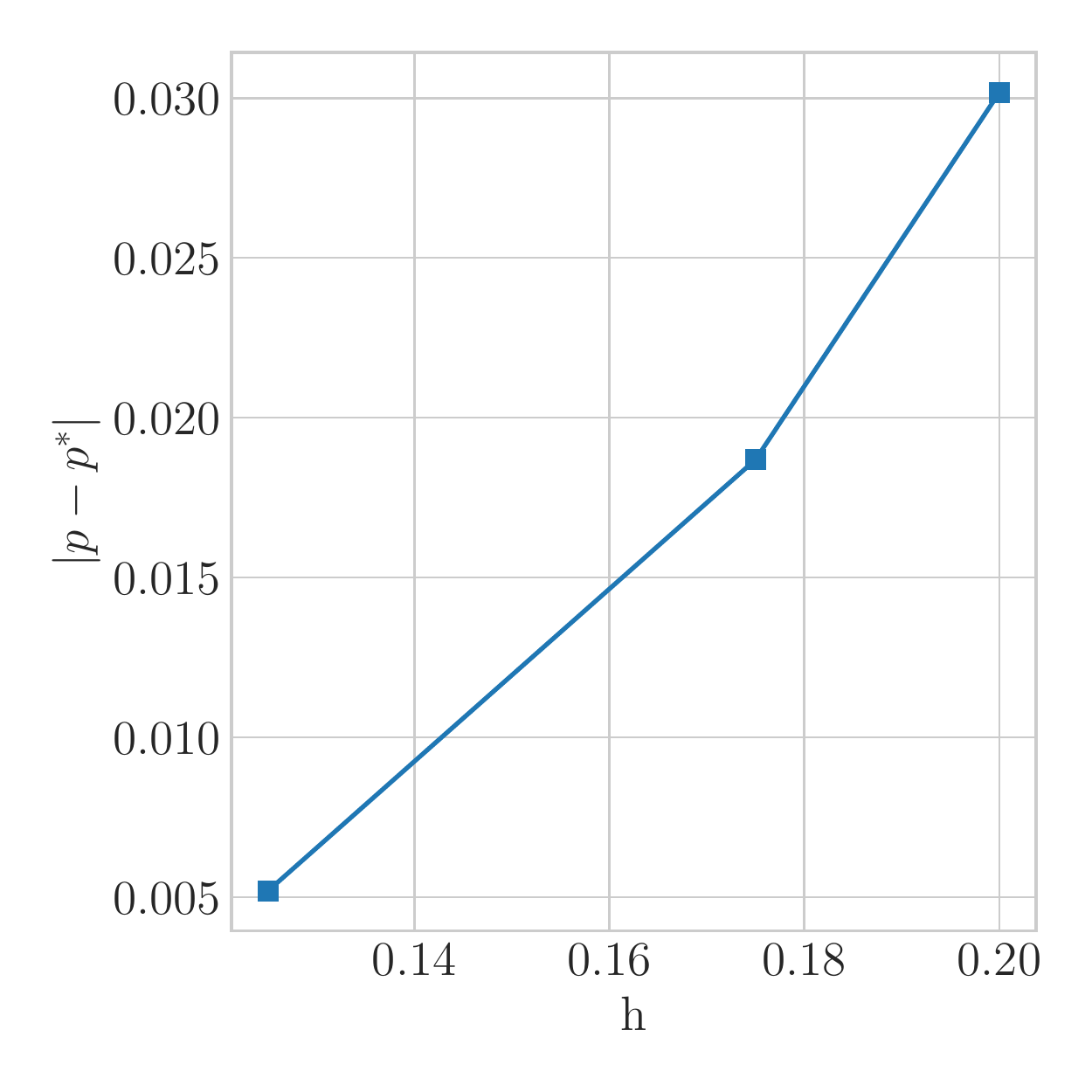}
\caption{Pressure convergence study.}
\label{fig:experiment:0144pressure}
\end{subfigure}
\begin{subfigure}[T]{0.3\textwidth}
\centering
\includegraphics[width=\textwidth]{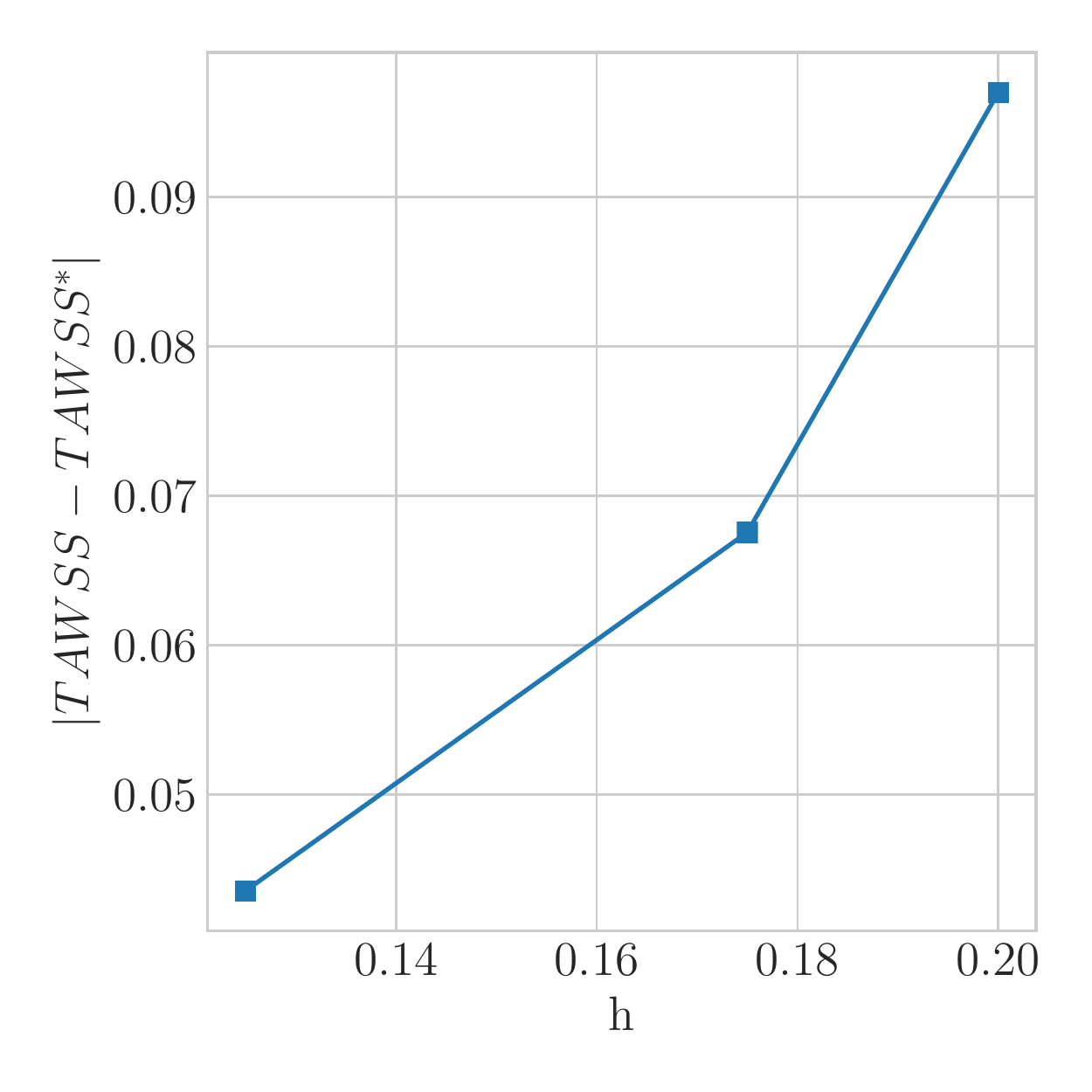}
\caption{WSS convergence study.}
\label{fig:experiment:0144wss}
\end{subfigure}
\begin{subfigure}[T]{0.3\textwidth}
\centering
\includegraphics[width=\textwidth]{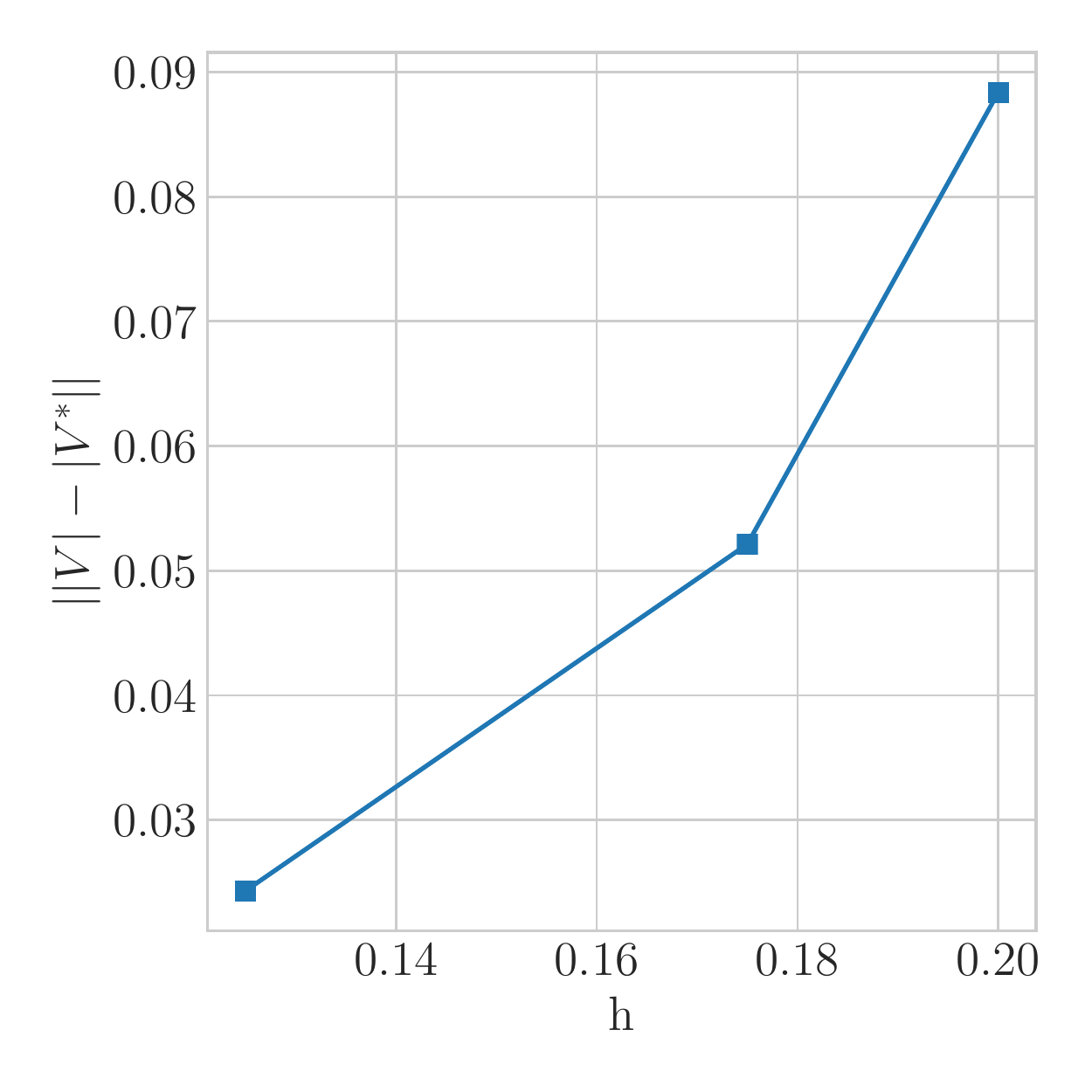}
\caption{Velocity convergence study.}
\label{fig:experiment:0144velocity}
\end{subfigure}
\end{subfigure}
\caption{AAA model anatomy (a) with boundary conditions (b) and mesh convergence analysis (c,d,e).}\label{fig:0144mesh}
\end{figure}

\noindent The second anatomy considered in this study includes the aorta and its main branches from an abdominal CT image of a patient with an abdominal aortic aneurysm (AAA, see Figure~\ref{fig:experiment:0144}), subject to the same aortic inflow used for the previous anatomy (see Figure~\ref{fig:exp:aortainflow}). RCR boundary condition parameters are reported in Table \ref{tab:experiment:tab0144rcr}.
For this anatomy, a family of $110$ geometries was generated through Monte Carlo sampling. The sample size was reduced as 110 models was sufficient for statistical convergence. We conducted a mesh convergence study, using the mean absolute error \eqref{eq:mae} and with meshes having roughly $500,000$, $700,000$ and $3,000,000$ elements and boundary layer mesh with 5 layers and compared these to a reference mesh with $7,000,000$ elements.
The $3,000,000$ element mesh showed a less than $0.5\%$, $5\%$ and $3\%$ mean error for the pressure, WSS magnitude and velocity magnitude, respectively (see Figure~\ref{fig:0144mesh}) and was subsequently used in all numerical experiments reported below.

\subsection{Left Coronary Artery model}

\begin{figure}[!ht]
\begin{subfigure}[T]{0.44\textwidth}
\centering
\begin{subfigure}[c]{1.0\textwidth}
\centering
\resizebox{1.0\textwidth}{!}{
\begin{tabular}{lccccc}
\hline
Vessel & $R_1$ & $C_1$ & $R_2$ & $C_2$ & $R_3$ \\
\hline
$LAD$ & 125,342 &  $4.54\cdot 10^{-7}$ & 203,681 & $3.68\cdot 10^{-6}$ & 62670\\
$LAD-D_1$ & 78,596 &  $7.26\cdot 10^{-7}$ & 127,720 & $5.87\cdot 10^{-6}$ & 39,298\\
$LCx$ & 90,578 &  $6.30\cdot 10^{-7}$ & 147,190 & $5.09\cdot 10^{-6}$ & 45,289\\
$LCx-OM_1$ & 38,155 &  $1.49\cdot 10^{-6}$ & 62,002 & $1.21\cdot 10^{-5}$ & 19,077 \\
$LCx-OM_2$ & 132,249 &  $4.31\cdot 10^{-7}$ & 214,905 & $3.49\cdot 10^{-6}$ & 66,124\\
$LCx-OM_3$ & 254,268 &  $2.24\cdot 10^{-7}$ & 413,186 & $1.81\cdot 10^{-6}$ & 127,134\\
\hline
\end{tabular}}
\caption{RCRCR boundary condition parameters for left coronary artery model, resistance in $dyne\cdot s/cm^5$, capacitance in $cm^5/dyne$.}\label{tab:experiment:tabcoronaryrcr}
\end{subfigure}

\vspace{5pt}

\begin{subfigure}[c]{1.0\textwidth}
\centering
\includegraphics[scale=0.5]{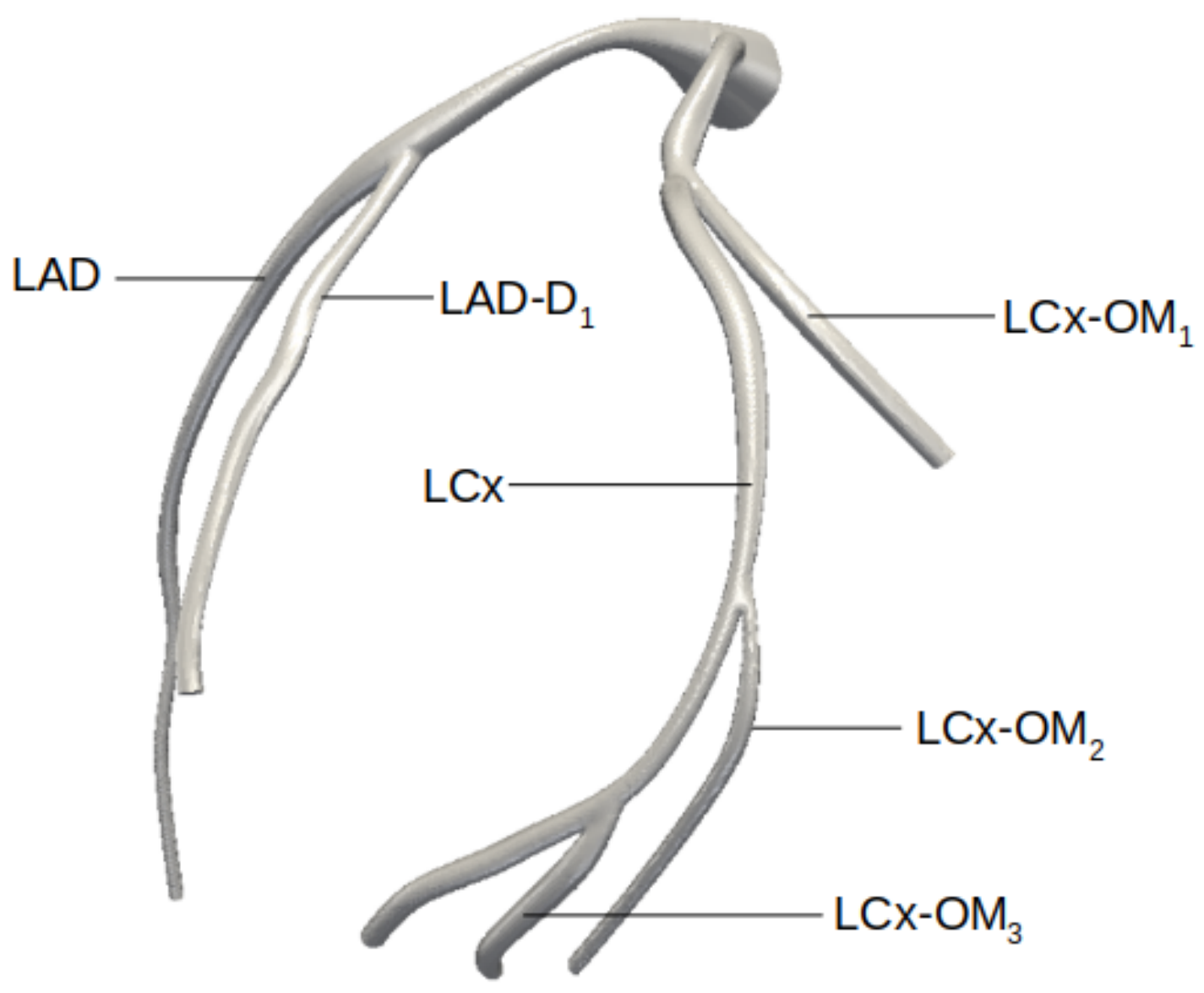}
\caption{Left coronary artery model anatomy.}\label{fig:experiment:coronary}
\end{subfigure}
\end{subfigure}
\hfill
\begin{subfigure}[T]{0.52\textwidth}
\centering

\begin{subfigure}[T]{0.48\textwidth}
\centering
\includegraphics[width=\textwidth]{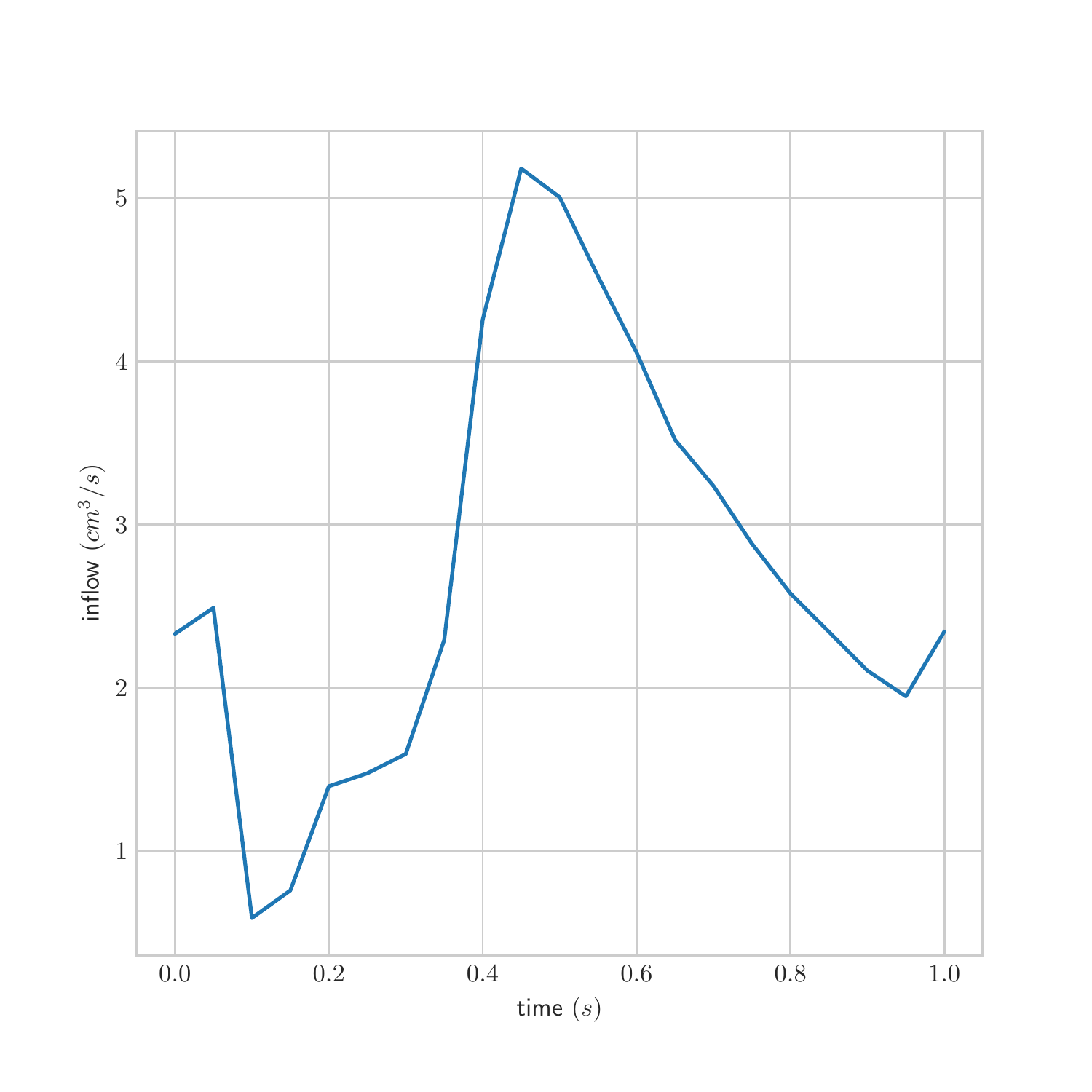}
\caption{Coronary inlet waveform}\label{fig:exp:coronaryinflow}
\end{subfigure}
\begin{subfigure}[T]{0.48\textwidth}
\centering
\includegraphics[width=\textwidth]{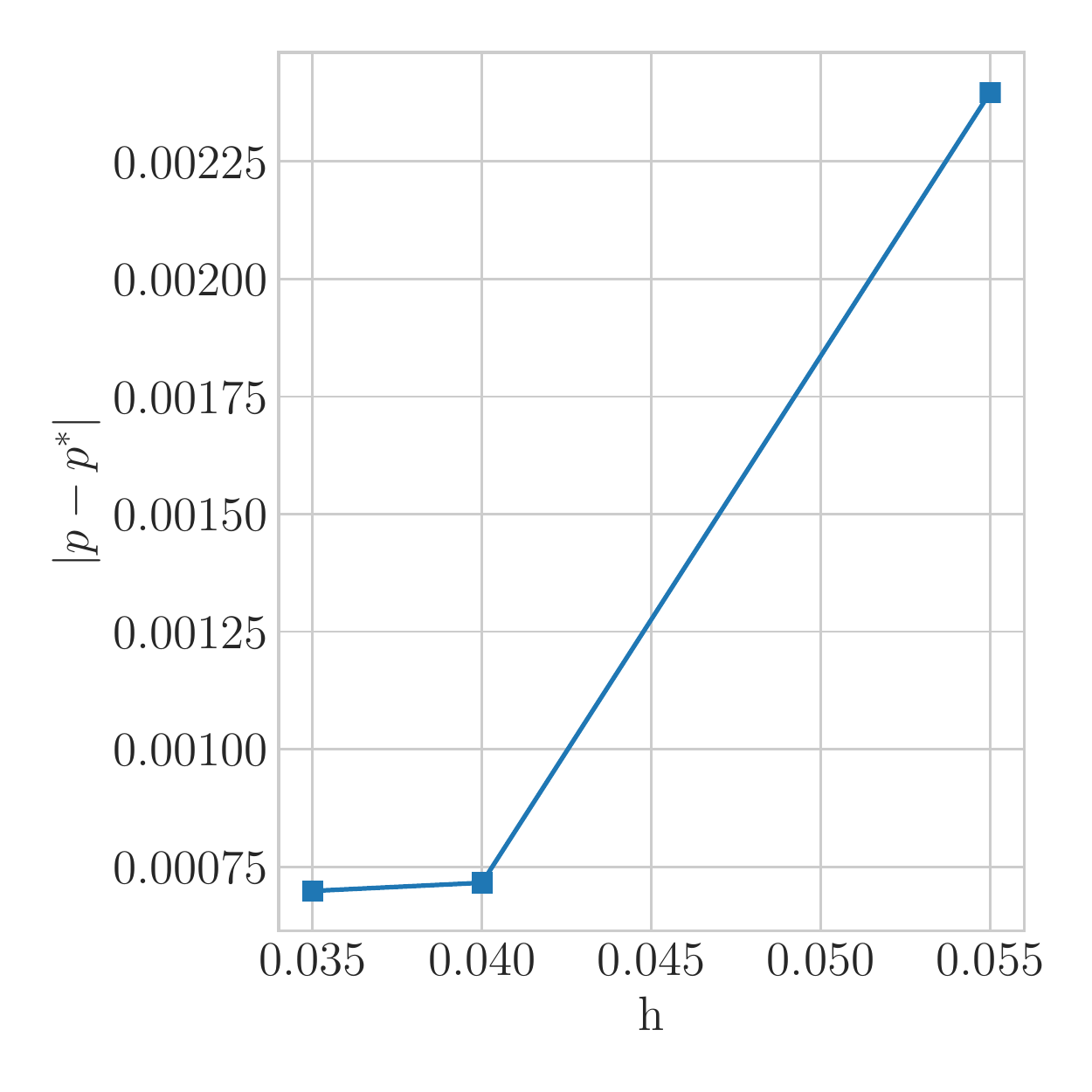}
\caption{Pressure convergence study.}\label{fig:experiment:coronarypressure}
\end{subfigure}

\begin{subfigure}[T]{0.48\textwidth}
\centering
\includegraphics[width=\textwidth]{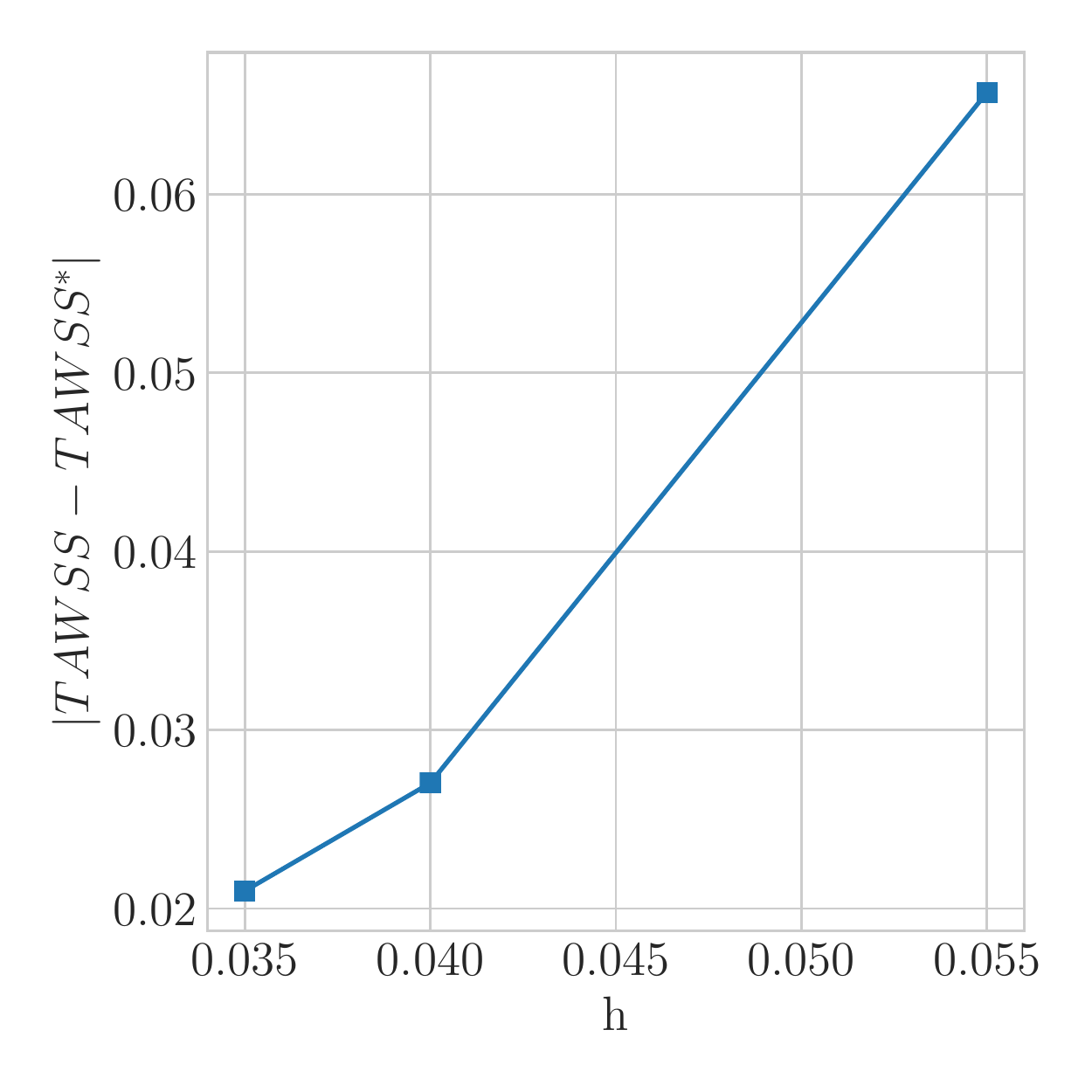}
\caption{WSS convergence study.}\label{fig:experiment:coronarywss}
\end{subfigure}
\begin{subfigure}[T]{0.48\textwidth}
\centering
\centering
\includegraphics[width=\textwidth]{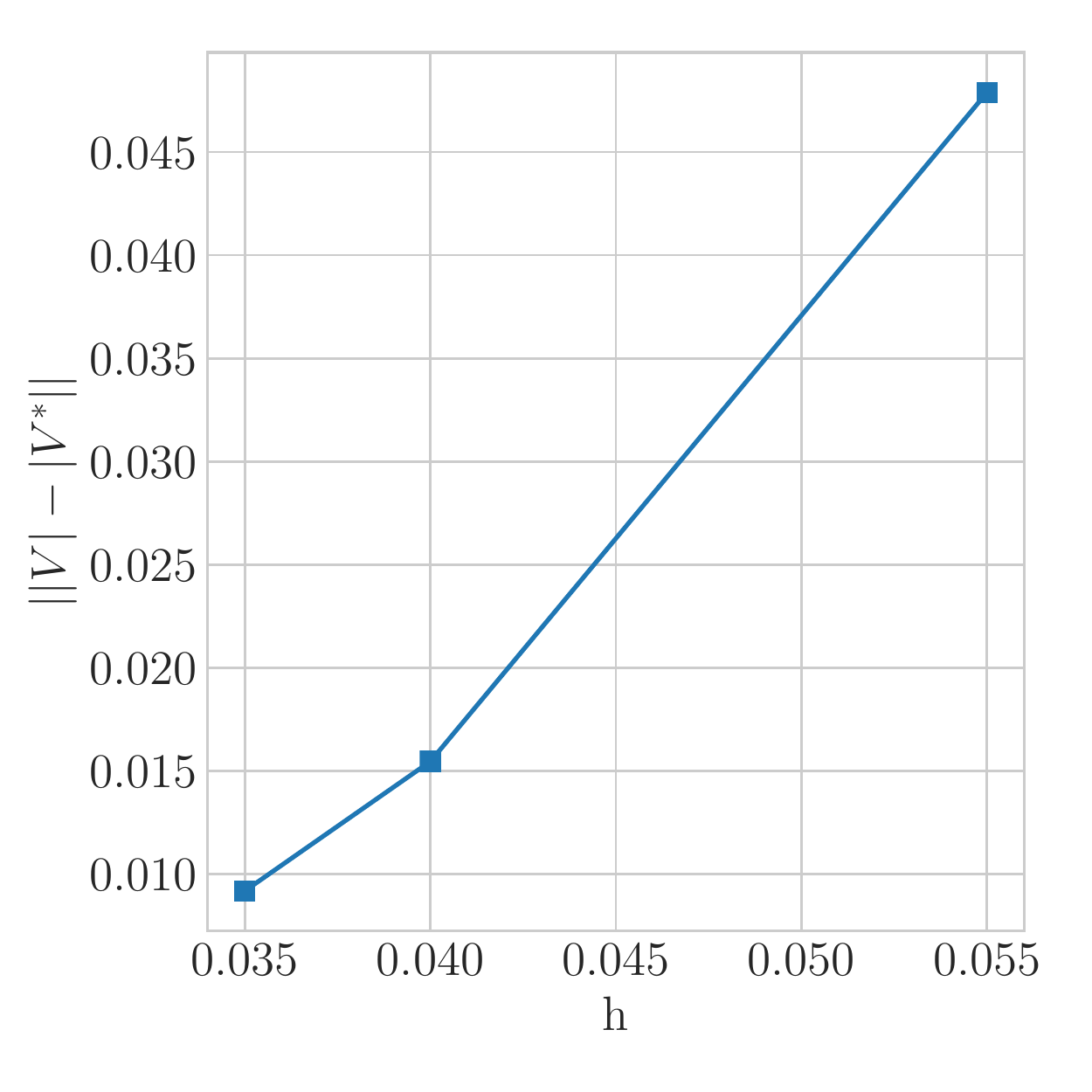}
\caption{Velocity convergence study.}\label{fig:experiment:coronaryvelocity}
\end{subfigure}
\end{subfigure}
\caption{Left coronary artery model with boundary conditions (a,c), lumen surface (b) and mesh convergence analysis (d,e,f). }\label{fig:coronarymesh}
\end{figure}

\noindent The third model we consider includes the left anterior descending (LAD) and left circumflex (LCx) coronary arteries extracted from a CT image volume, also studied in~\cite{seo19}.
Coronary lumped parameter boundary condition values were selected to produce physiological pressure ranges (Fig. \ref{tab:experiment:tabcoronaryrcr}).
The coronary simulations use a pulsatile coronary inflow waveform (Fig. \ref{fig:exp:coronaryinflow}).
A sample size of 110 models was used for the Monte Carlo trials.
We conducted a mesh convergence study, using the mean absolute error \eqref{eq:mae} and with meshes with roughly $500,000$, $1,000,000$ and $1,500,000$ elements and boundary layers with 5 layers and compared these to a reference mesh with $3,500,000$ elements. The $1,500,000$ element mesh showed a less than $0.075\%$, $3\%$ and $1\%$ mean absolute error for the pressure, WSS magnitude and velocity magnitude respectively (Fig. \ref{fig:coronarymesh}).
Subsequently, this mesh was used for all further experiments.

%% file: res_dropout.tex
\subsection{Comparison of Dropout Network Lumen Samples to Human Expert Segmentation}

Before commenting on the model results, we first investigate the statistical properties of the segmentations produced by our dropout network. In addition, we compare network samples and lumen segmentations produced by a number of expert SimVascular users. To do so, we selected four representative image volumes with vessel centerlines from the Vascular Model Repository~\cite{wilson13}, including a cerebrovascular anatomy imaged by MR, a coronary anatomy with aneurysms caused by Kawasaki disease imaged by CT, a coronary anatomy following bypass graft surgery imaged by CT, and an pulmonary anatomy imaged by MR.
Slices were selected at discrete intervals along the vessel centerlines, and each location was segmented by three individual SimVascular experts (see Figure~\ref{fig:modelconstruction}), resulting in a total of 290 segmentations per expert.
For the same slices, 50 neural network lumen samples were generated for various dropout probabilities, i.e., $p=0.9$, $p=0.7$ and $p=0.4$, and used for statistical analysis.

\begin{figure}[!ht]
\centering
\begin{subfigure}{0.24\textwidth}
\centering
\includegraphics[scale=0.4]{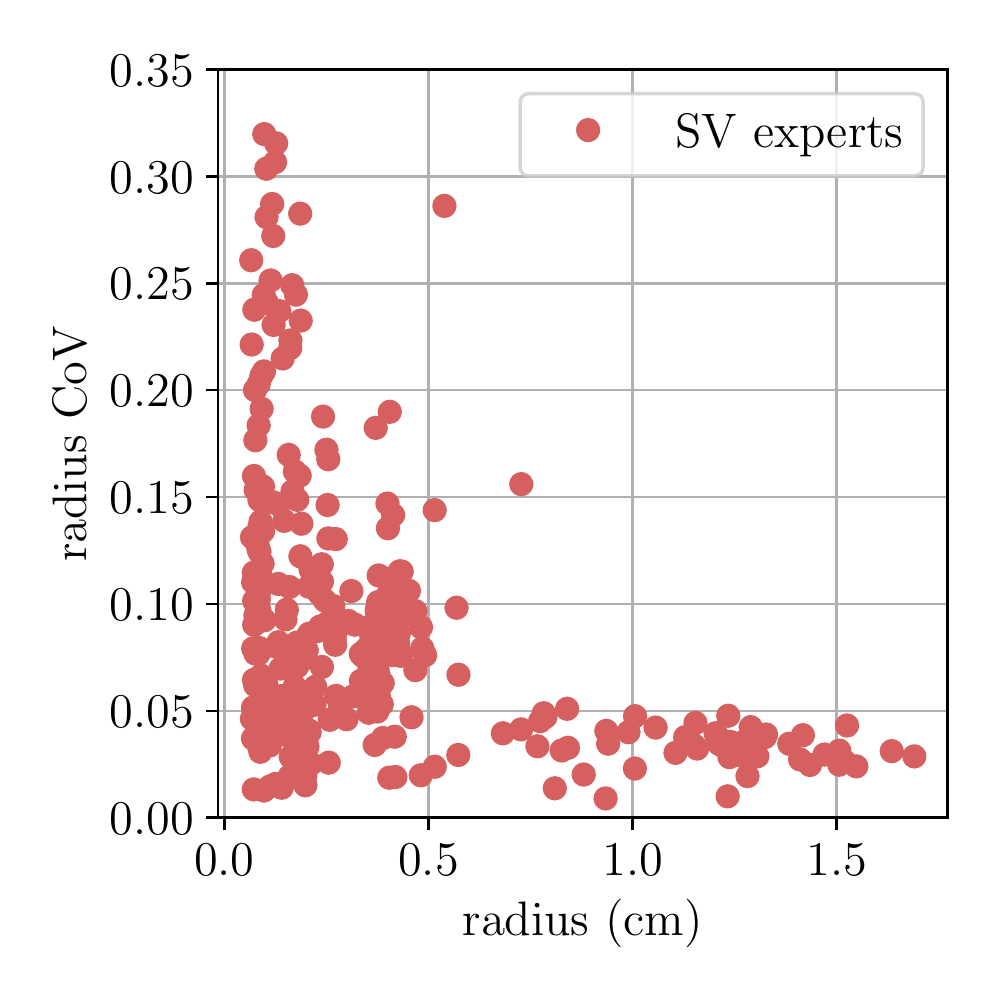}
\caption{SimVascular experts}
\label{figure:res:scatter_human}
\end{subfigure}
\begin{subfigure}{0.24\textwidth}
\centering\includegraphics[scale=0.4]{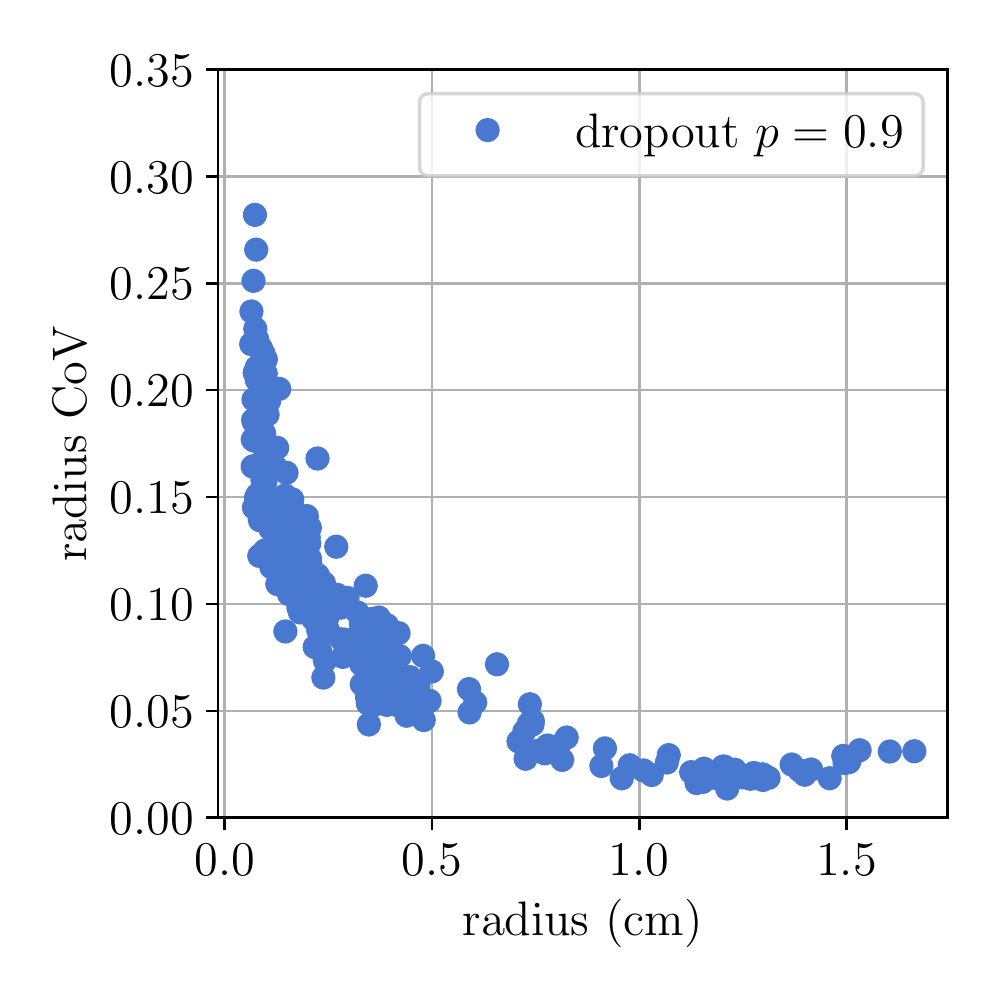}
\caption{CNN dropout $p=0.9$}
\label{figure:res:scatter_d1}
\end{subfigure}
\begin{subfigure}{0.24\textwidth}
\centering\includegraphics[scale=0.4]{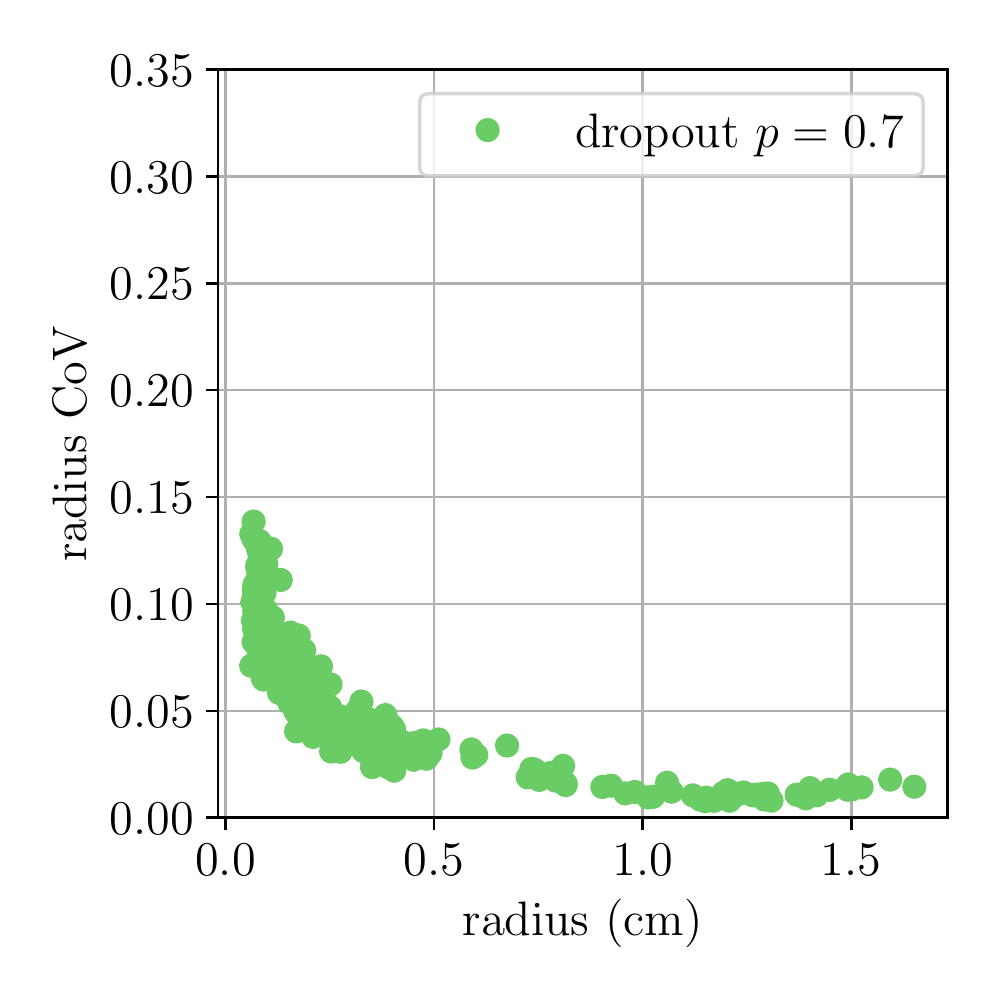}
\caption{CNN dropout $p=0.7$}
\label{figure:res:scatter_d3}
\end{subfigure}
\begin{subfigure}{0.24\textwidth}
\centering\includegraphics[scale=0.4]{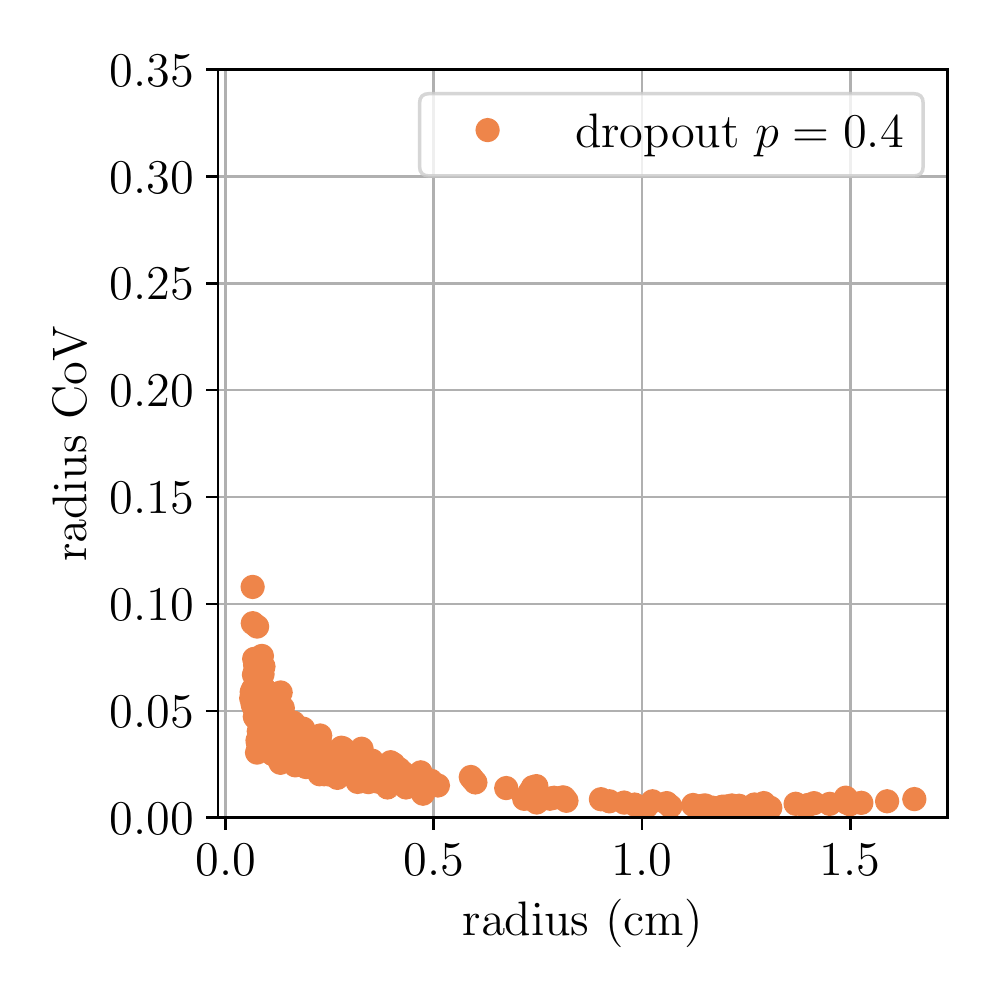}
\caption{CNN dropout $p=0.4$}
\label{figure:res:scatter_d6}
\end{subfigure}
\caption{Radius CoV against radius for SimVascular experts and GoogleNet network with different levels of dropout.}
\label{}
\end{figure}

Segmentation radius CoV observed for SimVascular expert users is separated into two classes, i.e., large vessels (r$>$0.4 cm), where the radius CoV is typically less than 5\%, and small vessels (r$\leq$0.4 cm), where the radius CoV is larger, with values as large as 30\% (see Figure~\ref{figure:res:scatter_human}).
The inverse relationship between radius CoV and lumen radius is explained by the fact there is a minimum error produced by human segmentation, due to limits in image resolution, acquisition noise, and expert image interpretation.
Similar trends are observed for neural network samples with varying dropout probabilities, where the radius CoV increases with the dropout probability (see Figures~\ref{figure:res:scatter_d1}-\ref{figure:res:scatter_d6}).
In addition, the network with dropout $p=0.9$ produces similar CoV to expert SimVascular users, whereas a dropout $p=0.4$ produces more precise segmentations. Thus, the dropout probability can be tuned to modulate the variability in the resulting segmentation.

\begin{figure}[!ht]
\centering
\begin{subfigure}{0.48\textwidth}
\centering
\includegraphics[scale=0.5]{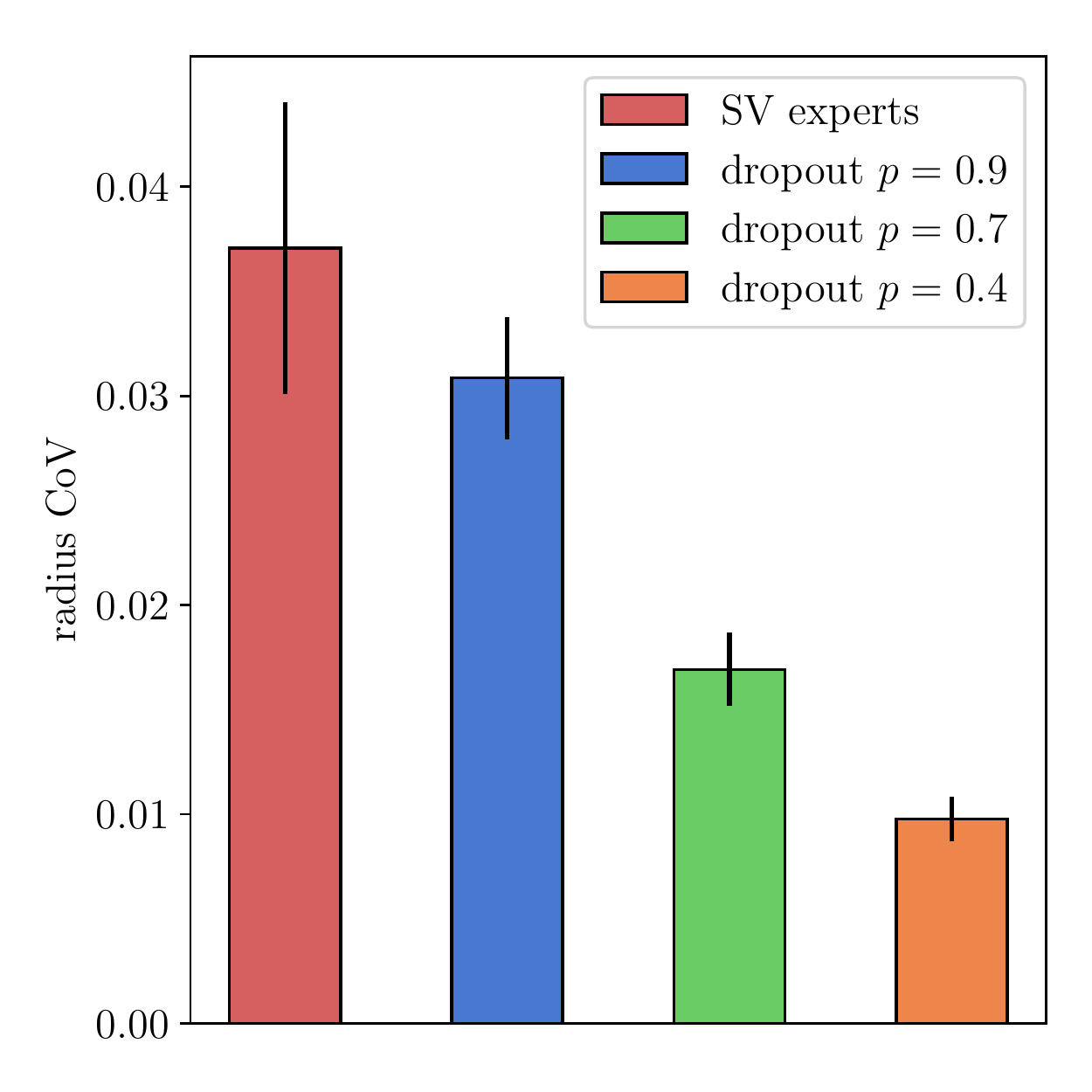}
\caption{Large vessels $r>0.4cm$}
\label{figure:res:rcov_bar_large}
\end{subfigure}
\begin{subfigure}{0.48\textwidth}
\centering\includegraphics[scale=0.5]{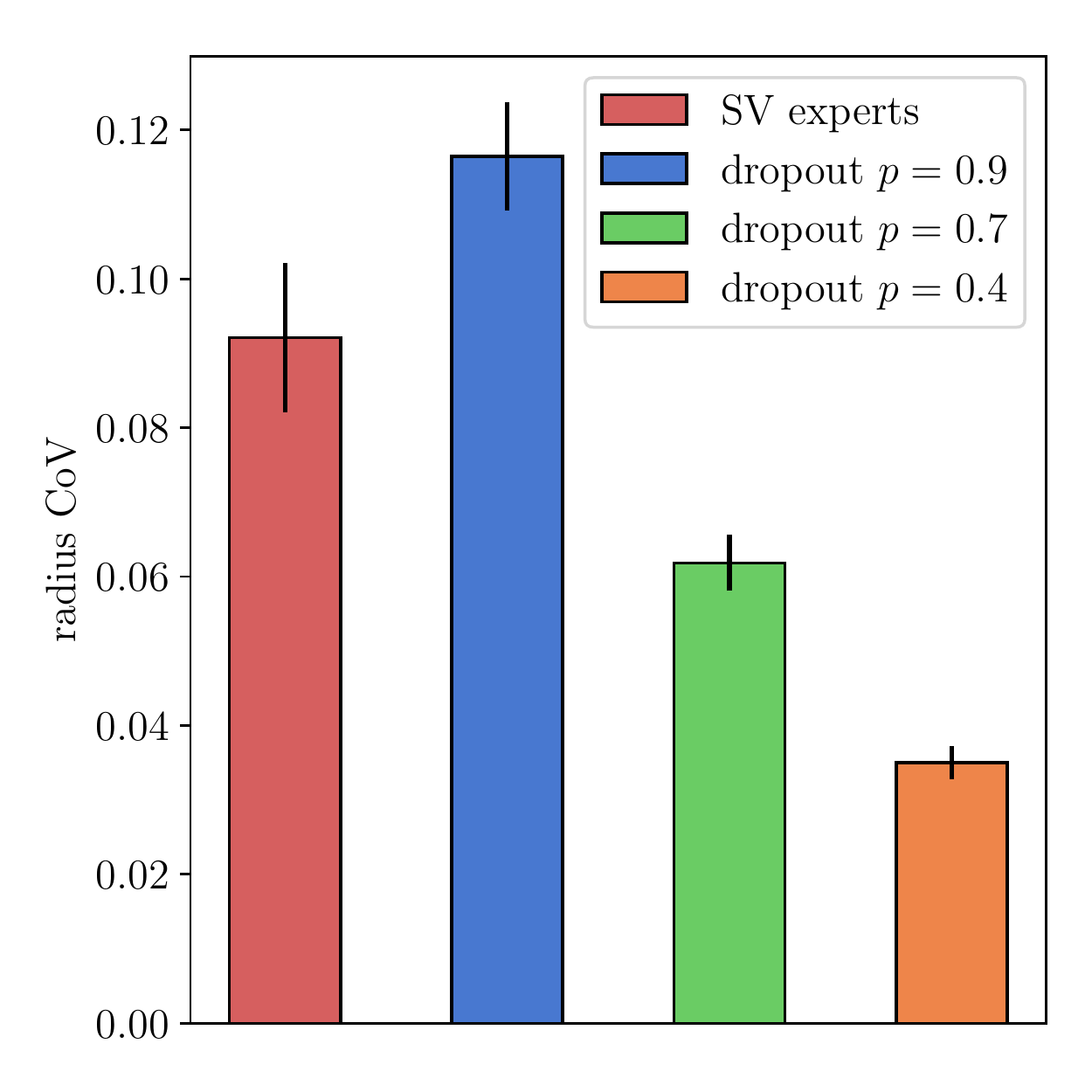}
\caption{Small vessels $r<0.4cm$}
\label{figure:res:rcov_bar_small}
\end{subfigure}
\caption{Comparison of mean radius CoV for large and small vessels between SimVascular experts and the proposed network, with varying dropout probability.
The distributions of CoV shown above were obtained by collecting one CoV for each cross-section $\mathbf{x}_i$.}
\label{}
\end{figure}

\begin{figure}[!ht]
\centering
\begin{subfigure}{0.2\textwidth}
\centering
\includegraphics[width=1\textwidth]{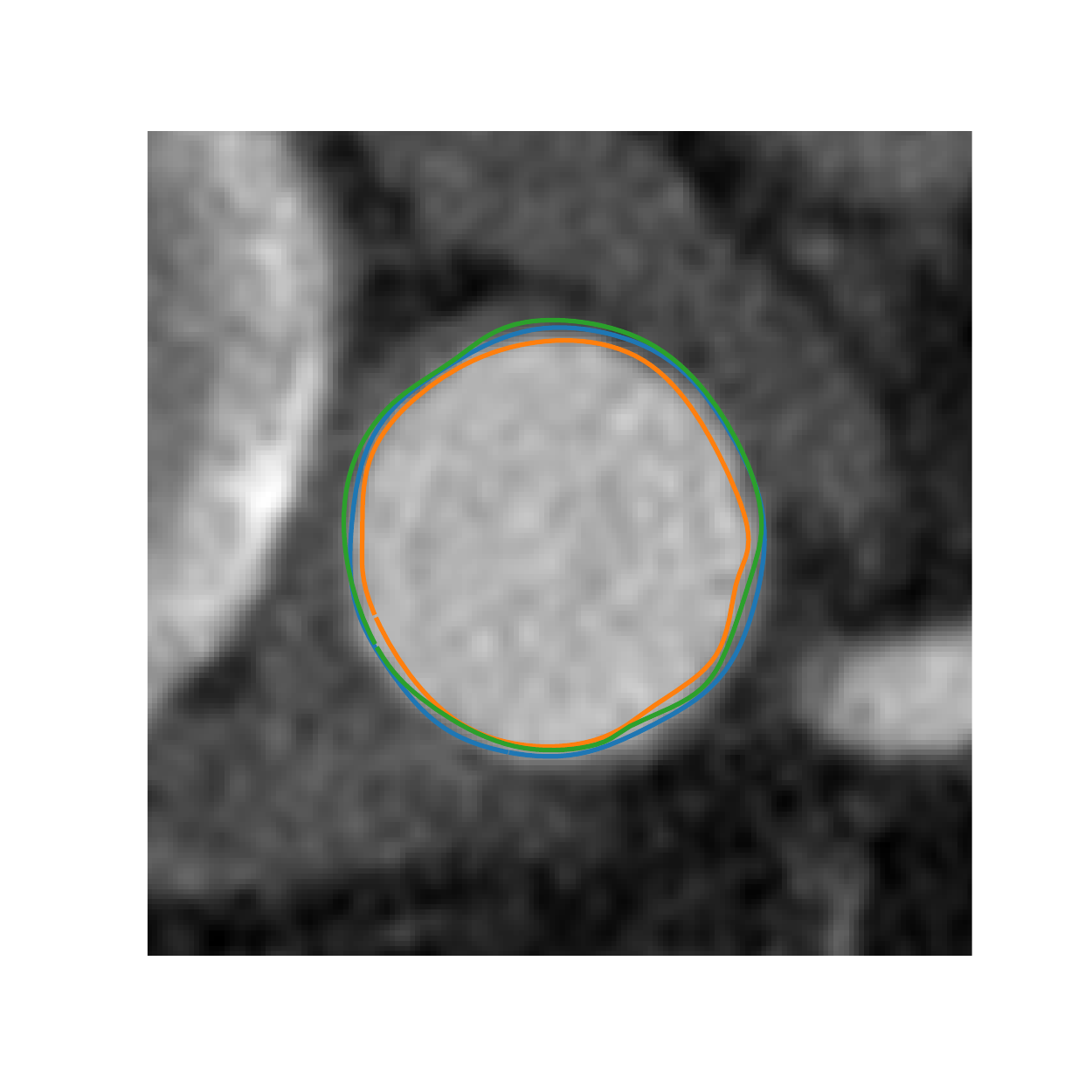}
\caption{SimVascular experts}
\label{figure:res:drop_seg_large_human}
\end{subfigure}
\begin{subfigure}{0.2\textwidth}
\centering\includegraphics[width=1\textwidth]{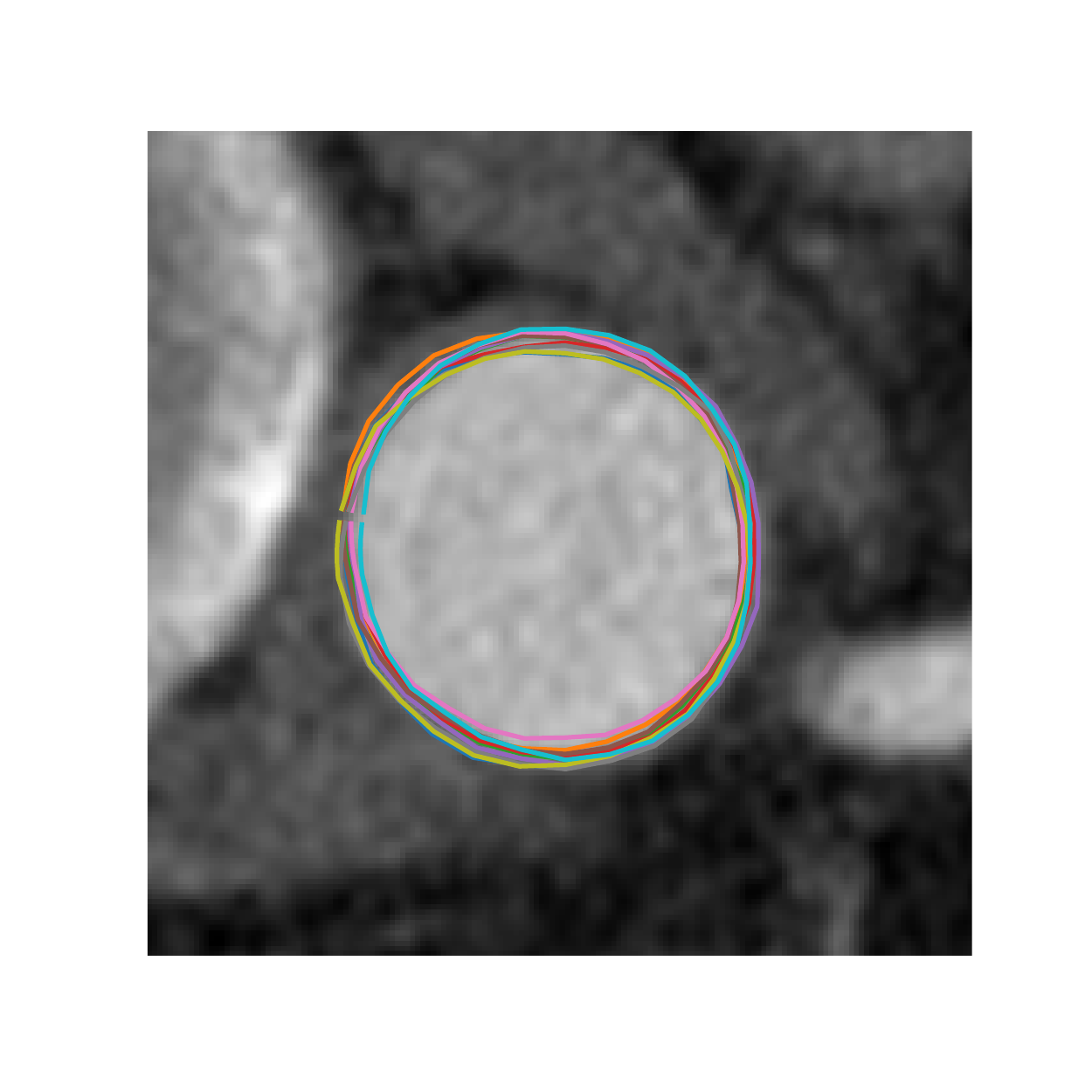}
\caption{CNN dropout $p=0.9$}
\label{figure:res:drop_seg_large_d1}
\end{subfigure}
\begin{subfigure}{0.2\textwidth}
\centering\includegraphics[width=1\textwidth]{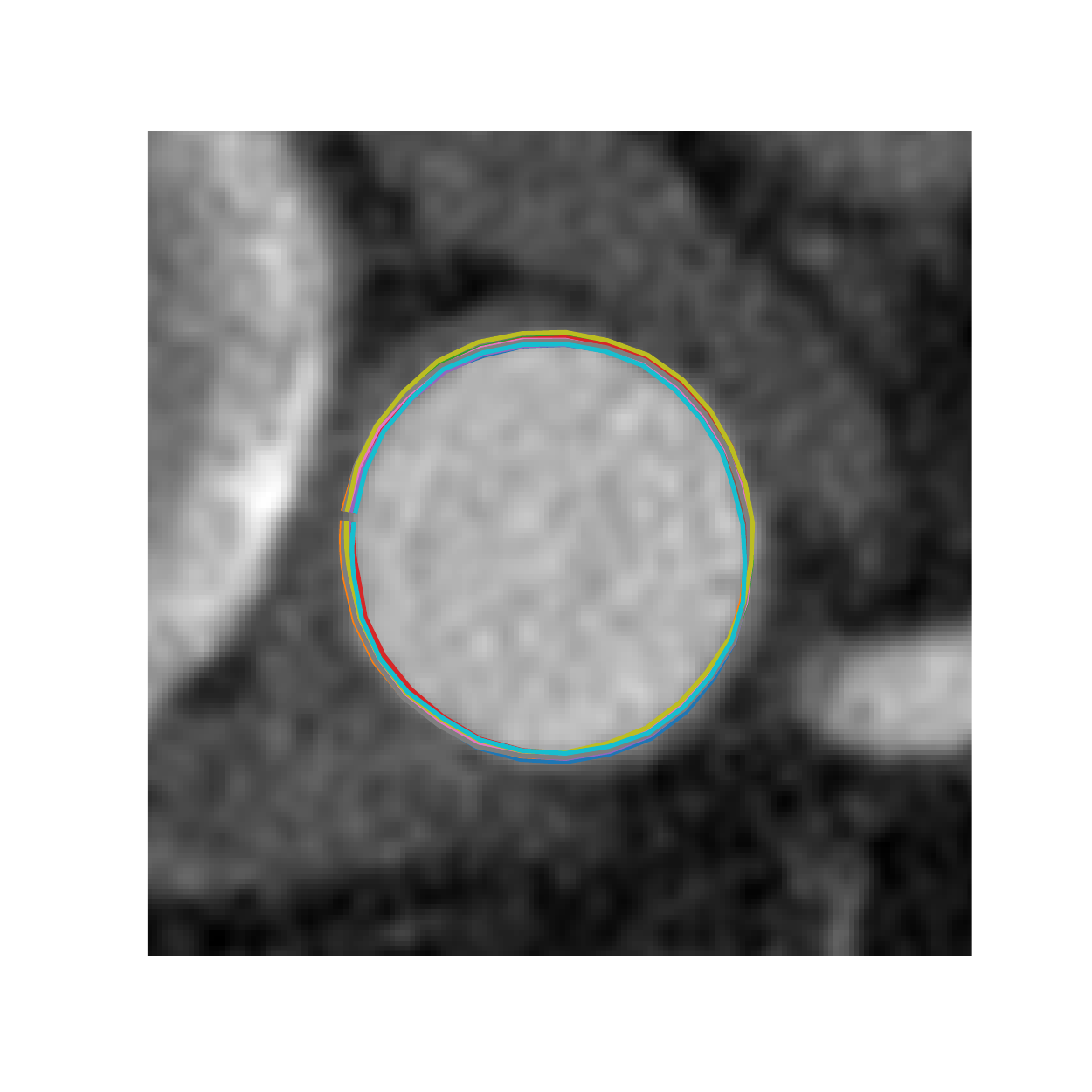}
\caption{CNN dropout $p=0.7$}
\label{figure:res:drop_seg_large_d3}
\end{subfigure}
\begin{subfigure}{0.2\textwidth}
\centering\includegraphics[width=1\textwidth]{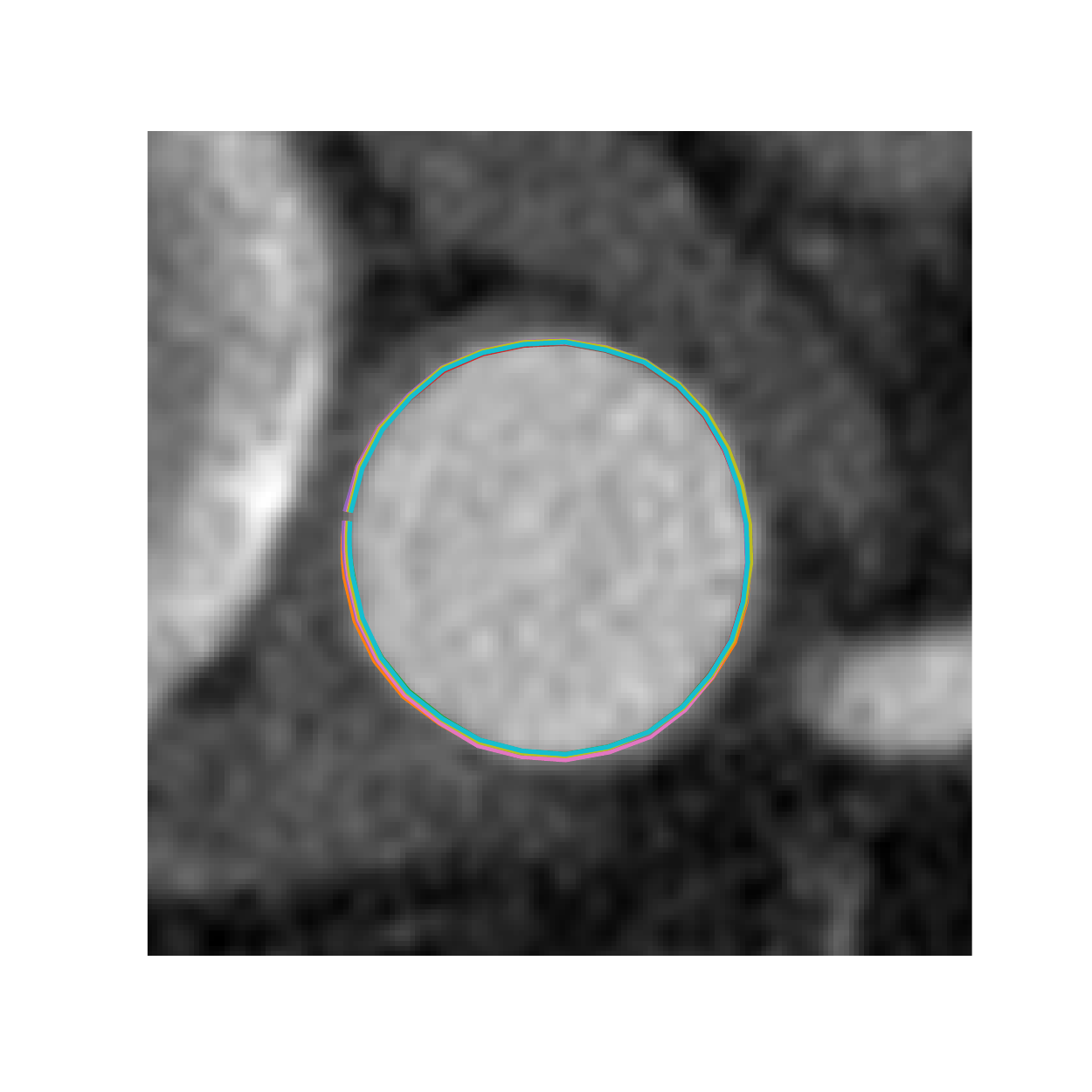}
\caption{CNN dropout $p=0.4$}
\label{figure:res:drop_seg_large_d6}
\end{subfigure}

\begin{subfigure}{0.2\textwidth}
\centering
\includegraphics[width=1\textwidth]{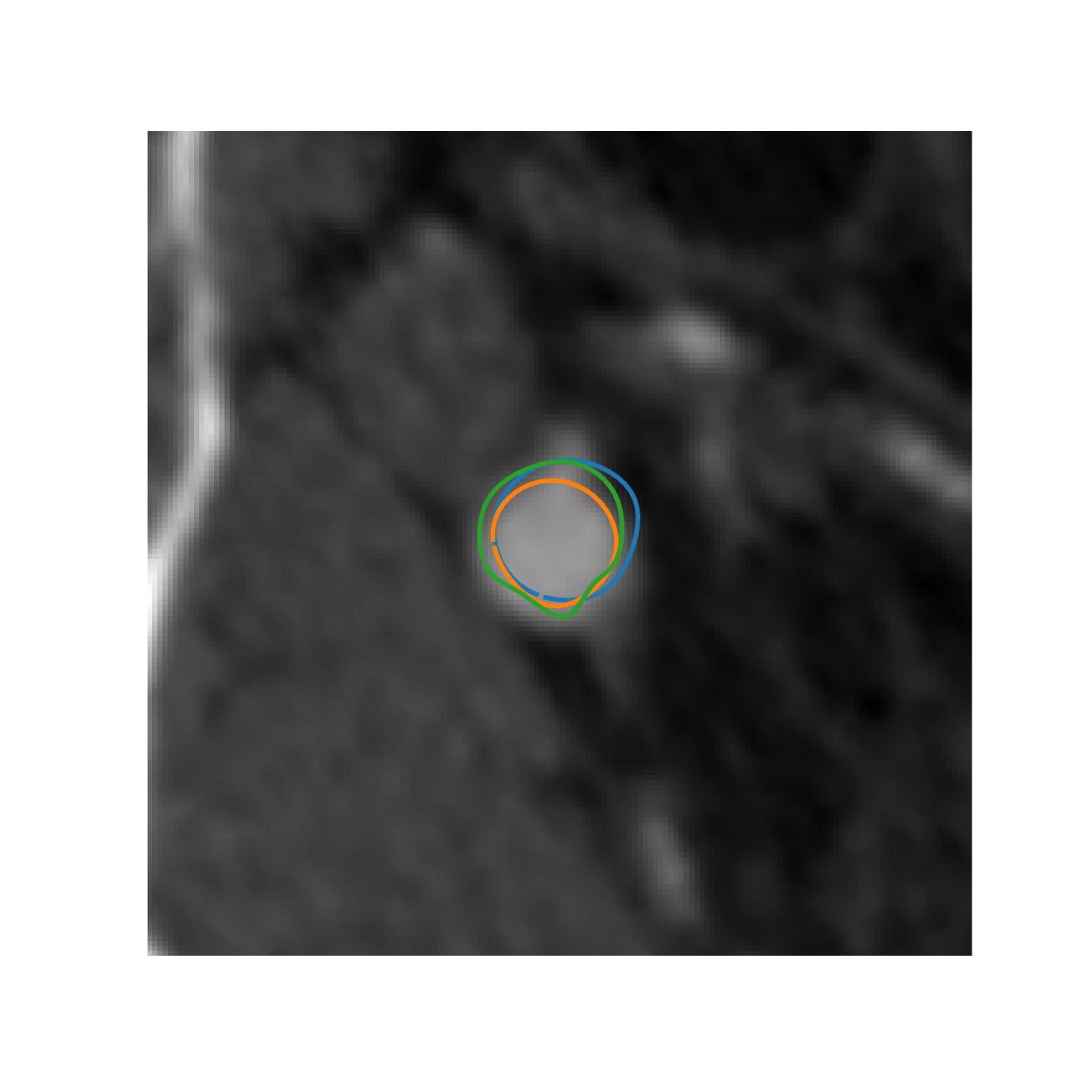}
\caption{SimVascular experts}
\label{figure:res:drop_seg_small_human}
\end{subfigure}
\begin{subfigure}{0.2\textwidth}
\centering\includegraphics[width=1\textwidth]{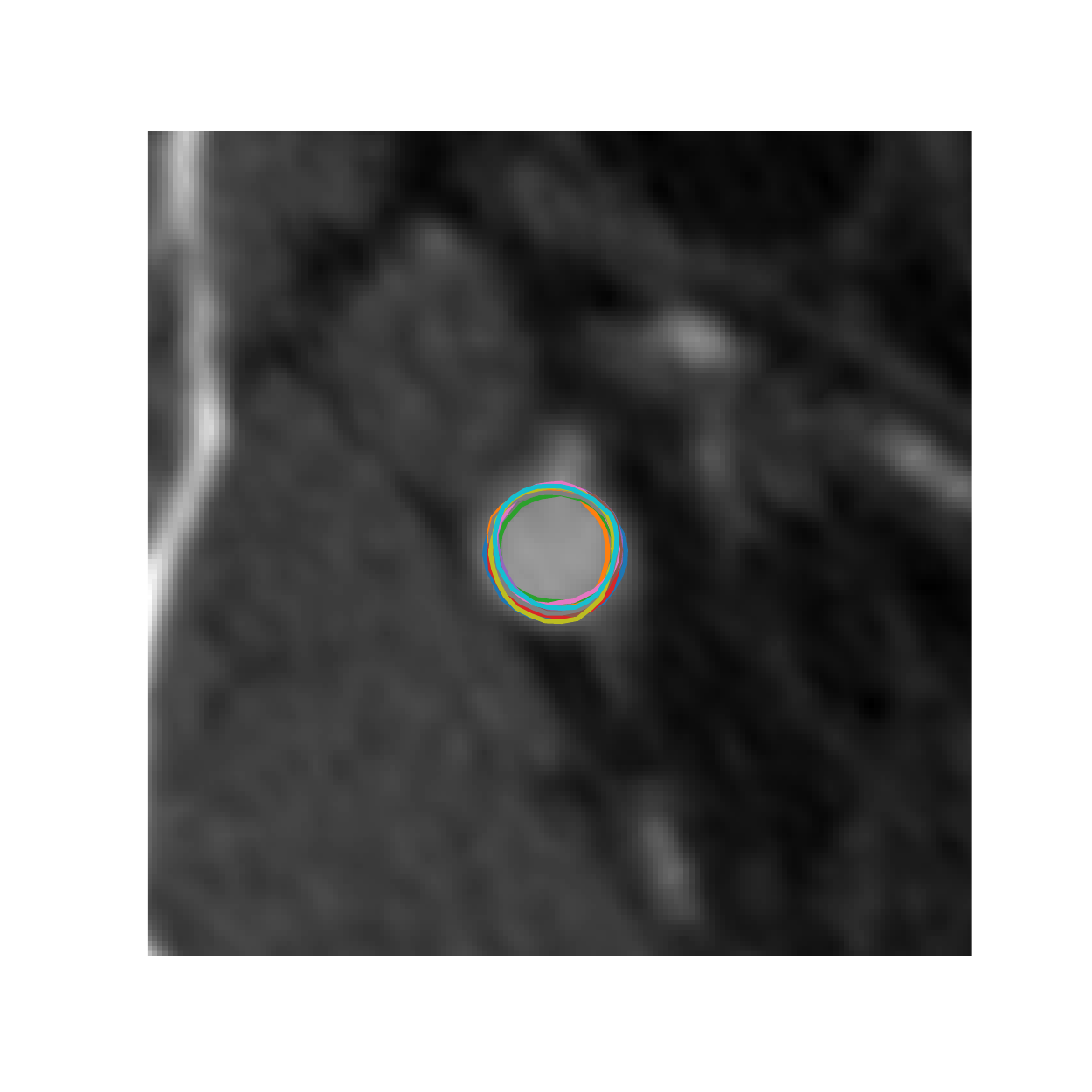}
\caption{CNN dropout $p=0.9$}
\label{figure:res:drop_seg_small_d1}
\end{subfigure}
\begin{subfigure}{0.2\textwidth}
\centering\includegraphics[width=1\textwidth]{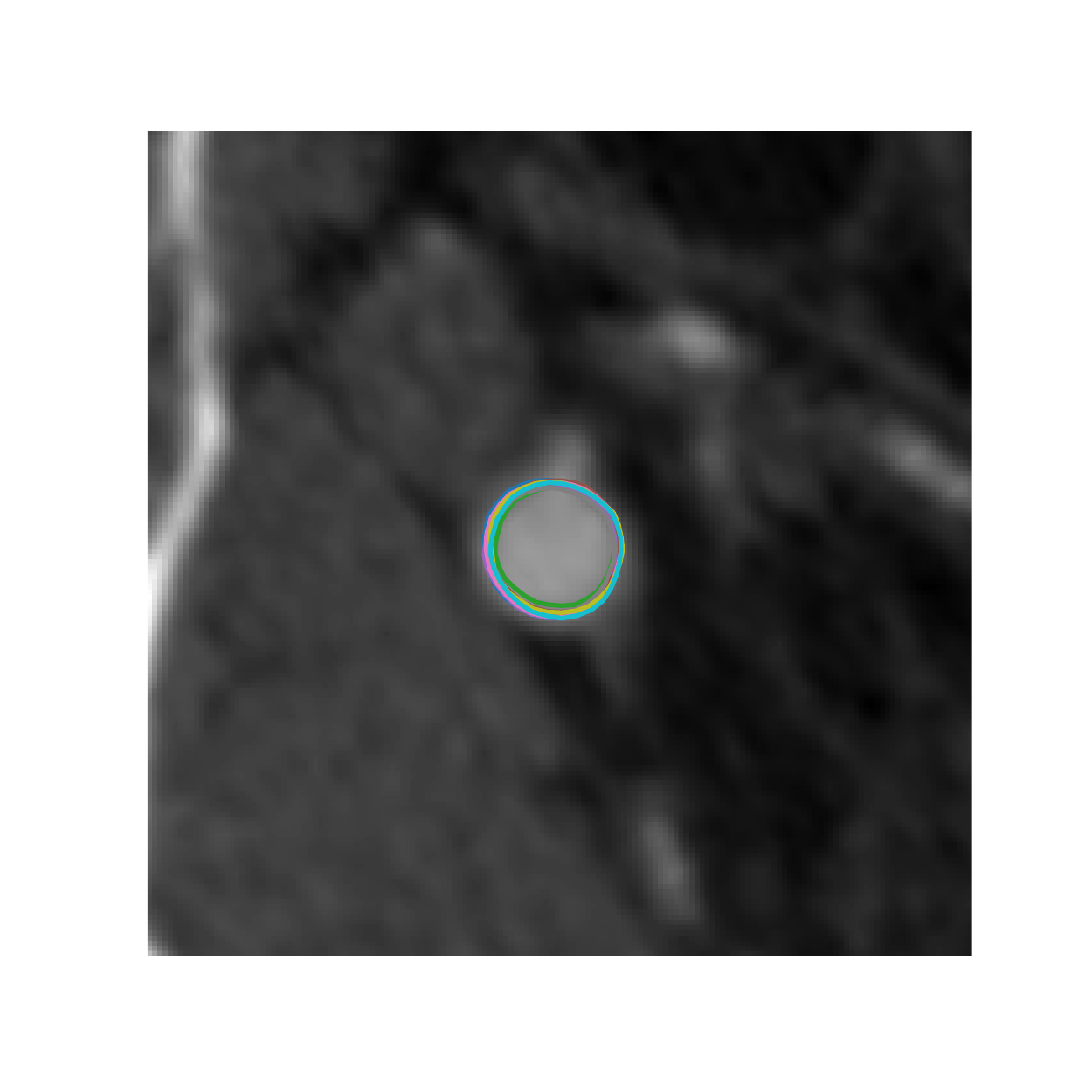}
\caption{CNN dropout $p=0.7$}
\label{figure:res:drop_seg_small_d3}
\end{subfigure}
\begin{subfigure}{0.2\textwidth}
\centering\includegraphics[width=1\textwidth]{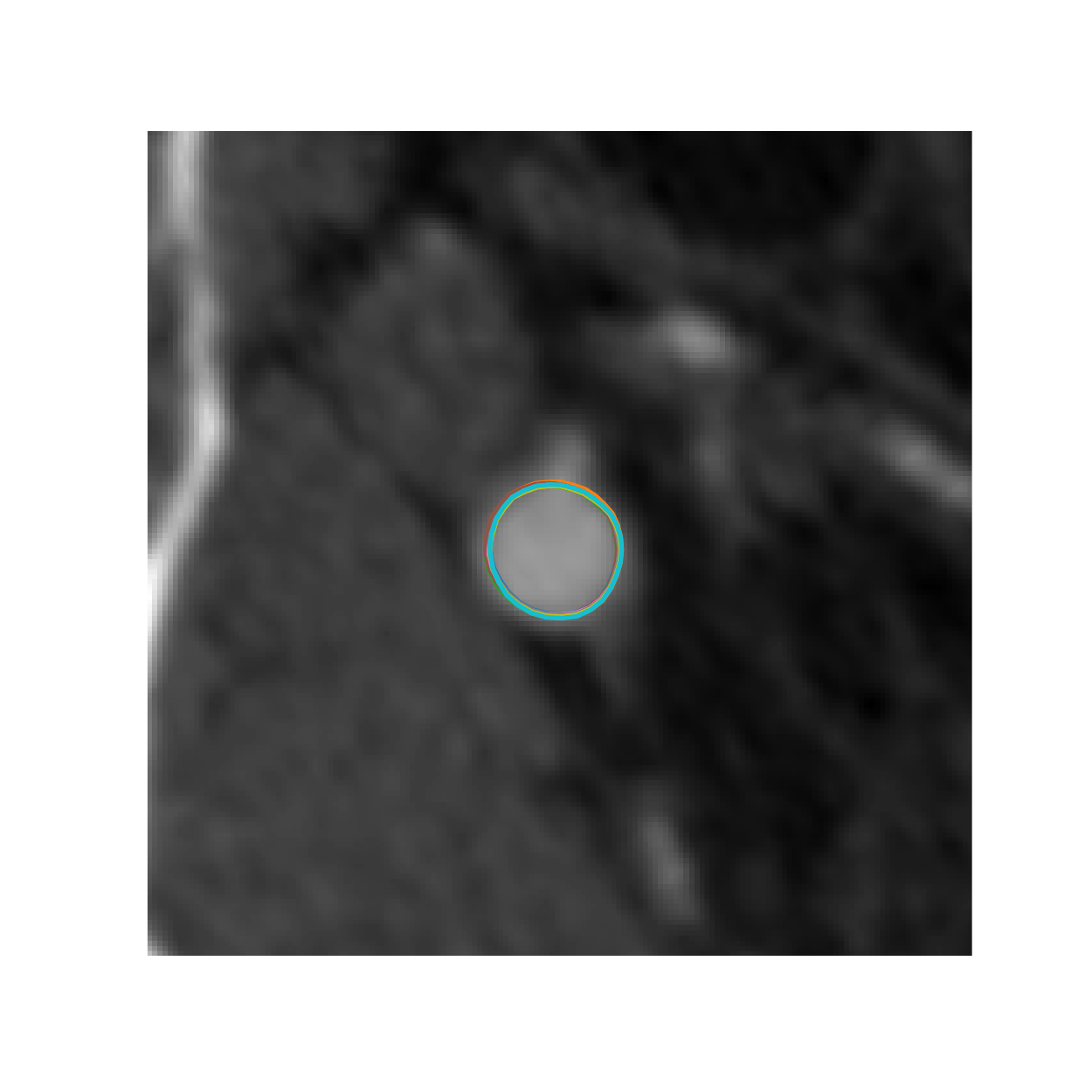}
\caption{CNN dropout $p=0.4$}
\label{figure:res:drop_seg_small_d6}
\end{subfigure}
\caption{Vessel lumen segmentation generated for large (top) and small (bottom) vessels by expert users and the proposed network with varying dropout probability.}
\label{}
\end{figure}

For large vessels, the network with dropout $p=0.9$ produces a CoV distribution compatible to that produced by an expert SimVascular user (see Figure~\ref{figure:res:rcov_bar_large}). However, for small vessels, the CoVs produced by the network are significantly larger (see Figure~\ref{figure:res:rcov_bar_small}). Conversely, dropout probabilities equal to $p=0.7$ and $p=0.4$ produce significantly lower CoVs than an expert user for both large and small vessels. In particular, a dropout probability of $p=0.4$ produces the smallest CoVs, showing direct proportionality between the dropout probability and the amount of segmentation uncertainty.
The overall lumen shape is captured rather well by expert users, but their segmentations exhibit deviations in local vessel radii (see Figures~\ref{figure:res:drop_seg_large_human}, \ref{figure:res:drop_seg_small_human}).
The same happens with our approach for a dropout probability equal to $p=0.9$ (see Figures~\ref{figure:res:drop_seg_large_d1}, \ref{figure:res:drop_seg_small_d1}).
For smaller $p$ the variability in the segmentations is reduced, and they all converge closely to the mean lumen profile (see Figures~\ref{figure:res:drop_seg_large_d1}-\ref{figure:res:drop_seg_large_d6} and \ref{figure:res:drop_seg_small_d1}-\ref{figure:res:drop_seg_small_d6}).
For the experiments in the remainder of this work a dropout probability of $0.4$ was used, consistent with the value used to train the original GoogLeNet network~\cite{szegedy_15}.

%% file: res_Bifurcation.tex
\subsection{Aorto-iliac Bifurcation Model}\label{section:paper3:0110}

\begin{figure}[!ht]
\centering
\begin{subfigure}[b]{0.4\textwidth}
\centering
\includegraphics[height=1.2\textwidth]{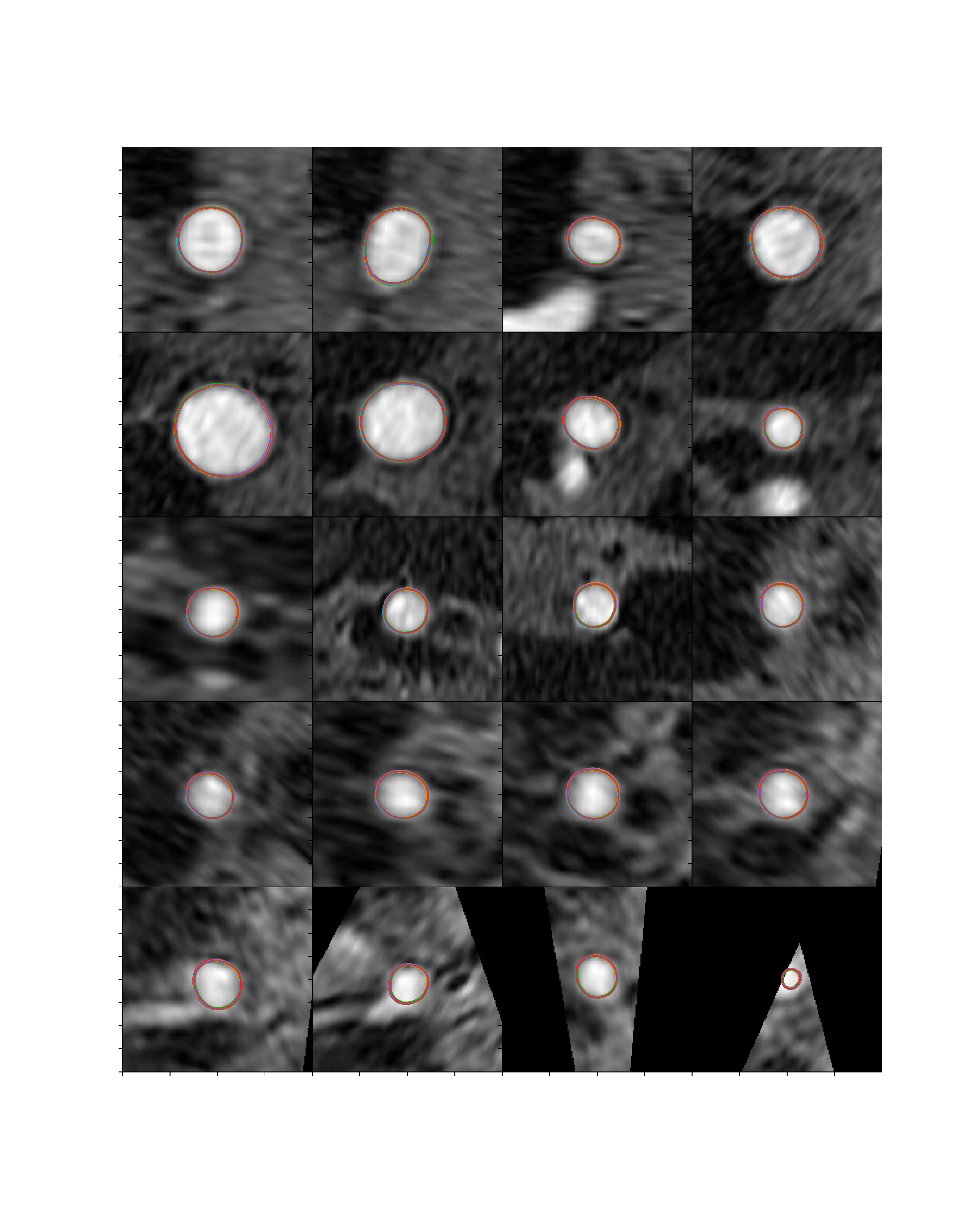}
\caption{Aorta and Iliacs case - Left Iliac}
\label{fig:results:seg0110iliac}
\end{subfigure}
\begin{subfigure}[b]{0.29\textwidth}
\centering\includegraphics[scale=0.5]{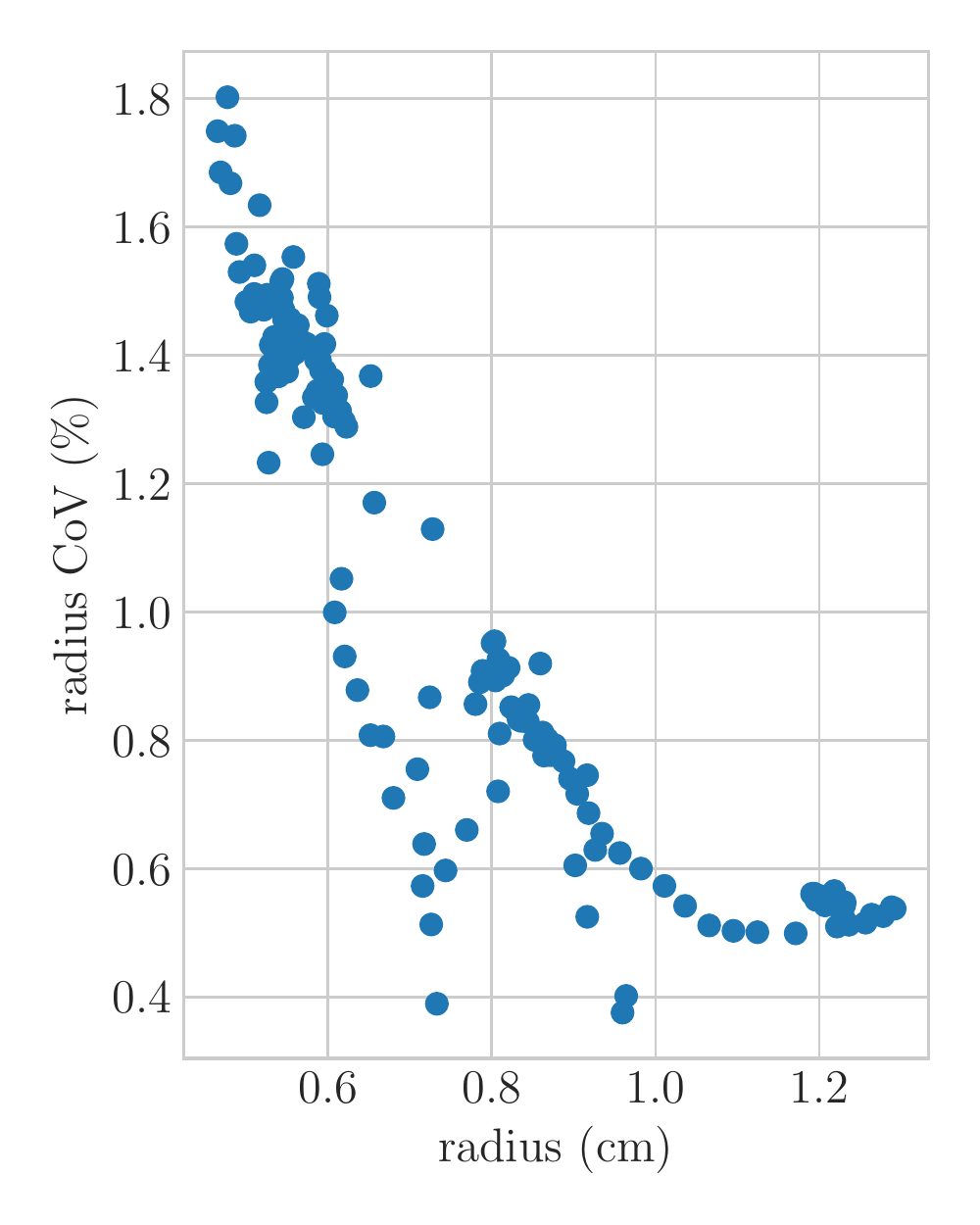}
\caption{Radius CoV}
\label{fig:results:rvrcv0110}
\end{subfigure}
\begin{subfigure}[b]{0.29\textwidth}
\centering\includegraphics[scale=0.5]{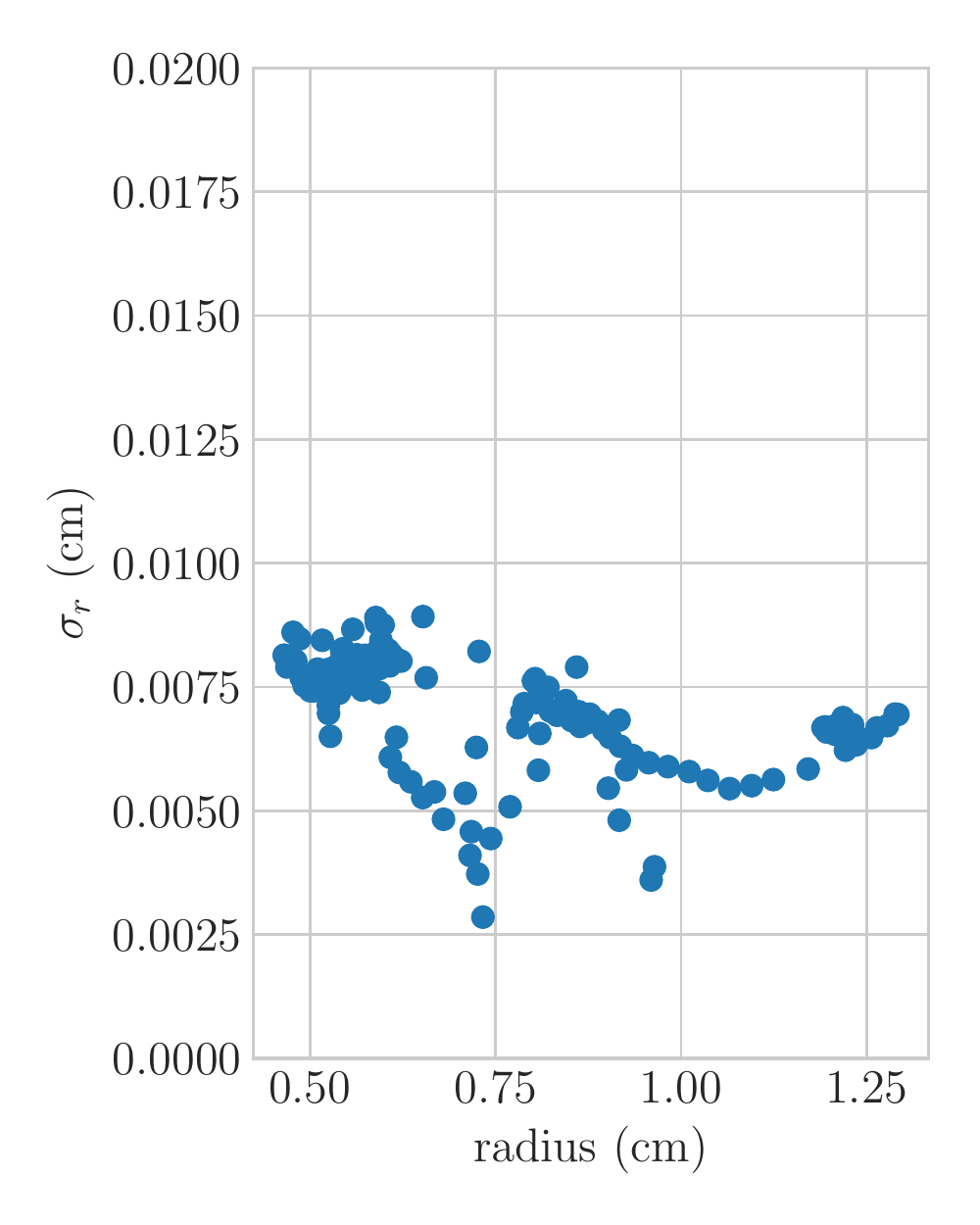}
\caption{Radius standard deviation}
\label{fig:results:rvrsig0110}
\end{subfigure}
\caption{Lumen segmentation samples and radius CoV/standard deviation for aorto-iliac bifurcation test case, computed over cross-sectional slices $\mathbf{x}_i,\,i=1,\dots,156$. }
\label{fig:results:rgraph0110}
\end{figure}

\noindent The lumen generated from our dropout network show good qualitative agreement with the depicted vessel lumen for the left iliac artery (Figure~\ref{fig:results:seg0110iliac}) and is able to correctly identify the relevant main branch even in the presence of surrounding tissue noise and branching vessels.

Variation in the segmentation radii appears to be limited, with standard deviation $\sigma_{r}$ between 0.005 cm and 0.01 cm (see Figure~\ref{fig:results:rvrcv0110} and Figure~\ref{fig:results:rvrsig0110}). In addition, variability in $\sigma_{r}$ appears to increase with decreasing vessel size, likely due to the typically poorer resolution of smaller vessels. The roughly constant $\sigma_{r}$ also results in a CoV that increases with decreasing vessel size, as typically seen for segmentations performed by expert operators~\cite{maher_20}, albeit with larger magnitude.

\begin{figure}[!ht]
\centering
\begin{subfigure}[b]{0.12\textwidth}
\centering\includegraphics[width=\textwidth]{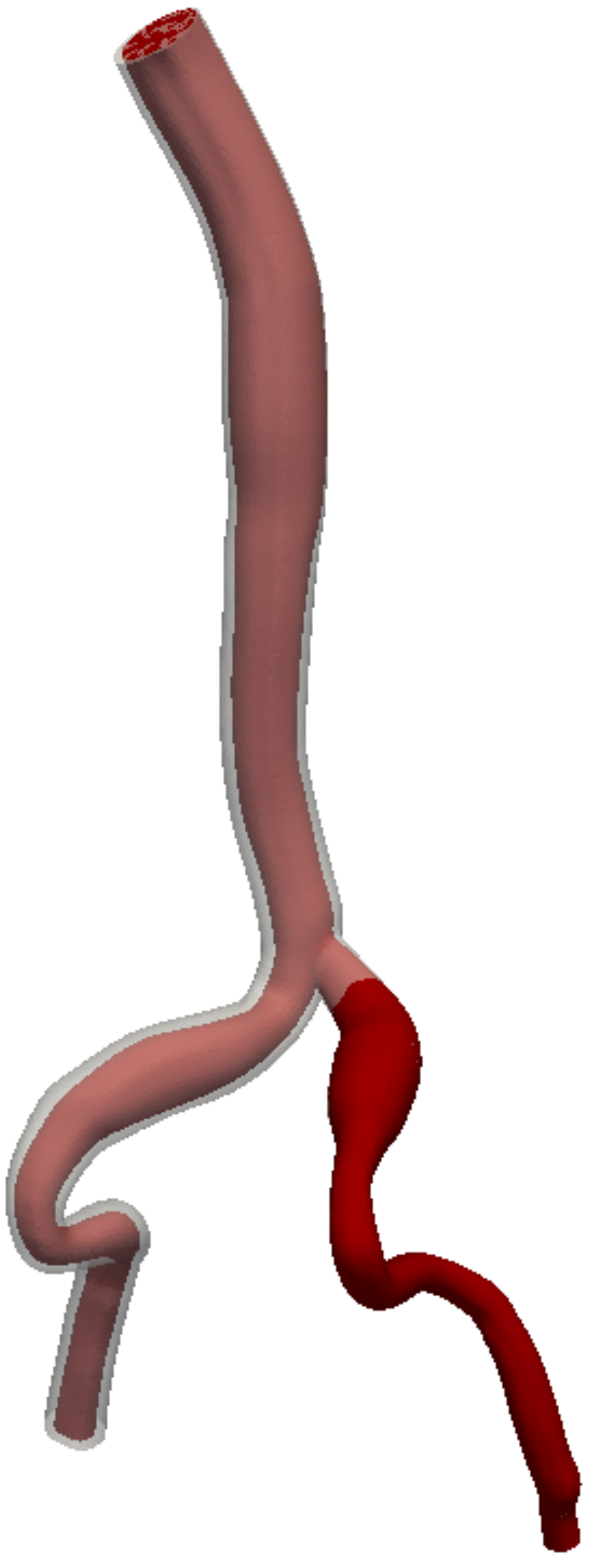}
\caption{$\lambda_1=16.86$}
\label{fig:results:0110pca1}
\end{subfigure}
$\quad\quad$
\begin{subfigure}[b]{0.12\textwidth}
\centering\includegraphics[width=\textwidth]{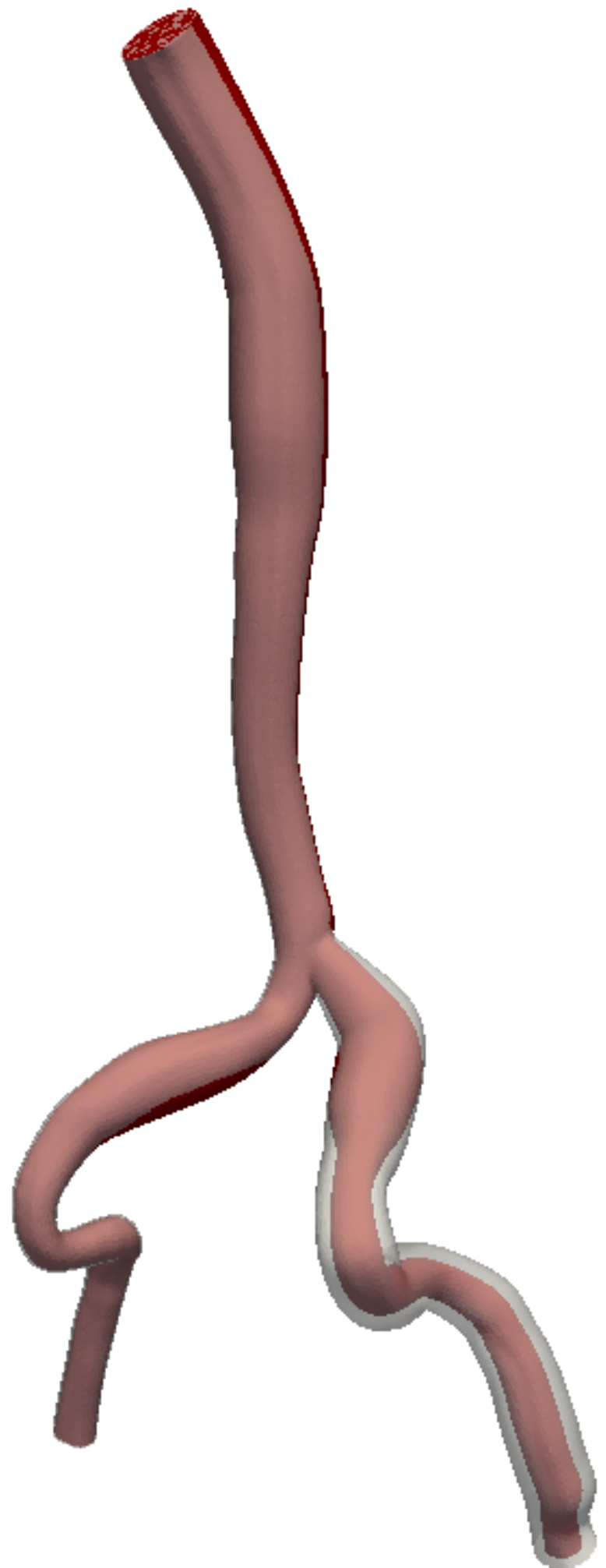}
\caption{$\lambda_2=13.80$}
\label{fig:results:0110pca2}
\end{subfigure}
$\quad\quad$
\begin{subfigure}[b]{0.12\textwidth}
\centering\includegraphics[width=\textwidth]{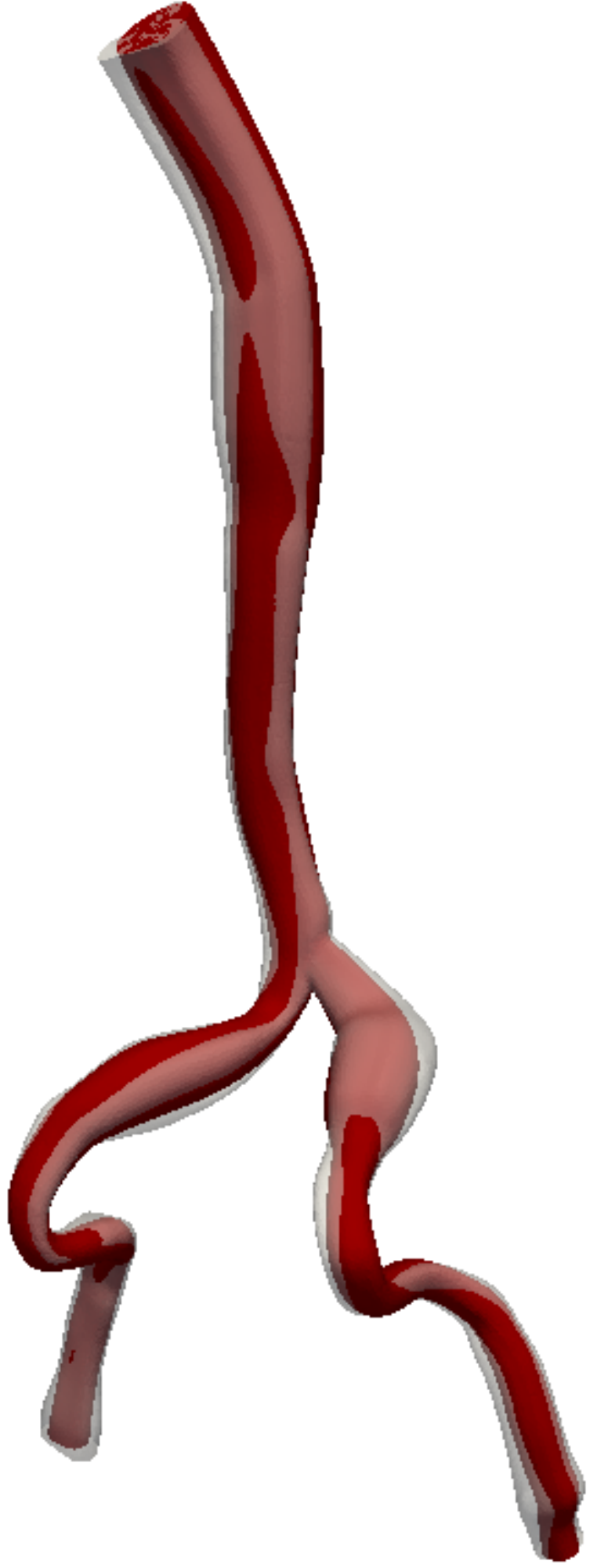}
\caption{$\lambda_{19}=3.00$}
\label{fig:results:0110pca3}
\end{subfigure}
$\quad\quad$
\begin{subfigure}[b]{0.12\textwidth}
\centering\includegraphics[width=\textwidth]{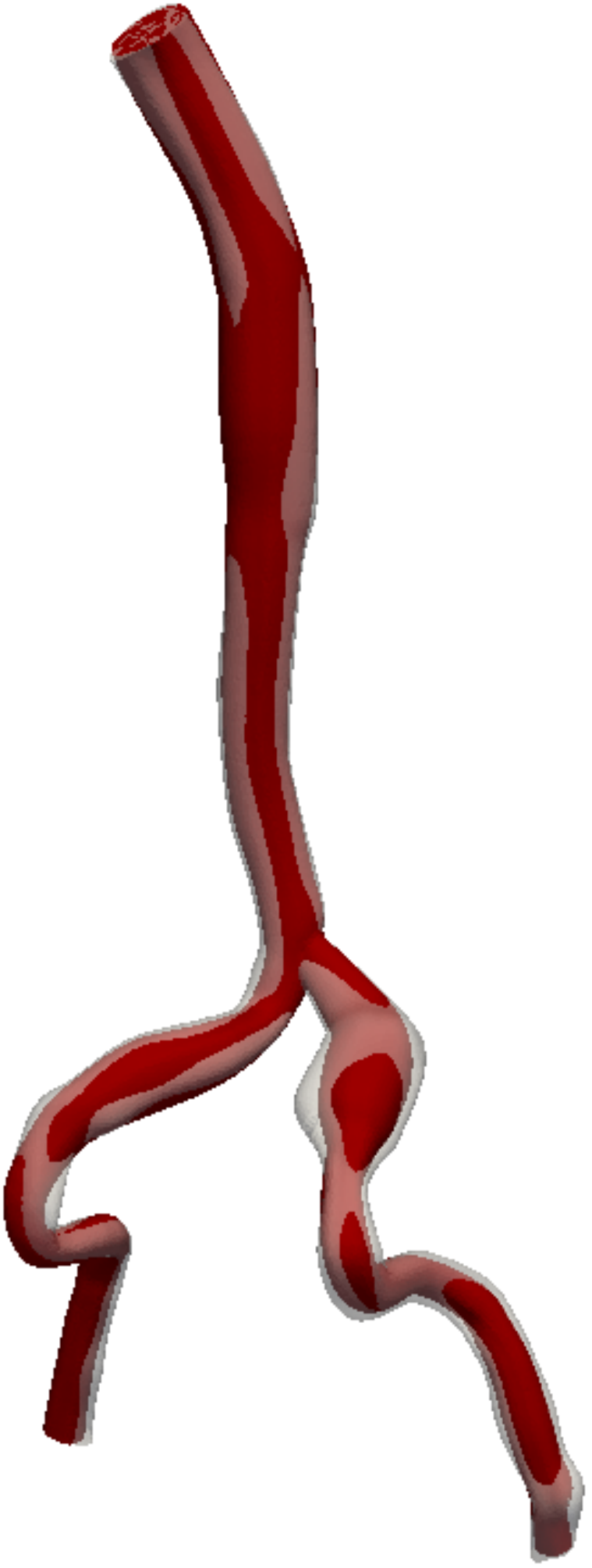}
\caption{$\lambda_{20}=2.94$}
\label{fig:results:0110pca4}
\end{subfigure}
\begin{subfigure}[b]{0.3\textwidth}
\centering\includegraphics[width=\textwidth]{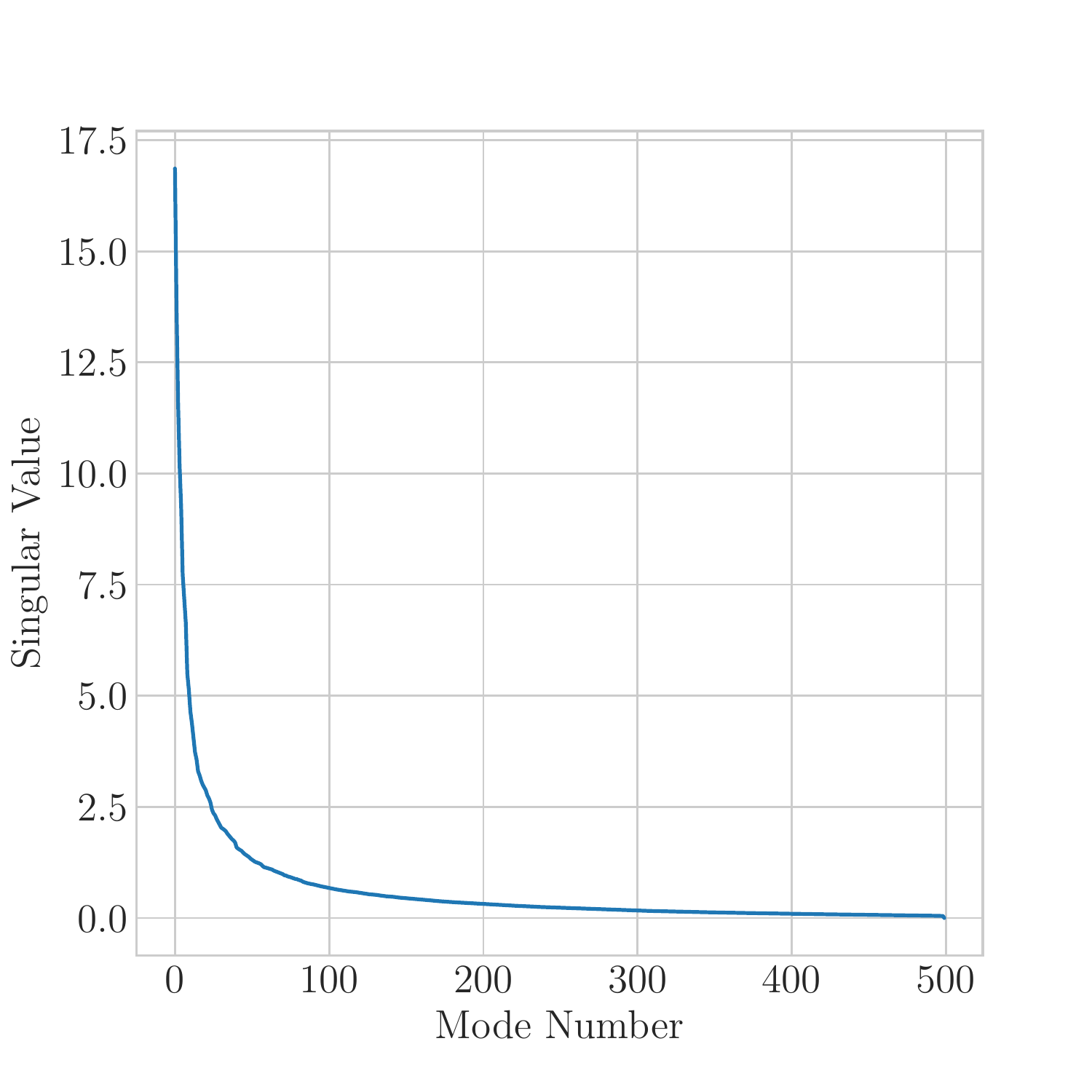}
\caption{Singular values}
\label{fig:results:0110pca4}
\end{subfigure}
\caption{PCA modes overlayed on mean aorto-iliac bifurcation model geometry. }\label{fig:0110pca}
\end{figure}

Dominant modes from PCA appear to perturb whole vessels, such as the right iliac or large segments of the aorta (Figures~\ref{fig:results:0110pca1}-\ref{fig:results:0110pca4}) and to increase with the reduction in the vessel radius. Conversely, higher modes involve local geometrical changes such as the bulbous region at the proximal end of the left iliac. These latter modes, however, contribute less, being associated with much smaller singular values (e.g. the 19th and 20th mode have singular values that are a factor of approximately 5.7 smaller than the first mode). Despite generating segmentations independently, this confirms that the geometric variability produced by the proposed dropout network is distributed across the entire model, and increases with small vessel radii.

\begin{figure}[!ht]
\centering
\includegraphics[width=0.65\textwidth]{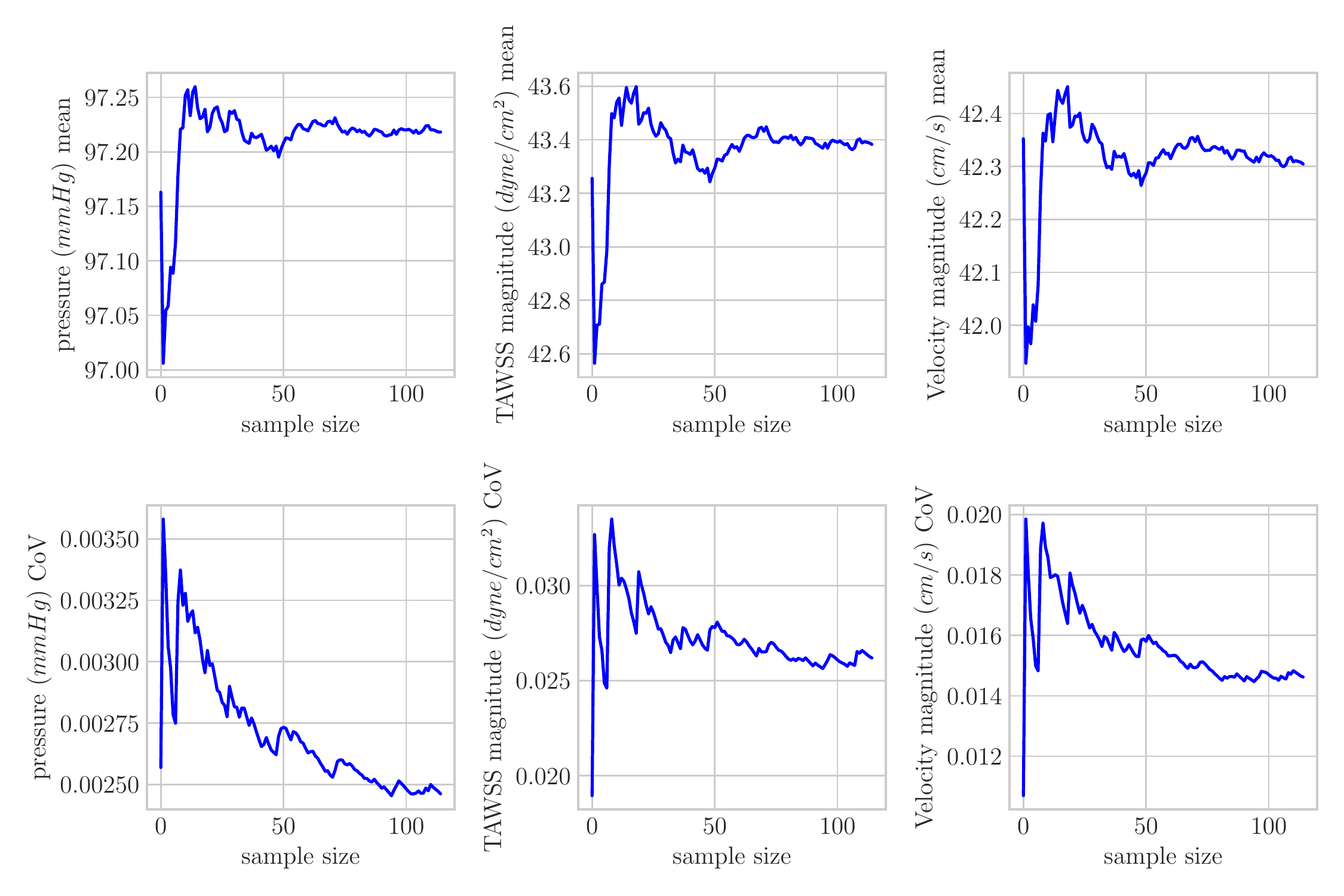}
\caption{Monte Carlo moment traces for aorto-iliac bifurcation model QoIs.}
\label{fig:results:mvconvergeaorta0110}
\end{figure}

Convergence of Monte Carlo statistics such as the mean, standard deviation and CoV for the pressure, TAWSS and velocity magnitude integrated over the aorta appears to be satisfactory (Figure~\ref{fig:results:mvconvergeaorta0110}), with 95\% confidence intervals for the mean showing faster convergence for the pressure, then velocity magnitude and finally TAWSS.

\begin{figure}[!ht]
\centering
\begin{subfigure}[b]{0.48\textwidth}
\centering\includegraphics[width=\textwidth]{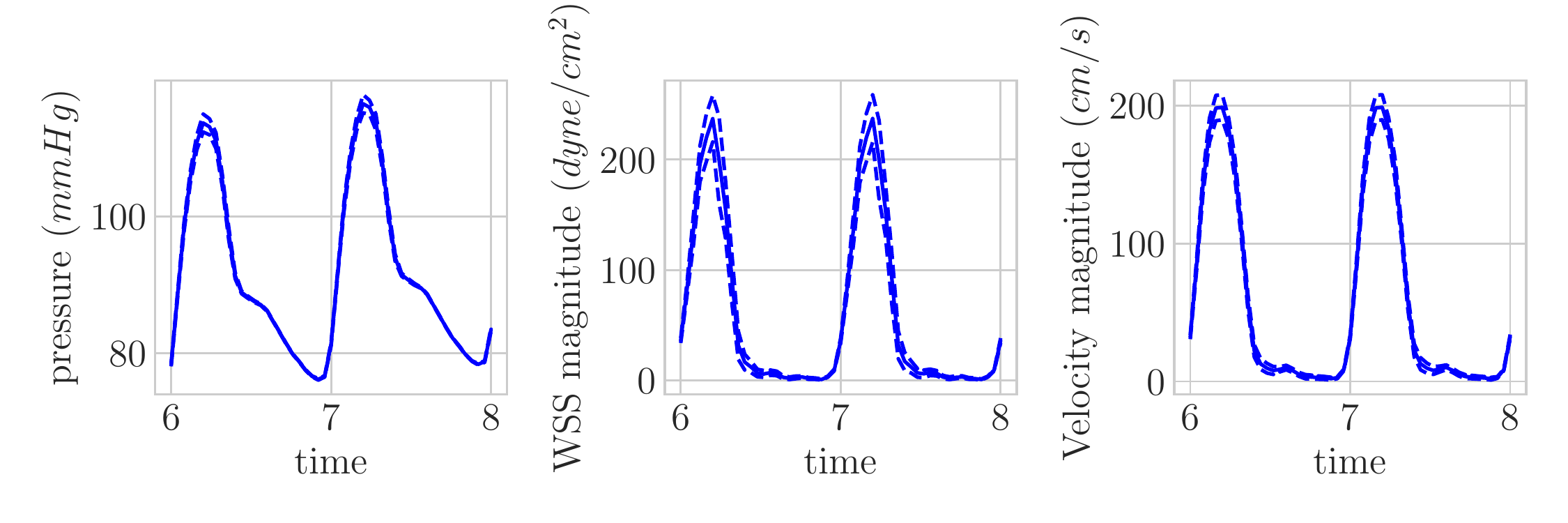}
\caption{Aorta}
\label{fig:results:outlet0110aorta}
\end{subfigure}
\begin{subfigure}[b]{0.48\textwidth}
\centering\includegraphics[width=\textwidth]{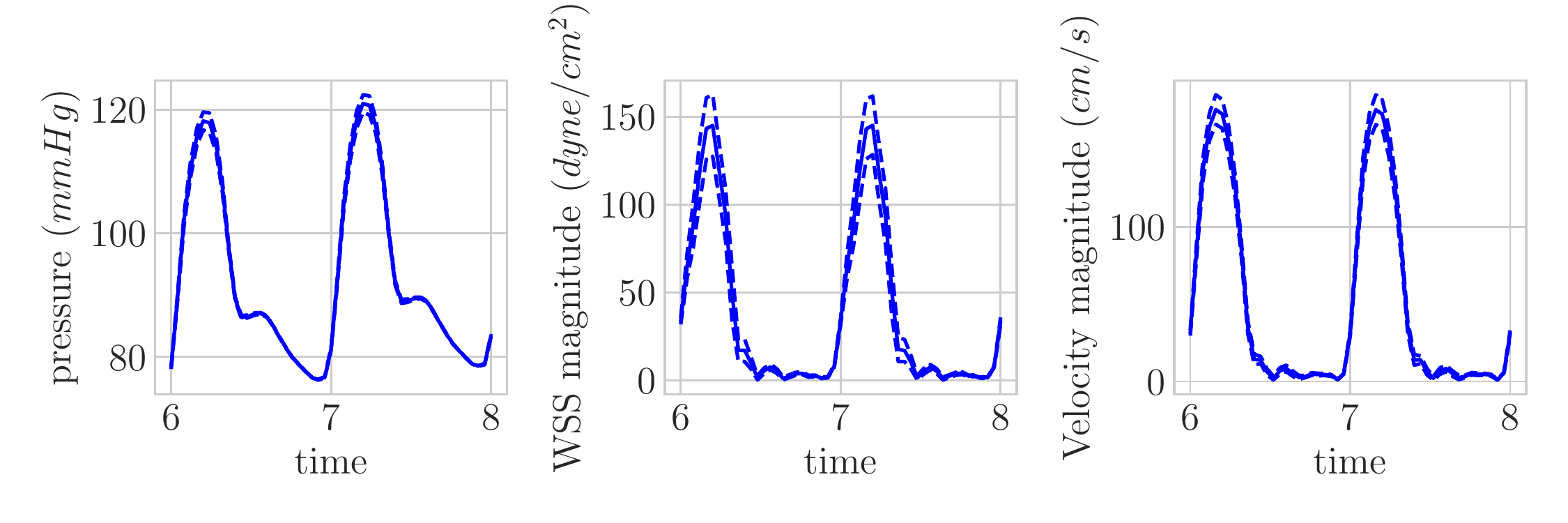}
\caption{Left Iliac}
\label{fig:results:outlet0110iliac}
\end{subfigure}
\caption{Outlet QoIs and $\pm 2\sigma$ interval for aorta-iliac bifurcation model.}
\end{figure}

The larger sensitivity of TAWSS to geometric uncertainty can be observed from their time histories over the last two cardiac cycles, as shown in Figures~\ref{fig:results:outlet0110aorta}, \ref{fig:results:outlet0110iliac}.
Pressure variability tends to decrease towards the distal end of the vessel, whereas TAWSS and velocity magnitude show an opposite trend. This relates to an increase in the wall shear stress and velocity after the bifurcation which, for this model, amplifies the effect of the geometric uncertainty for these two QoIs. Pressure uncertainty instead depends on the variability of the vessel resistance which cumulates the contributions of each uncertain segmentation along the vessel.

CoVs were found to be approximately 0.4\%, 1.5\% and 3\% for pressure, velocity magnitude and TAWSS, respectively. Thus, CoVs for TAWSS and velocity magnitude are roughly a factor of 10 and 5 larger compared to the pressure CoV, highlighting the increased sensitivity of TAWSS and velocity to geometry variation (see Table~\ref{tab:results:0110cv}).

%
%

\begin{table}[!ht]
\centering
\caption{Monte Carlo sample mean, coefficient of variation (CoV) and 95\% relative confidence interval for all QoIs in the aorto-iliac bifurcation model. $n$ indicates the number of cross-sectional slices for the associated vessel.}\label{tab:results:0110conf}
\resizebox{\textwidth}{!}{
\begin{tabular}{lcc | lcc}
\toprule
{\bf Path} & {\bf Aorta} $(n=95)$ & {\bf left iliac} $(n=61)$ & {\bf Path} & {\bf Aorta} & {\bf left iliac}\\
\midrule
{\bf Radius mean} [cm] & 0.84 & 0.61 & {\bf TAWSS mean} [dyne/cm$^2$] & 40.24 & 46.55\\ 
{\bf Radius CoV} & 0.006 & 0.011 & {\bf TAWSS CoV} & 0.027 & 0.030\\
{\bf Radius conf.} & 0.0012 & 0.0020 & {\bf TAWSS conf.} & 0.0050 & 0.0056\\
\midrule
{\bf Pressure mean} [mmHg] & 98.43 & 96.02 & {\bf Velocity mean} [cm/s] & 42.84 & 41.78\\
{\bf Pressure CoV} & 0.003 & 0.002 & {\bf Velocity CoV} & 0.014 & 0.019\\
{\bf Pressure conf.} & 0.0005 & 0.0004 & {\bf Velocity conf.} & 0.0026 & 0.0035\\
\bottomrule
\end{tabular}}
\label{tab:results:0110cv}
\end{table}

\begin{figure}[!ht]
\centering
\begin{subfigure}[b]{0.48\textwidth}
\centering\includegraphics[width=\textwidth]{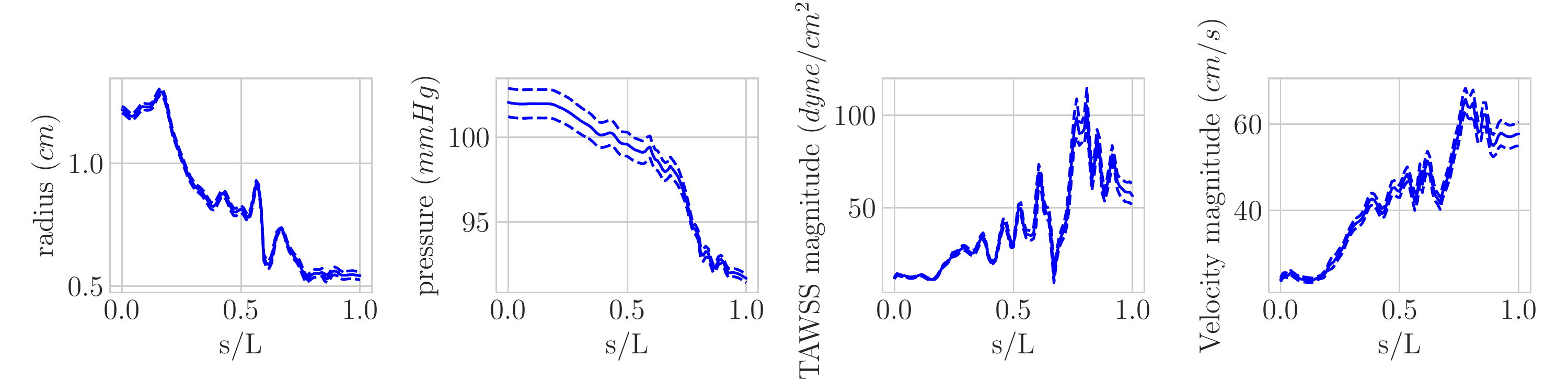}
\caption{Aorta}
\label{fig:results:timeavgaorta0110}
\end{subfigure}
\begin{subfigure}[b]{0.48\textwidth}
\centering\includegraphics[width=\textwidth]{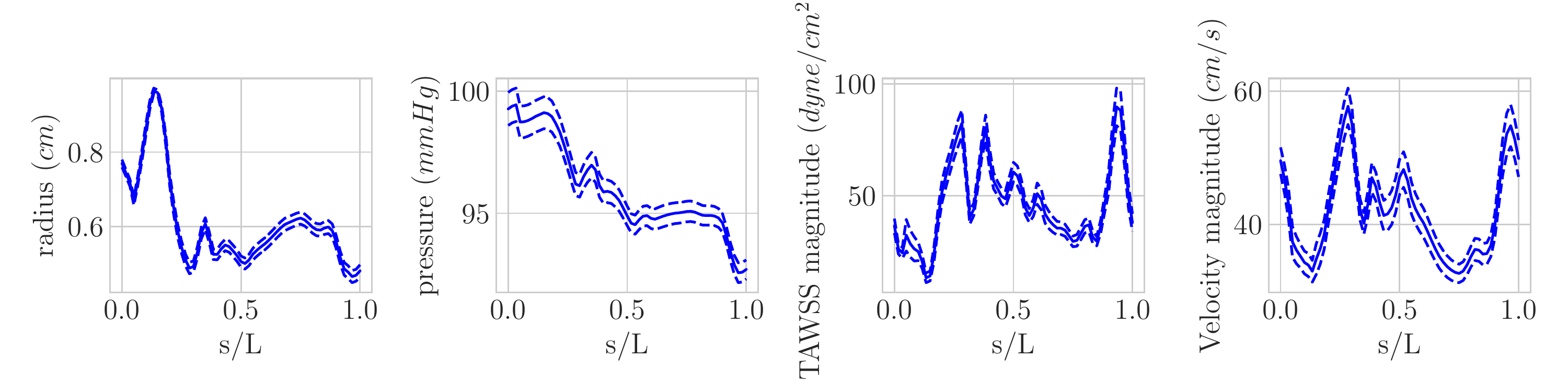}
\caption{left Iliac}
\label{fig:results:timeavgiliac0110}
\end{subfigure}
\caption{Time averaged QoIs and $\pm 2 \sigma$ interval for aorto-iliac bifurcation model, plotted along the vessel centerline.}
\end{figure}



\begin{figure}[!ht]
\centering
\begin{subfigure}[b]{0.15\textwidth}
\centering\includegraphics[width=\textwidth]{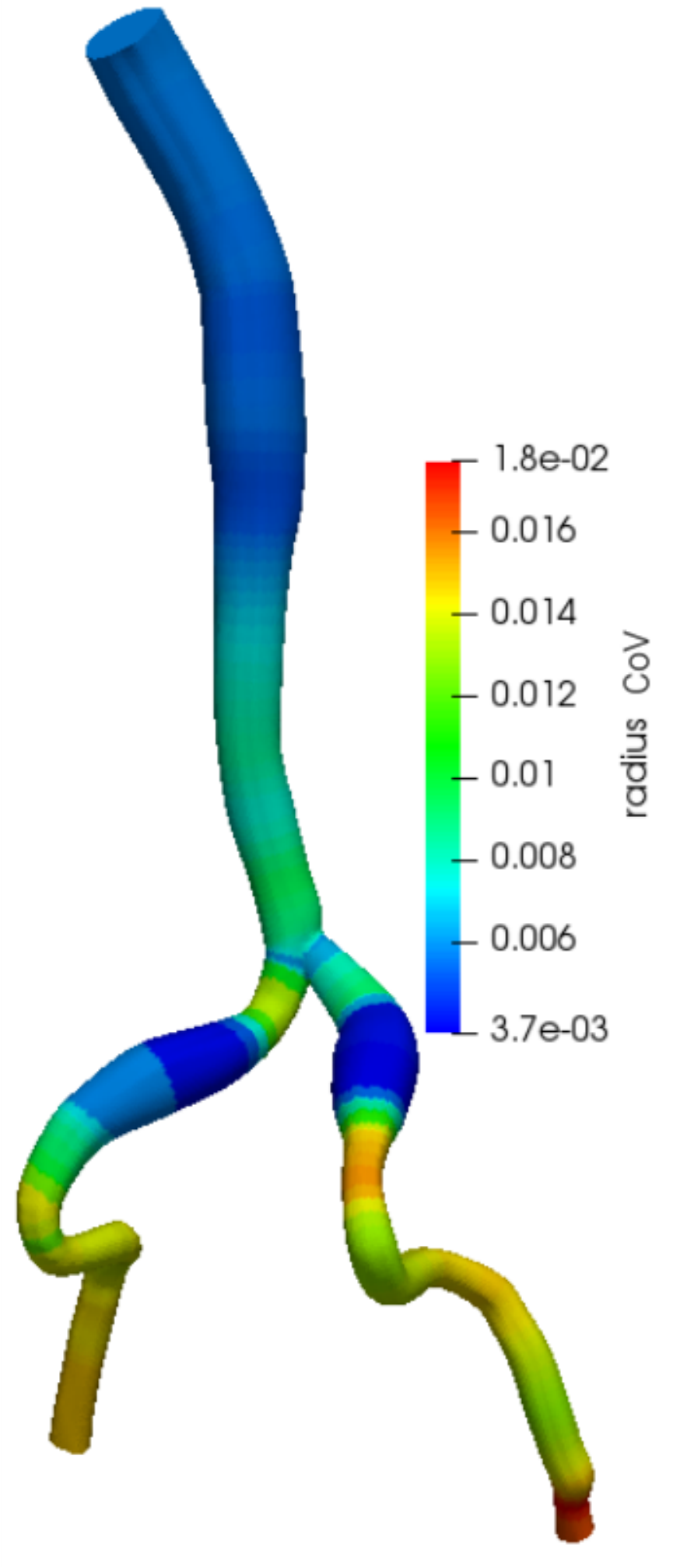}
\caption{}
\label{fig:results:01103dradius}
\end{subfigure}
$\quad\quad$
\begin{subfigure}[b]{0.15\textwidth}
\centering\includegraphics[width=\textwidth]{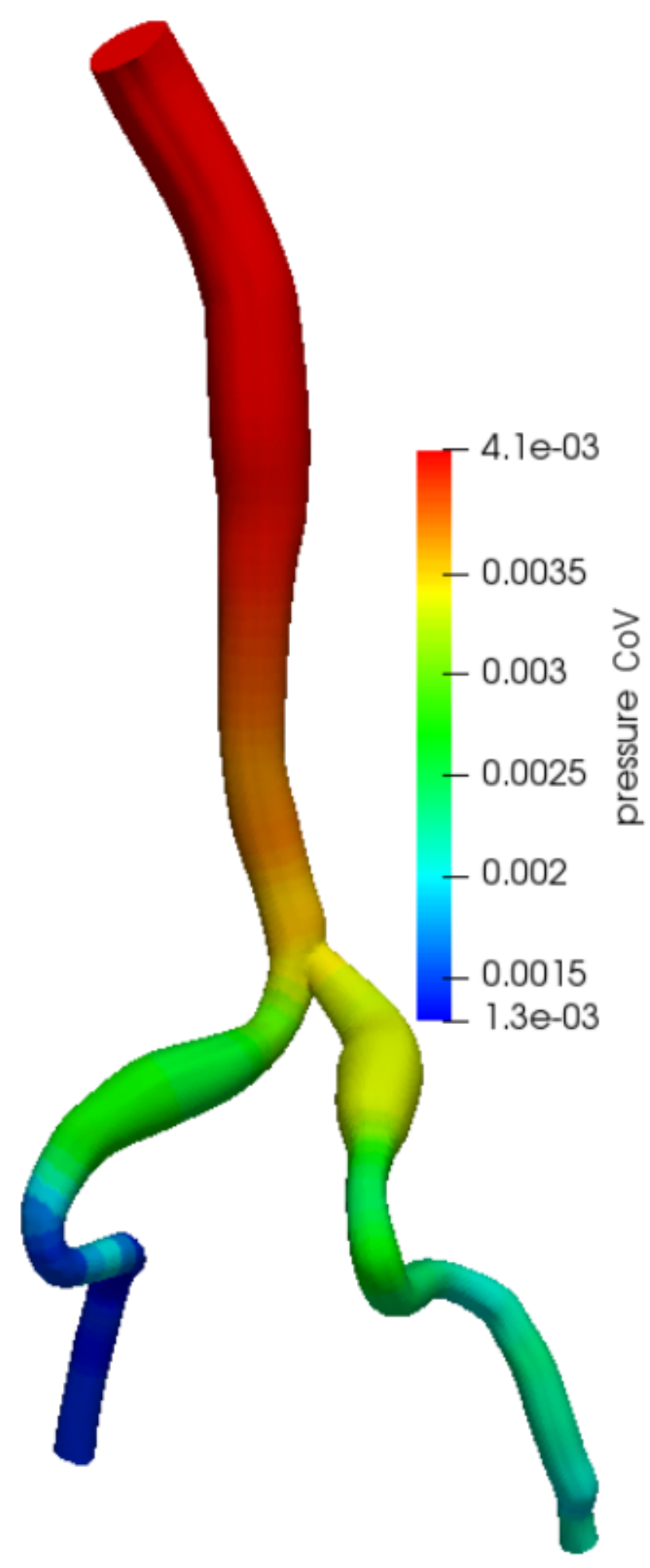}
\caption{}
\label{fig:results:01103dpressure}
\end{subfigure}
$\quad\quad$
\begin{subfigure}[b]{0.15\textwidth}
\centering\includegraphics[width=\textwidth]{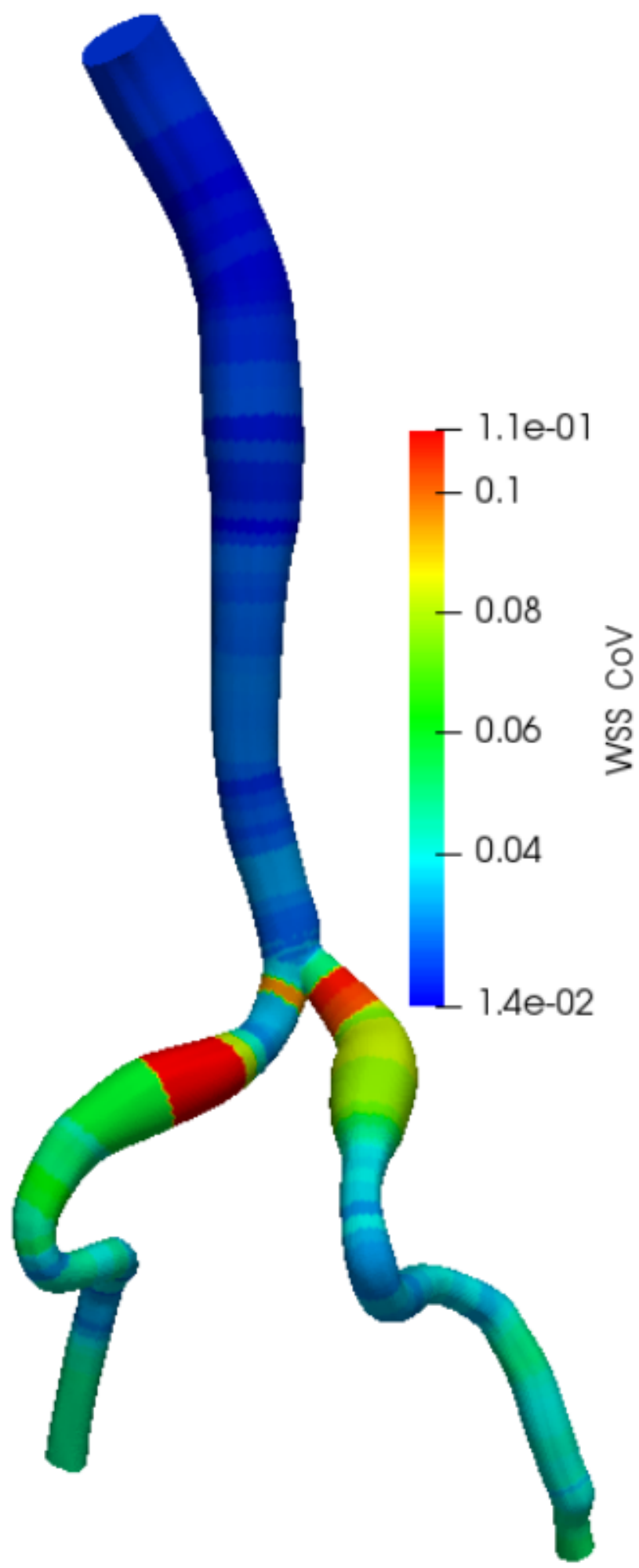}
\caption{}
\label{fig:results:01103dwss}
\end{subfigure}
$\quad\quad$
\begin{subfigure}[b]{0.15\textwidth}
\centering\includegraphics[width=\textwidth]{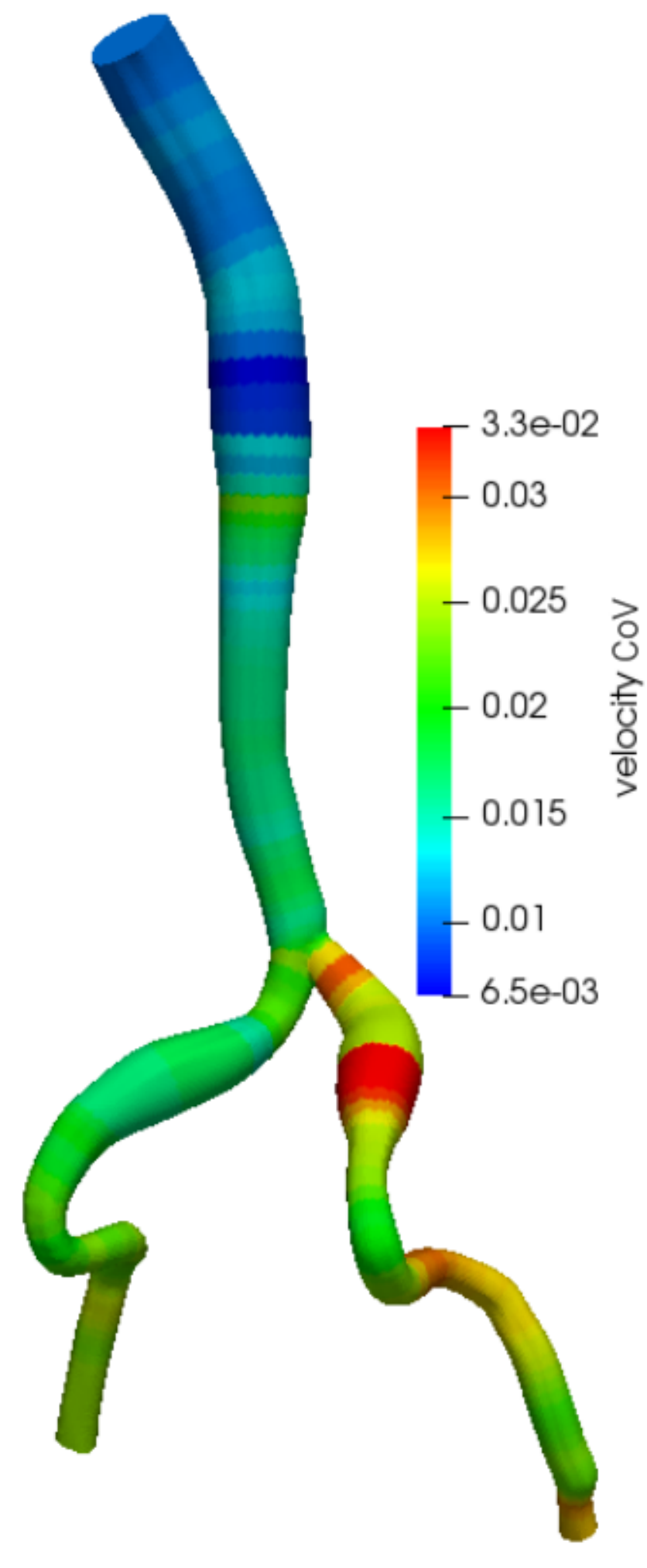}
\caption{}
\label{fig:results:01103dvelocity}
\end{subfigure}
\caption{Nearest neighbor interpolation of cross-sectional time-averaged CoVs for aorto-iliac bifurcation model.}\label{fig:01103d}
\end{figure}

%% file: res_AAA.tex
\subsection{Abdominal Aortic Aneurysm Model}\label{section:paper3:0144}

\begin{figure}[!ht]
\centering
\begin{subfigure}[b]{0.4\textwidth}
\centering
\includegraphics[height=1.0\textwidth]{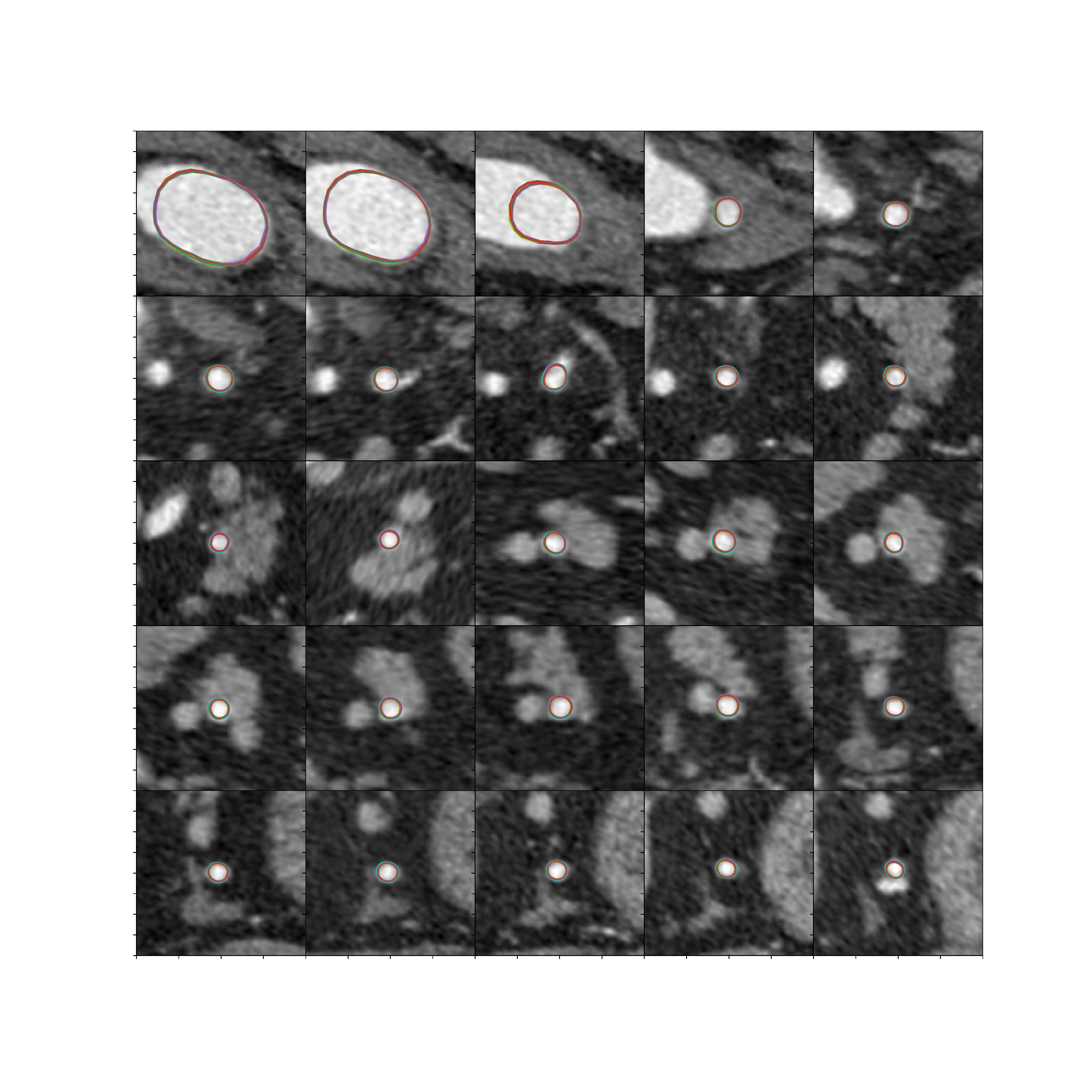}
\caption{Abdominal Aortic Aneurysm Case - Celiac Splenic}\label{fig:results:seg0144celiacsplenic}
\end{subfigure}
\begin{subfigure}[b]{0.29\textwidth}
\centering\includegraphics[scale=0.5]{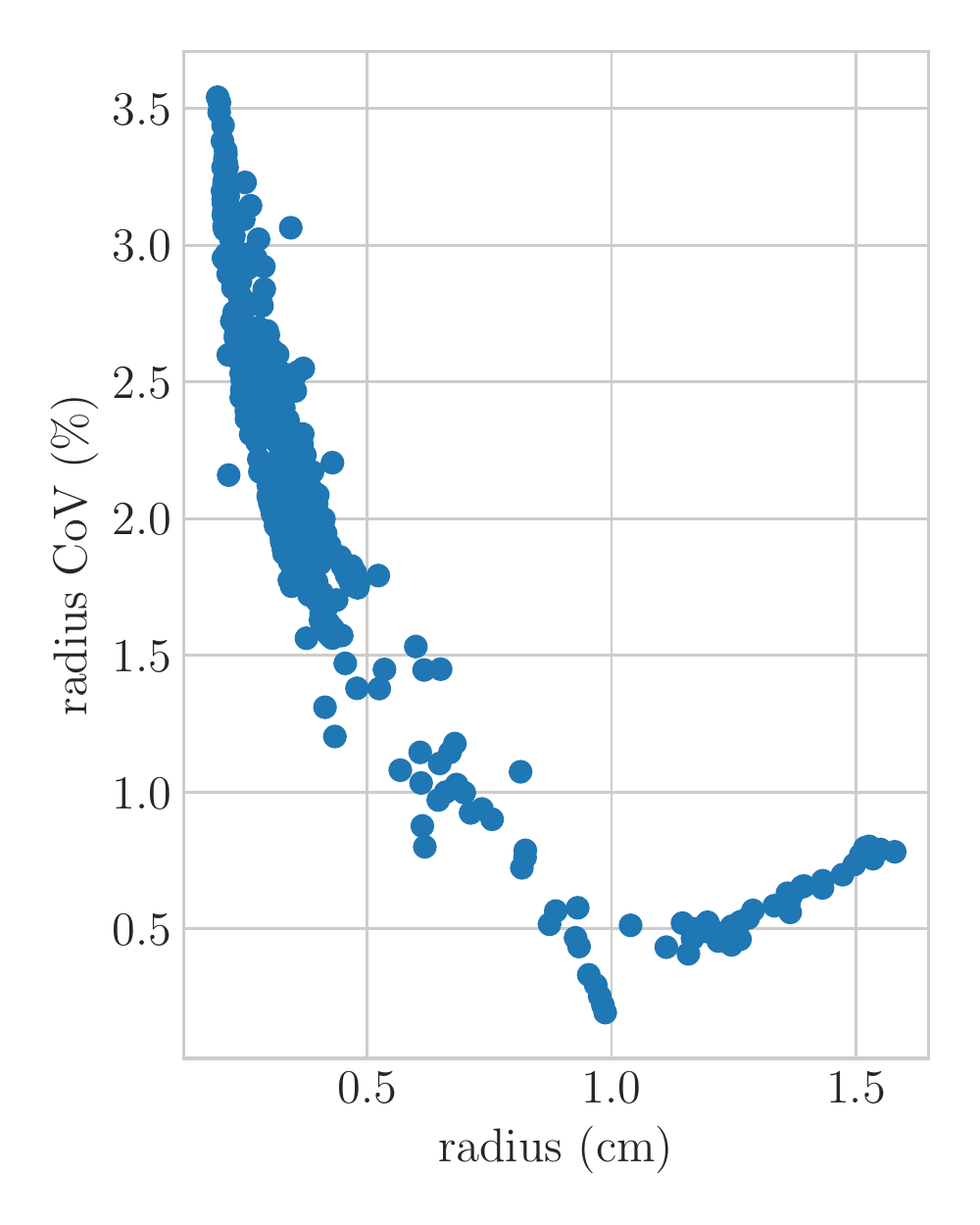}
\caption{Radius CoV}
\label{fig:results:rvrcv0144}
\end{subfigure}
\begin{subfigure}[b]{0.29\textwidth}
\centering\includegraphics[scale=0.5]{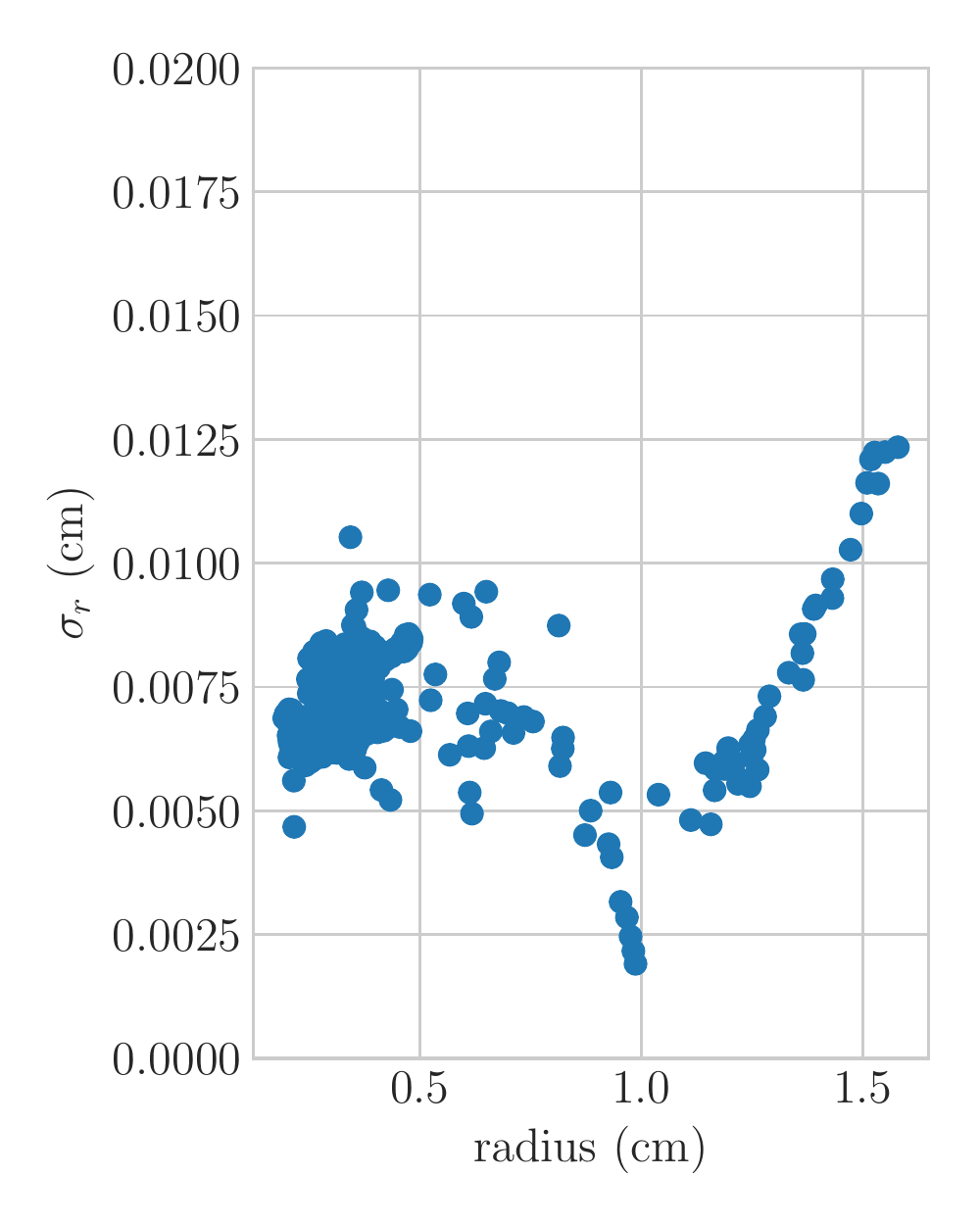}
\caption{Radius standard deviation}
\label{fig:results:rvrsig0144}
\end{subfigure}
\caption{Lumen segmentation samples and radius CoV/standard deviation for abdominal aortic aneurysm test case, computed over cross-sectional slices $\mathbf{x}_i,\,i=1,\dots,581$.}
\label{fig:results:rgraph0144}
\end{figure}

\begin{figure}[!ht]
\centering
\begin{subfigure}[b]{0.125\textwidth}
\centering\includegraphics[width=\textwidth]{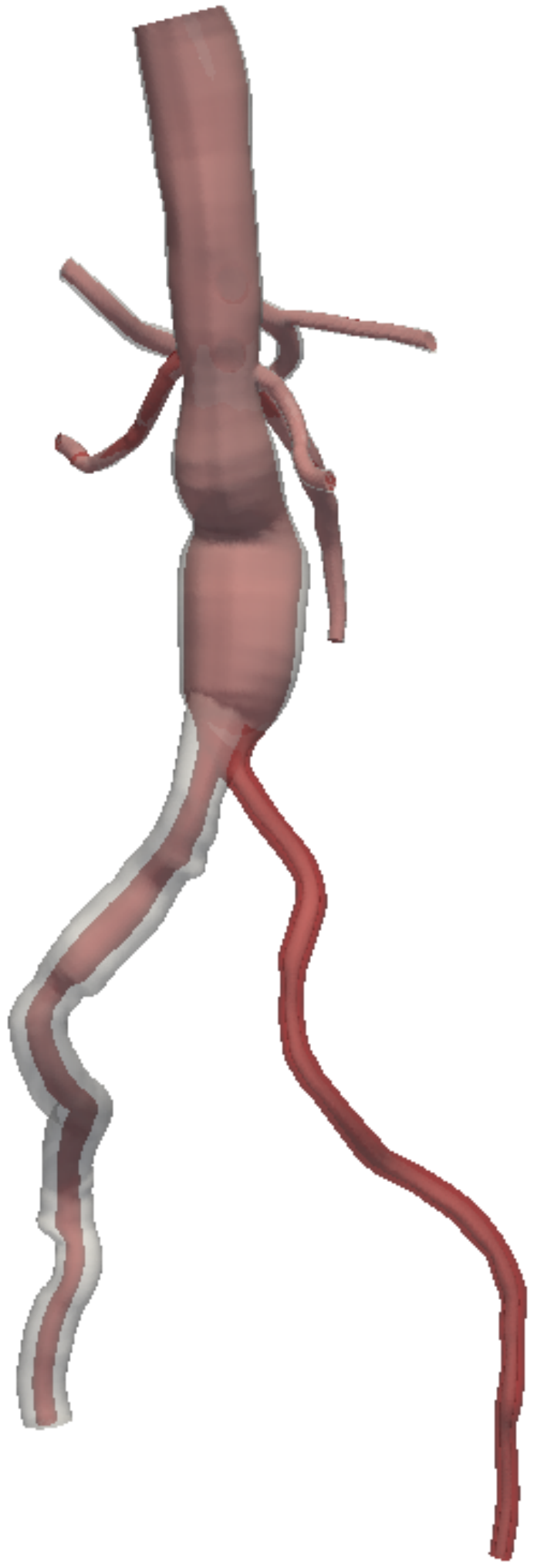}
\caption{$\lambda_1=24.60$}
\label{fig:results:0144pca1}
\end{subfigure}
$\quad\quad$
\begin{subfigure}[b]{0.12\textwidth}
\centering\includegraphics[width=\textwidth]{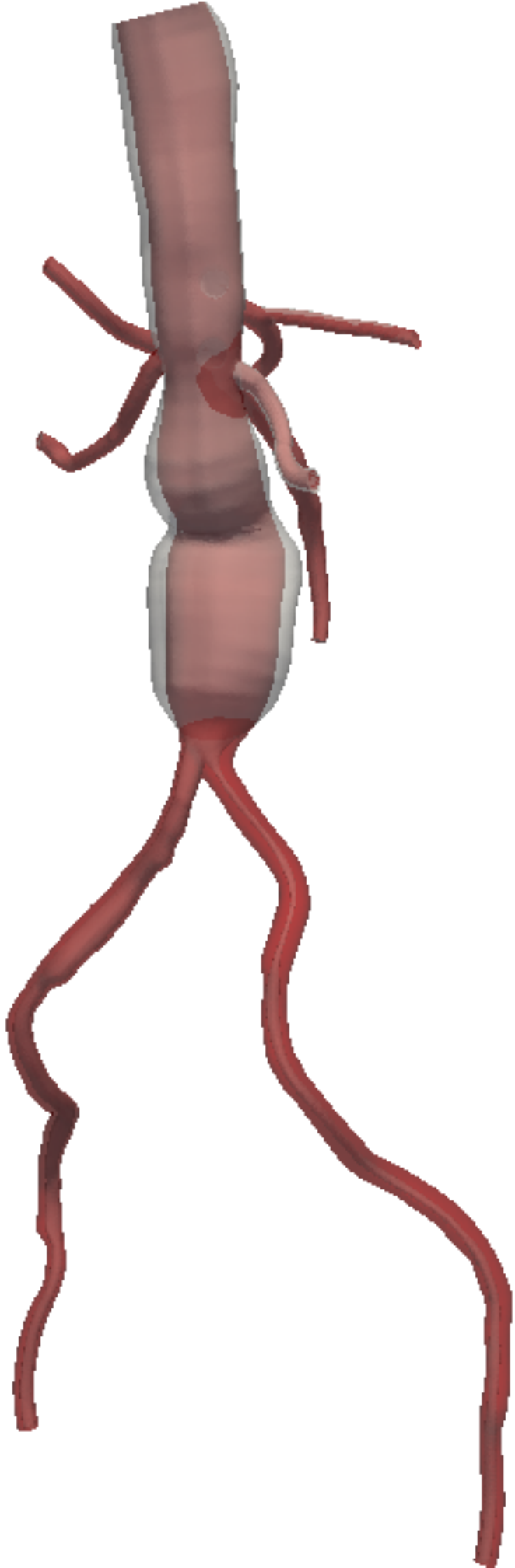}
\caption{$\lambda_2=21.41$}
\label{fig:results:0144pca2}
\end{subfigure}
$\quad\quad$
\begin{subfigure}[b]{0.12\textwidth}
\centering\includegraphics[width=\textwidth]{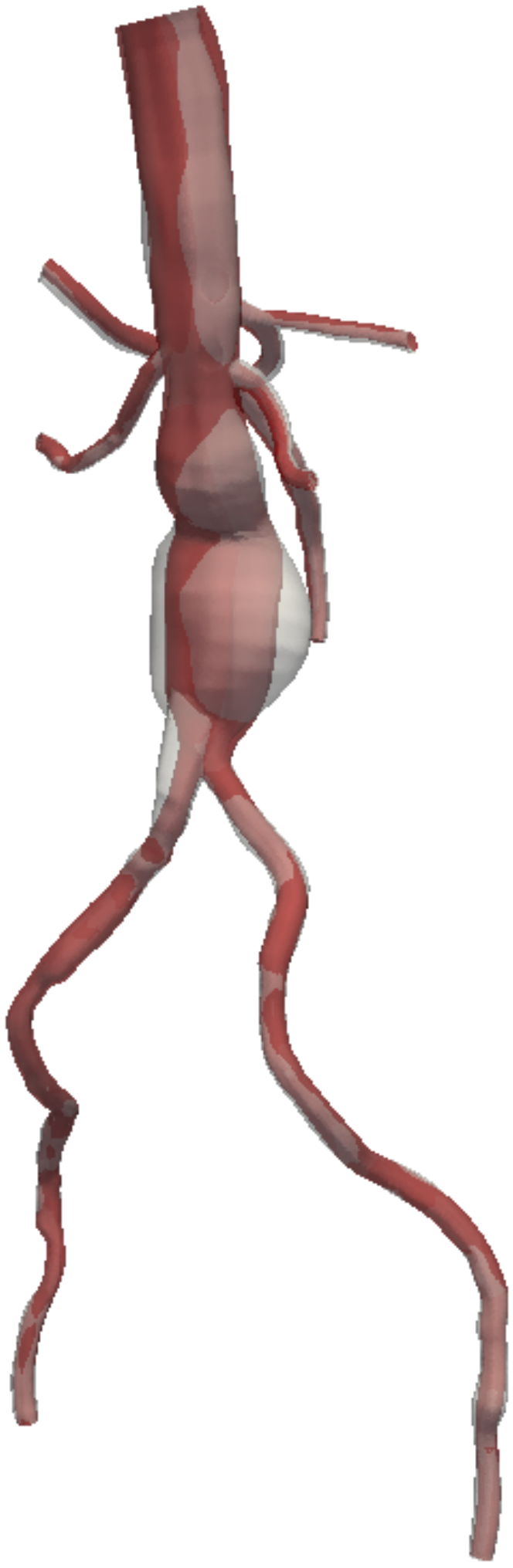}
\caption{$\lambda_{19}=6.28$}
\label{fig:results:0144pca3}
\end{subfigure}
$\quad\quad$
\begin{subfigure}[b]{0.12\textwidth}
\centering\includegraphics[width=\textwidth]{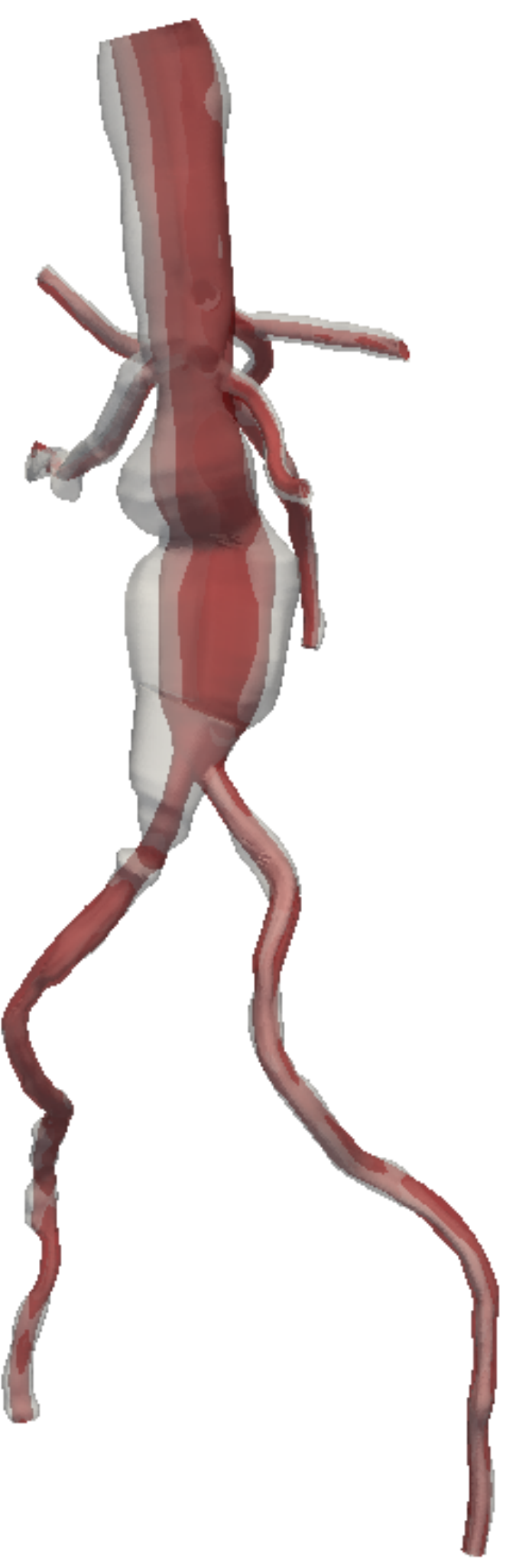}
\caption{$\lambda_{20}=5.94$}
\label{fig:results:0144pca4}
\end{subfigure}
\begin{subfigure}[b]{0.35\textwidth}
\centering\includegraphics[width=\textwidth]{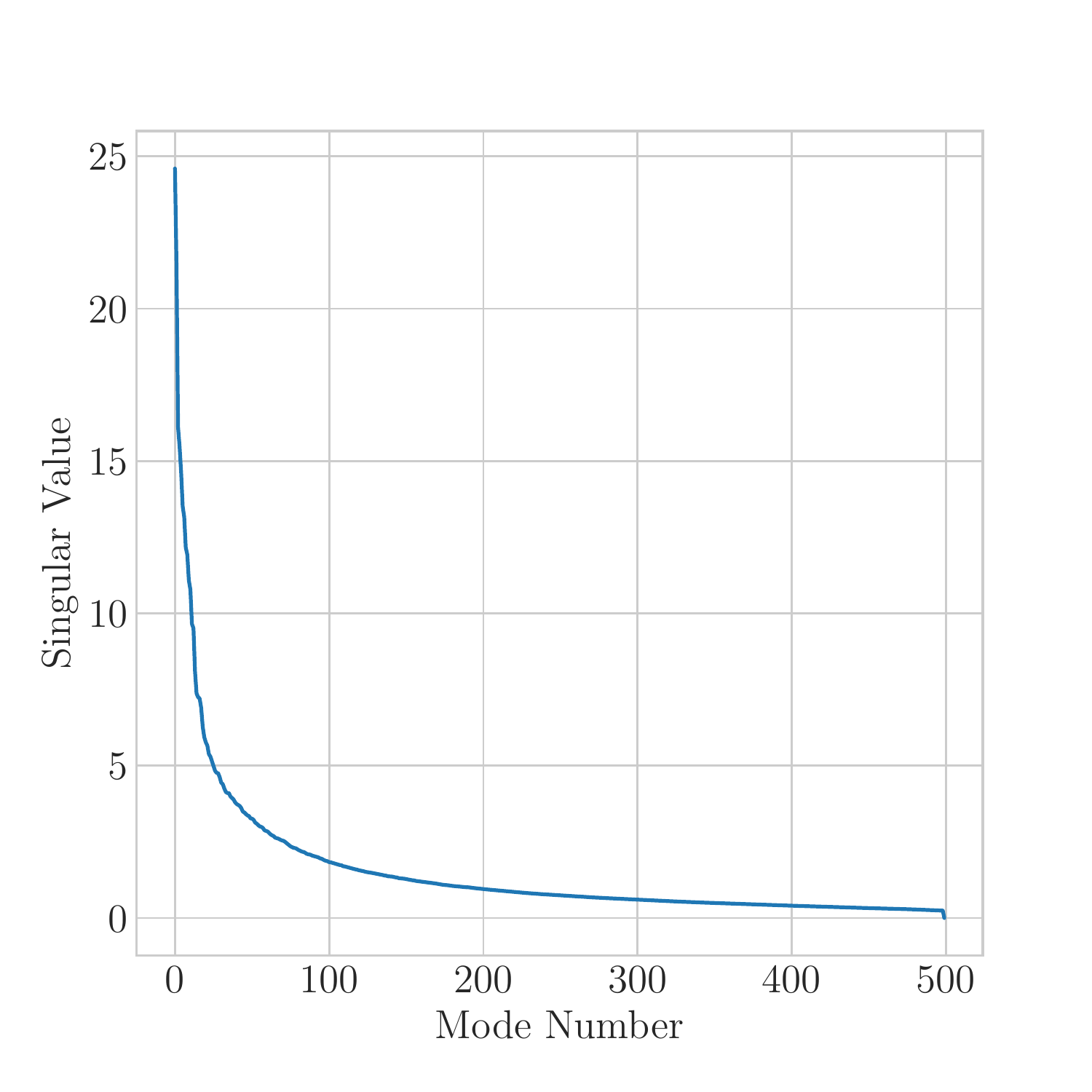}
\caption{Singular Values}
\label{fig:results:0144pca4}
\end{subfigure}
\caption{PCA modes overlayed on mean abdominal aortic aneurysm model geometry.}\label{fig:0144pca}
\end{figure}

The lumen generated by the network agrees well with the vessel lumen images, even in the presence of noise and bifurcations (see Figure~\ref{fig:results:seg0144celiacsplenic}). Values of $\sigma_{r}$ from the dropout network are generally in the range of 0.005-0.01 cm, with some outliers, particularly for larger vessel lumens. Segmentation of small vessels is affected by increased uncertainty, with radius CoV of 3\%, versus 1\% for the largest vessels.
PCA shape analysis shows the first two modes affecting large scale model features such as the aortic aneurysm as well as the entire aorta and iliac branches (see Figure~\ref{fig:results:0144pca1}). The 19\textsuperscript{th} and 20\textsuperscript{th} modes (with singular values a factor of four smaller than the first mode) are instead associated with local features, like celiac branches or by asymmetric aneurysm perturbations (see Figure~\ref{fig:results:0144pca3}).

\begin{figure}[!ht]
	\centering
	\includegraphics[scale=0.475]{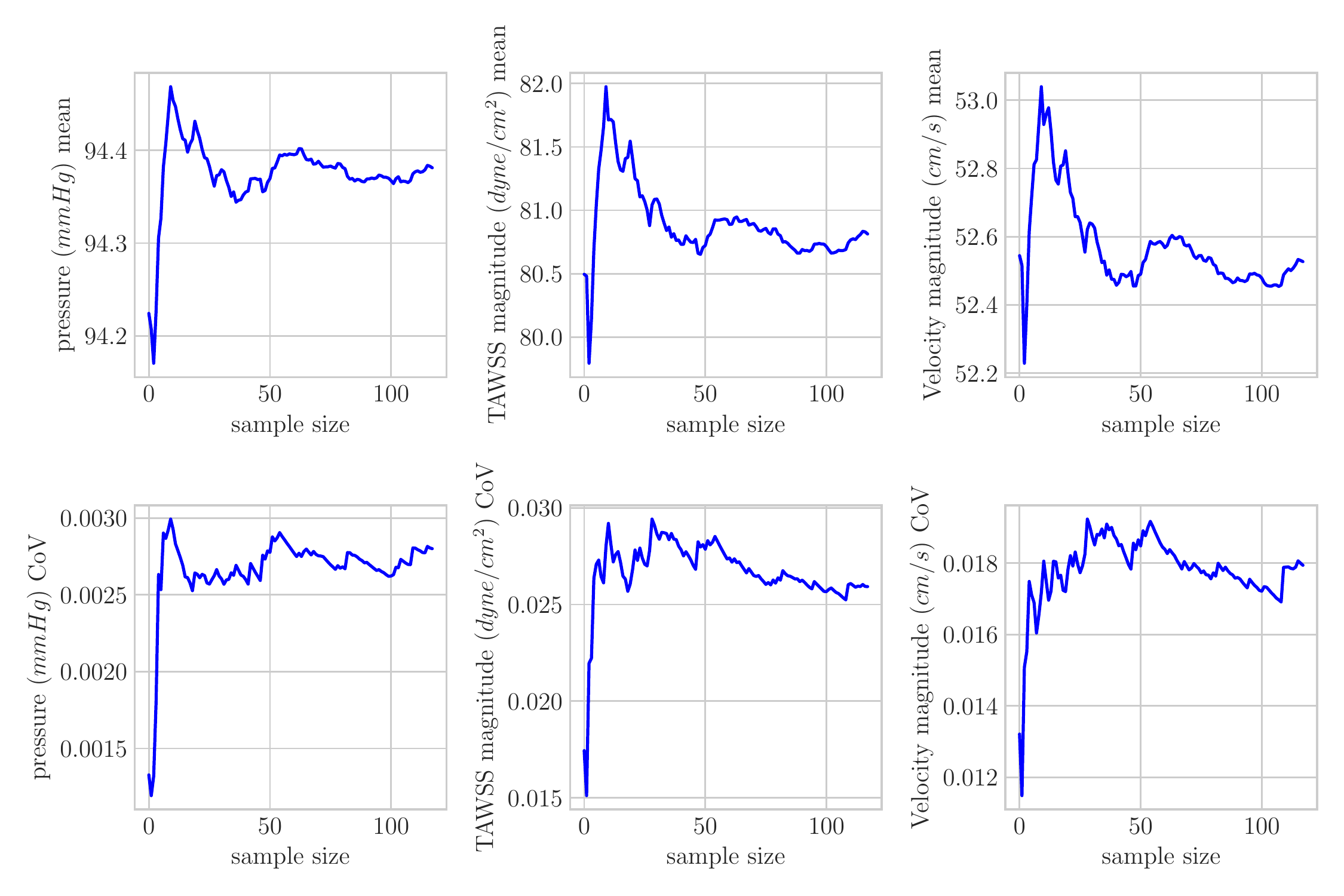}
	\caption{Monte carlo convergence - Abdominal Aortic Aneurysm case}
	\label{fig:results:mvconvergeaorta0144}
\end{figure}

\begin{figure}[!ht]
\centering
\begin{subfigure}[b]{0.48\textwidth}
\centering\includegraphics[width=\textwidth]{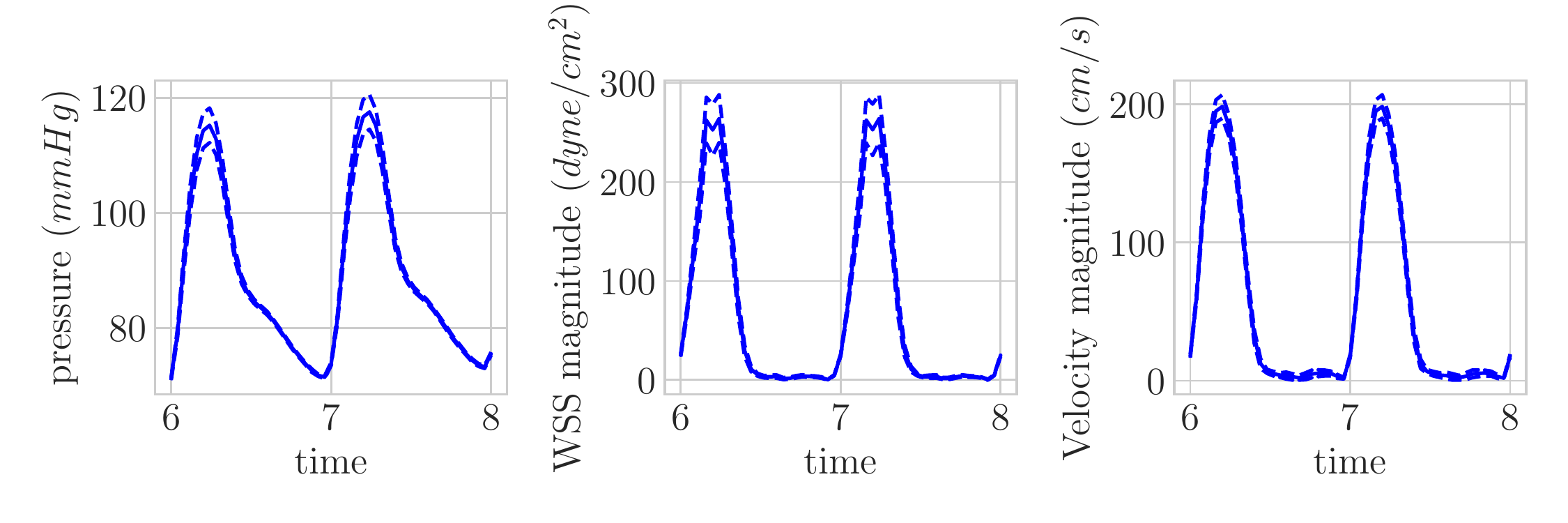}
\caption{Aorta}
\label{fig:results:outlet0144aorta}
\end{subfigure}
\begin{subfigure}[b]{0.48\textwidth}
\centering\includegraphics[width=\textwidth]{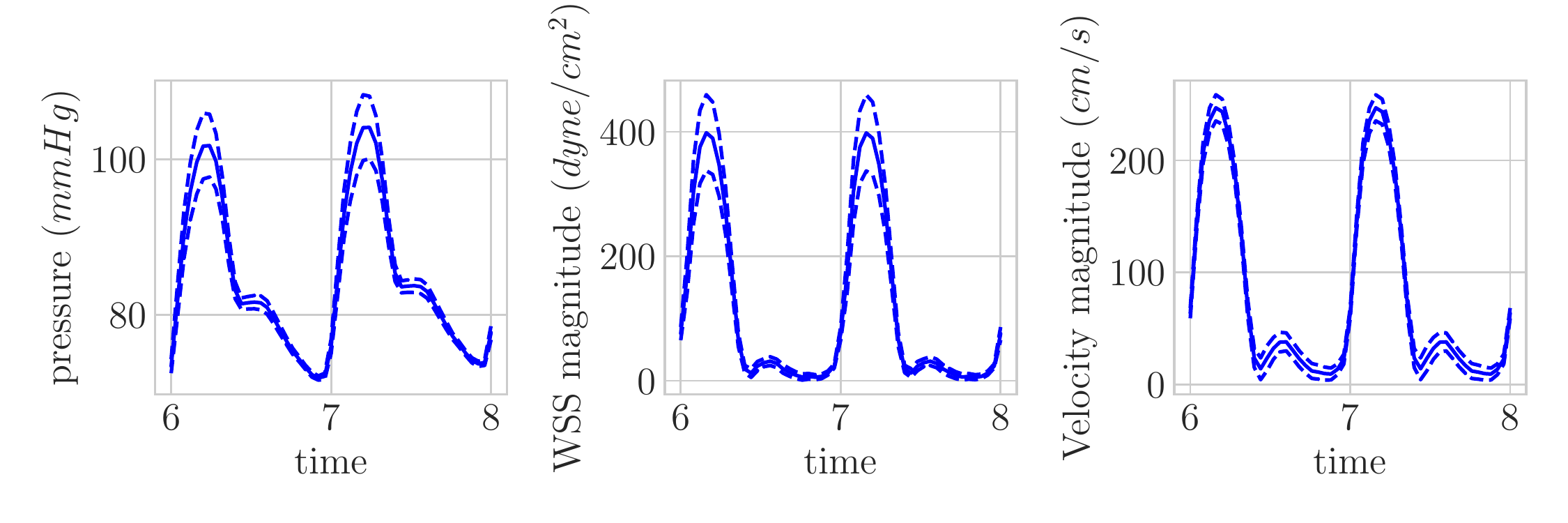}
\caption{Celiac Hepatic}
\label{fig:results:outlet0144celiachepatic}
\end{subfigure}
	
\begin{subfigure}[b]{0.48\textwidth}
\centering\includegraphics[width=\textwidth]{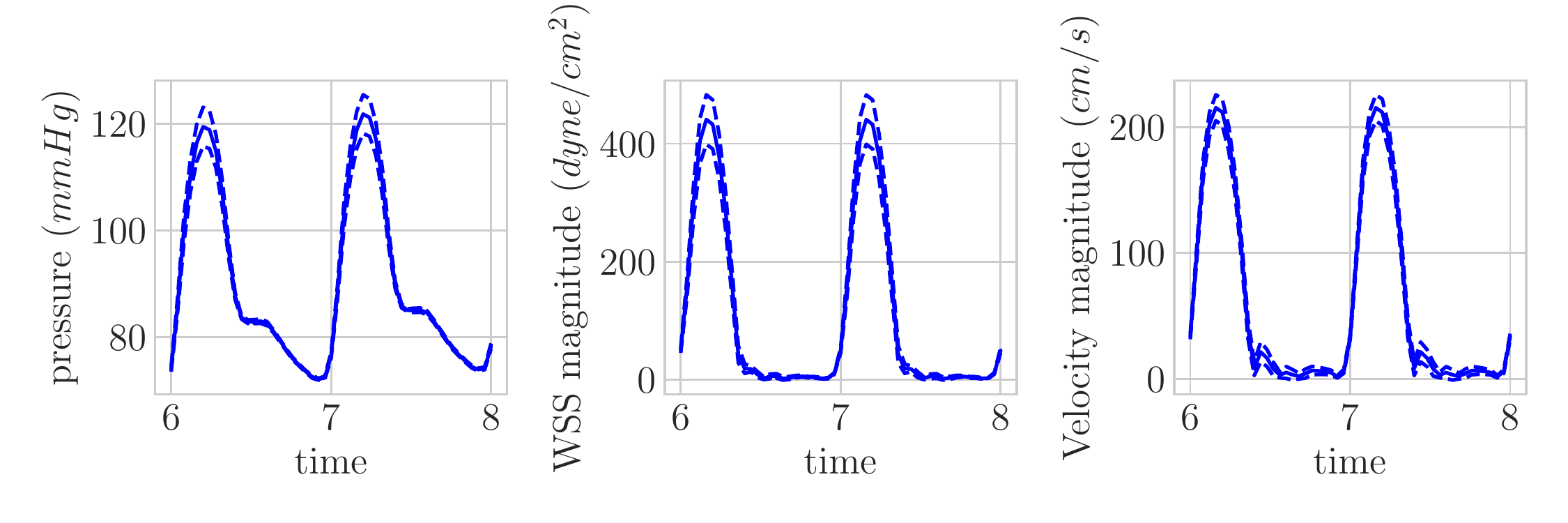}
\caption{Celiac Splenic}
\label{fig:results:outlet0144celiacsplenic}
\end{subfigure}
\begin{subfigure}[b]{0.48\textwidth}
\centering\includegraphics[width=\textwidth]{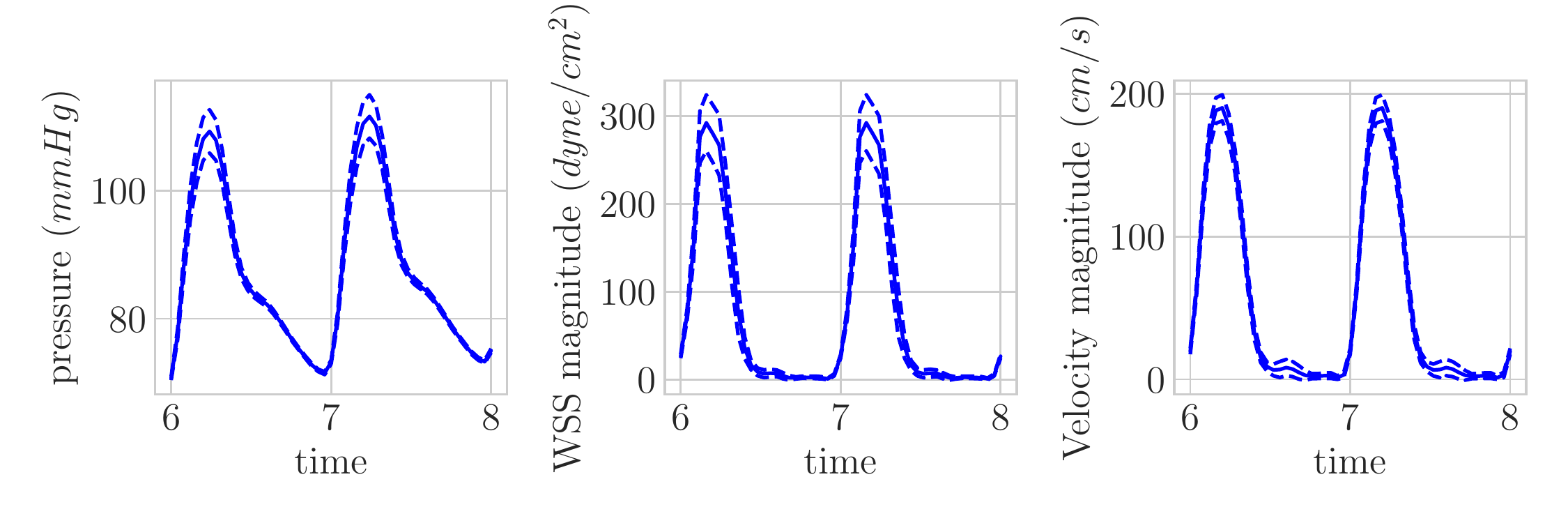}
\caption{Left external Iliac}
\label{fig:results:outlet0144liliac}
\end{subfigure}
	
\begin{subfigure}[b]{0.48\textwidth}
\centering\includegraphics[width=\textwidth]{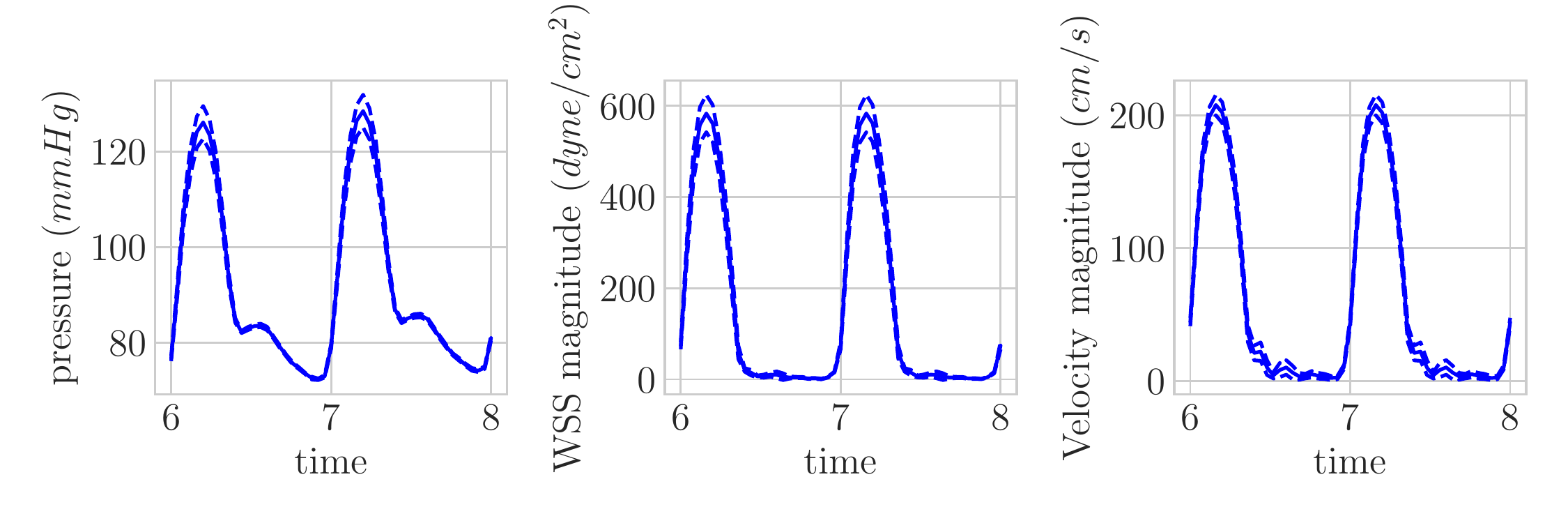}
\caption{Left Renal}
\label{fig:results:outlet0144lrenal}
\end{subfigure}
\begin{subfigure}[b]{0.48\textwidth}
\centering\includegraphics[width=\textwidth]{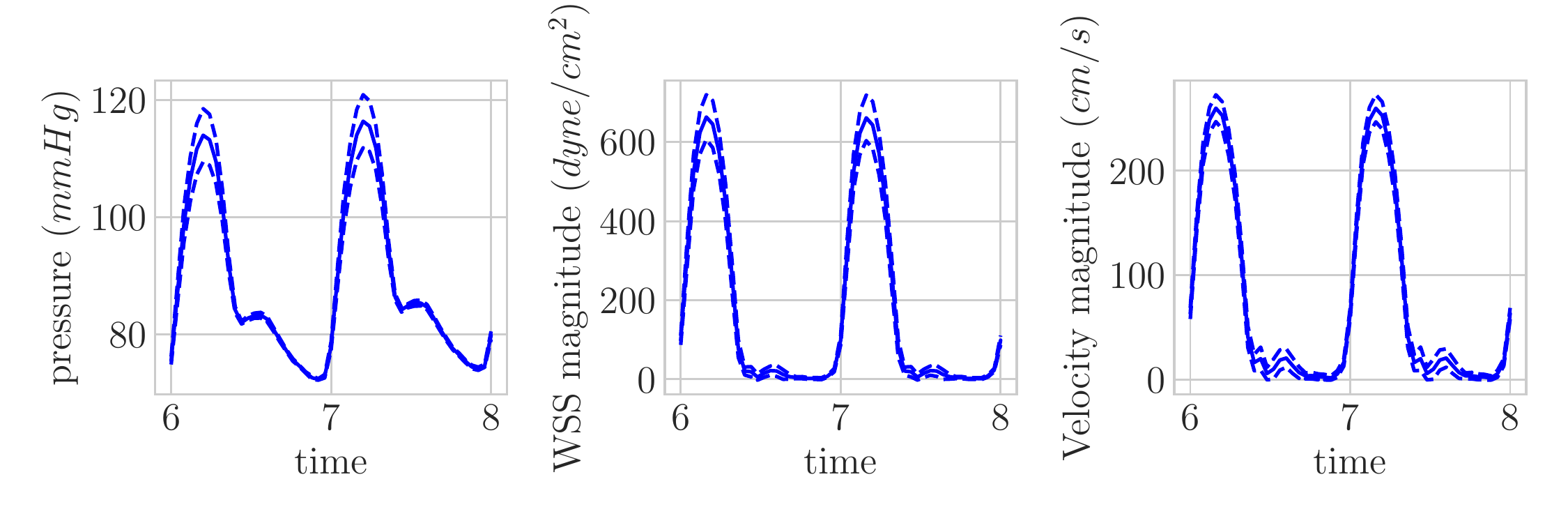}
\caption{Right Renal}
\label{fig:results:outlet0144rrenal}
\end{subfigure}
	
\begin{subfigure}[b]{0.48\textwidth}
\centering\includegraphics[width=\textwidth]{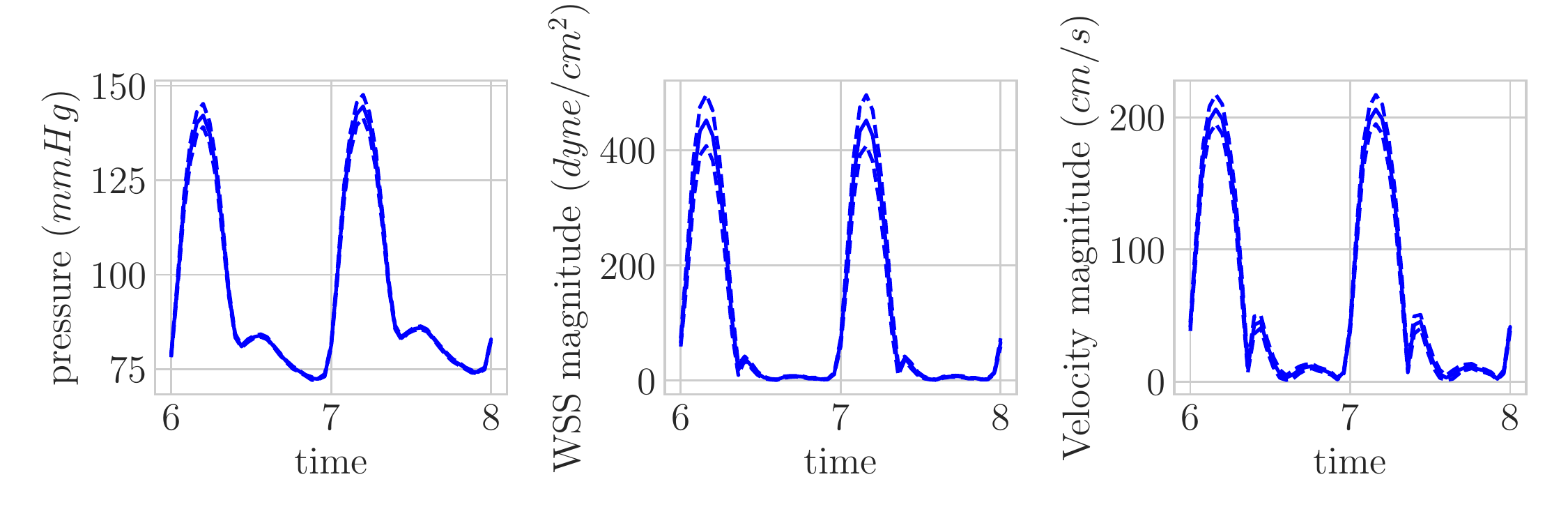}
\caption{SMA}
\label{fig:results:outlet0144sma}
\end{subfigure}
\caption{Outlet QoIs and $\pm 2 \sigma$ interval for abdominal aortic aneurysm model.}
\end{figure}

\begin{figure}[!ht]
	\centering
	\begin{subfigure}[b]{0.48\textwidth}
		\centering\includegraphics[width=\textwidth]{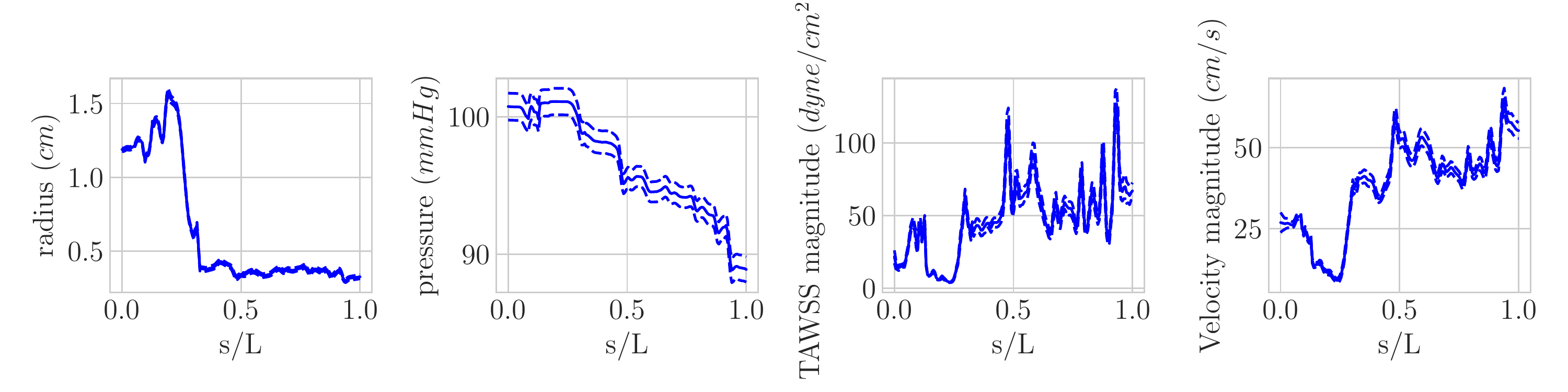}
		\caption{Aorta}
		\label{fig:results:timeavgaorta0144}
	\end{subfigure}
	\begin{subfigure}[b]{0.48\textwidth}
		\centering\includegraphics[width=\textwidth]{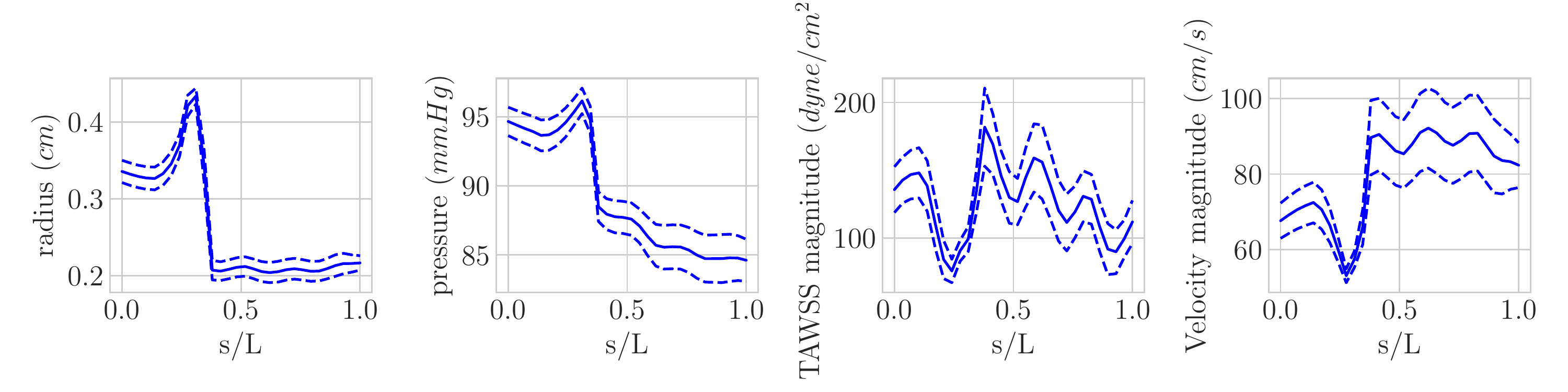}
		\caption{Celiac Hepatic}
		\label{fig:results:timeavghepatic0144}
	\end{subfigure}
	
	\begin{subfigure}[b]{0.48\textwidth}
		\centering\includegraphics[width=\textwidth]{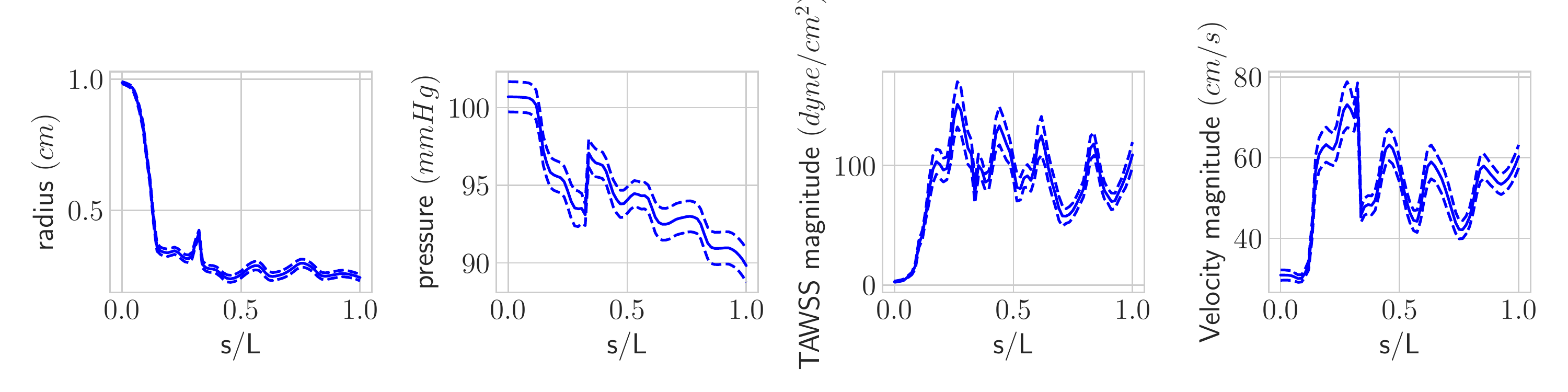}
		\caption{Celiac Splenic}
		\label{fig:results:timeavgsplenic0144}
	\end{subfigure}
	\begin{subfigure}[b]{0.48\textwidth}
		\centering\includegraphics[width=\textwidth]{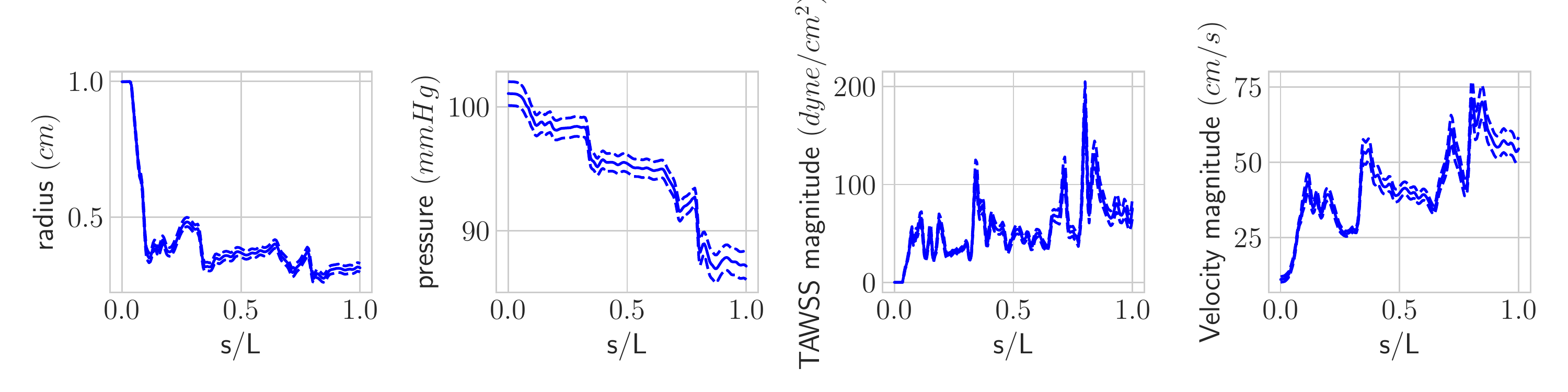}
		\caption{Left external Iliac}
		\label{fig:results:timeavgiliac0144}
	\end{subfigure}
	
	\begin{subfigure}[b]{0.48\textwidth}
		\centering\includegraphics[width=\textwidth]{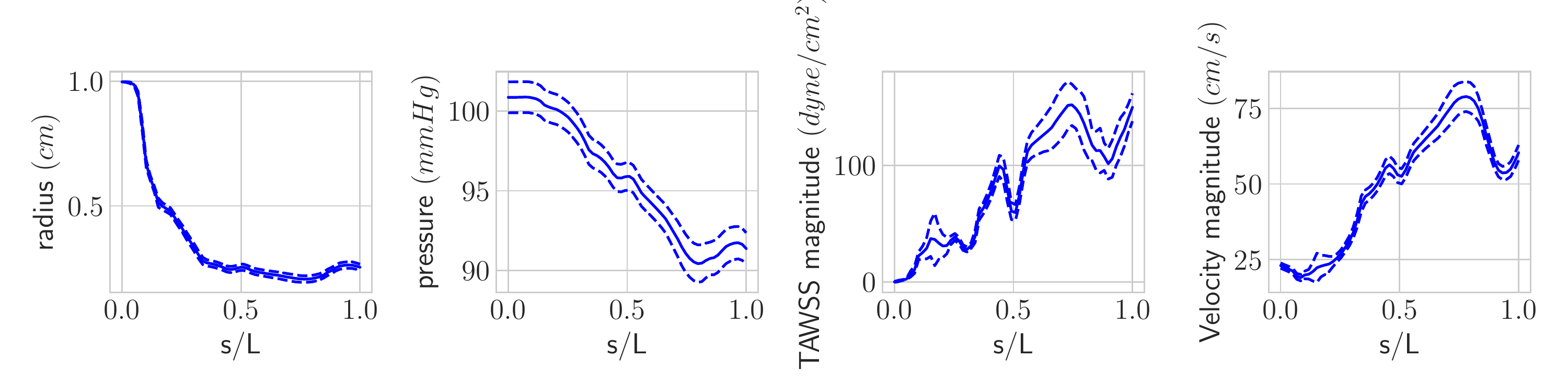}
		\caption{Left Renal}
		\label{fig:results:timeavglrenal0144}
	\end{subfigure}
	\begin{subfigure}[b]{0.48\textwidth}
		\centering\includegraphics[width=\textwidth]{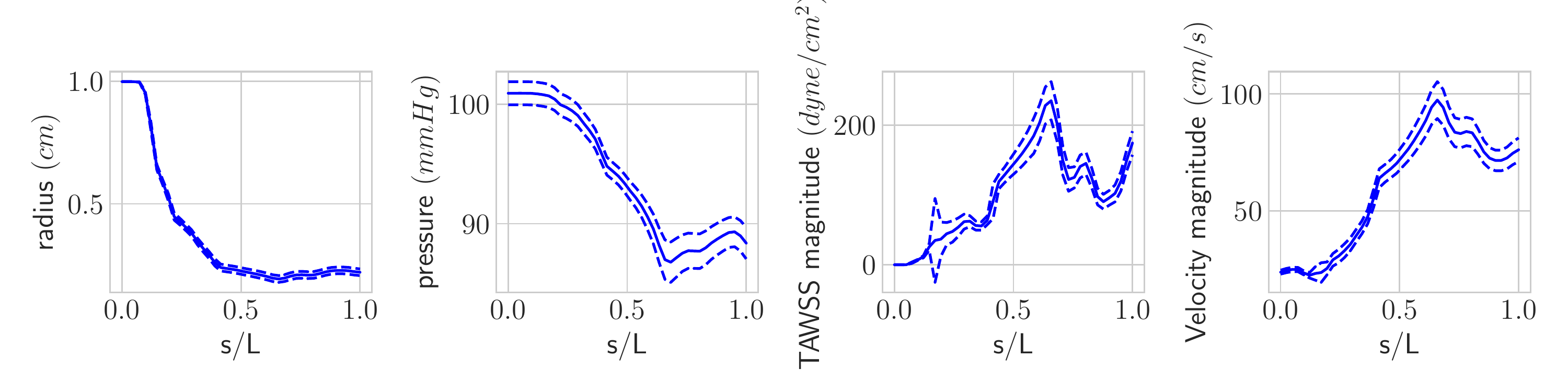}
		\caption{Right Renal}
		\label{fig:results:timeavgrrenal0144}
	\end{subfigure}
	
	\begin{subfigure}[b]{0.48\textwidth}
		\centering\includegraphics[width=\textwidth]{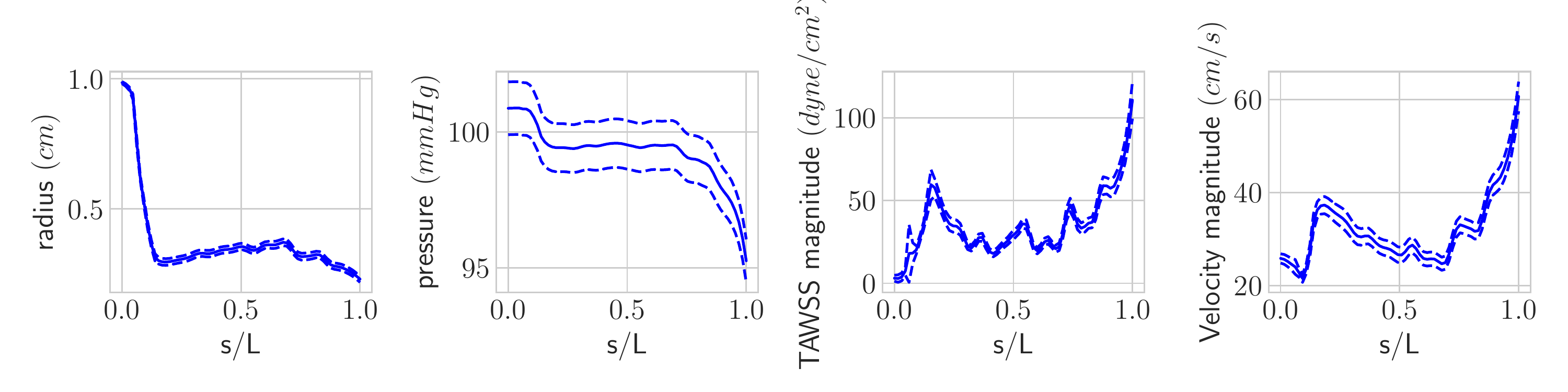}
		\caption{SMA}
		\label{fig:results:timeavgsma0144}
	\end{subfigure}
	\caption{Time averaged QoIs and $\pm 2 \sigma$ interval for abdominal aortic aneurysm model, plotted along the vessel centerline.}
\end{figure}

The relative confidence intervals of the Monte Carlo estimates for the mean pressure, TAWSS and velocity magnitude were found to be within the ranges 0.07-0.8\%, 0.44-0.79\% and 0.28-0.65\%, indicating a satisfactory convergence, particularly compared to the observed CoV for the same vessels (see Table~\ref{tab:results:0144converge} and Figure~\ref{fig:results:mvconvergeaorta0144}).
Outlet profiles show increased variability in branch vessels for all QoIs with respect to the Aorta (see Figures~\ref{fig:results:outlet0144aorta}-\ref{fig:results:outlet0144sma}). 
Time average flow results show increasing TAWSS and velocity magnitude variability towards the distal end of the vessel (see Figures~\ref{fig:results:timeavgaorta0144}-\ref{fig:results:timeavgsma0144}).

\begin{figure}[!ht]
	\centering
	\begin{subfigure}[b]{0.23\textwidth}
		\centering\includegraphics[width=\textwidth]{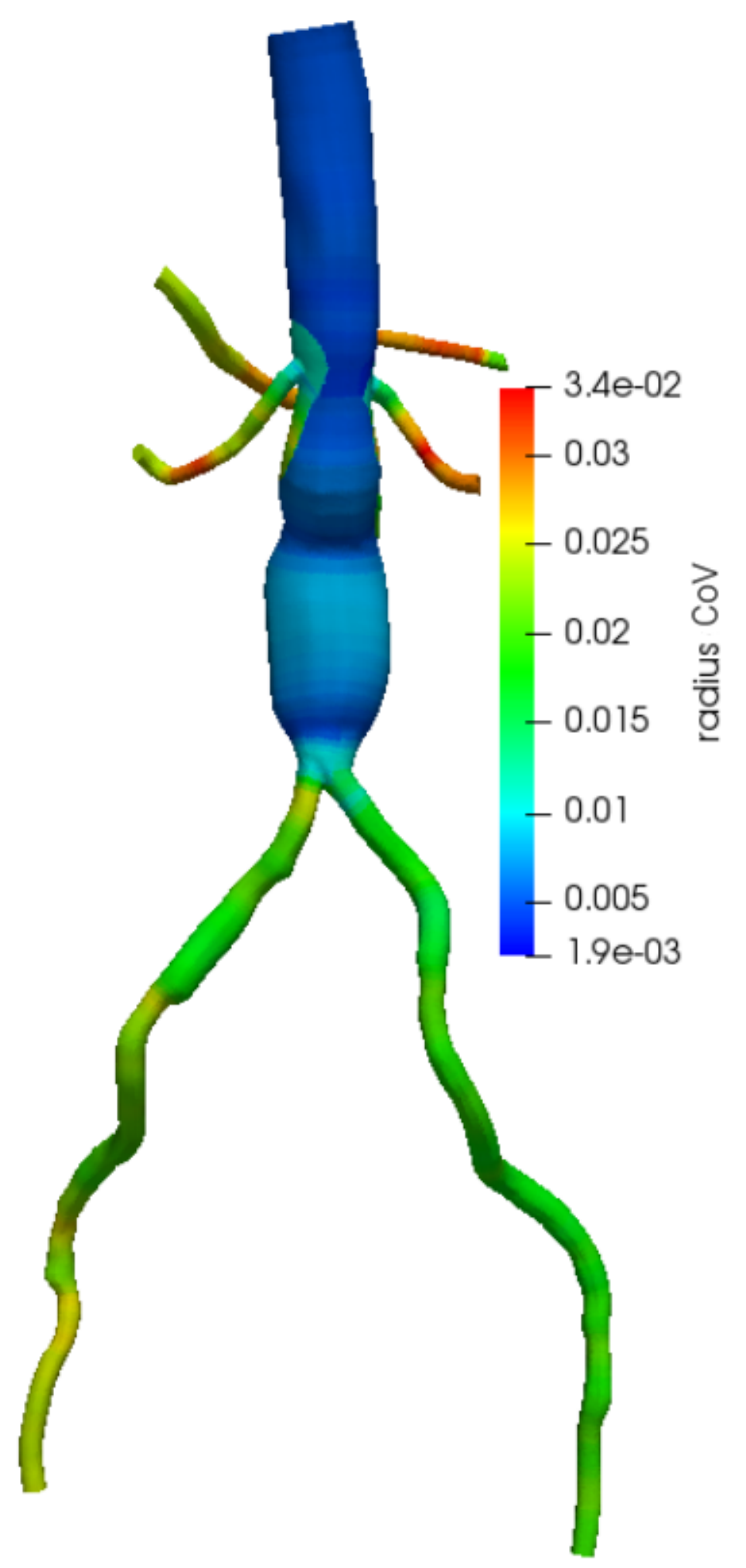}
		\caption{}
		\label{fig:results:01443dradius}
	\end{subfigure}
	\begin{subfigure}[b]{0.24\textwidth}
		\centering\includegraphics[width=\textwidth]{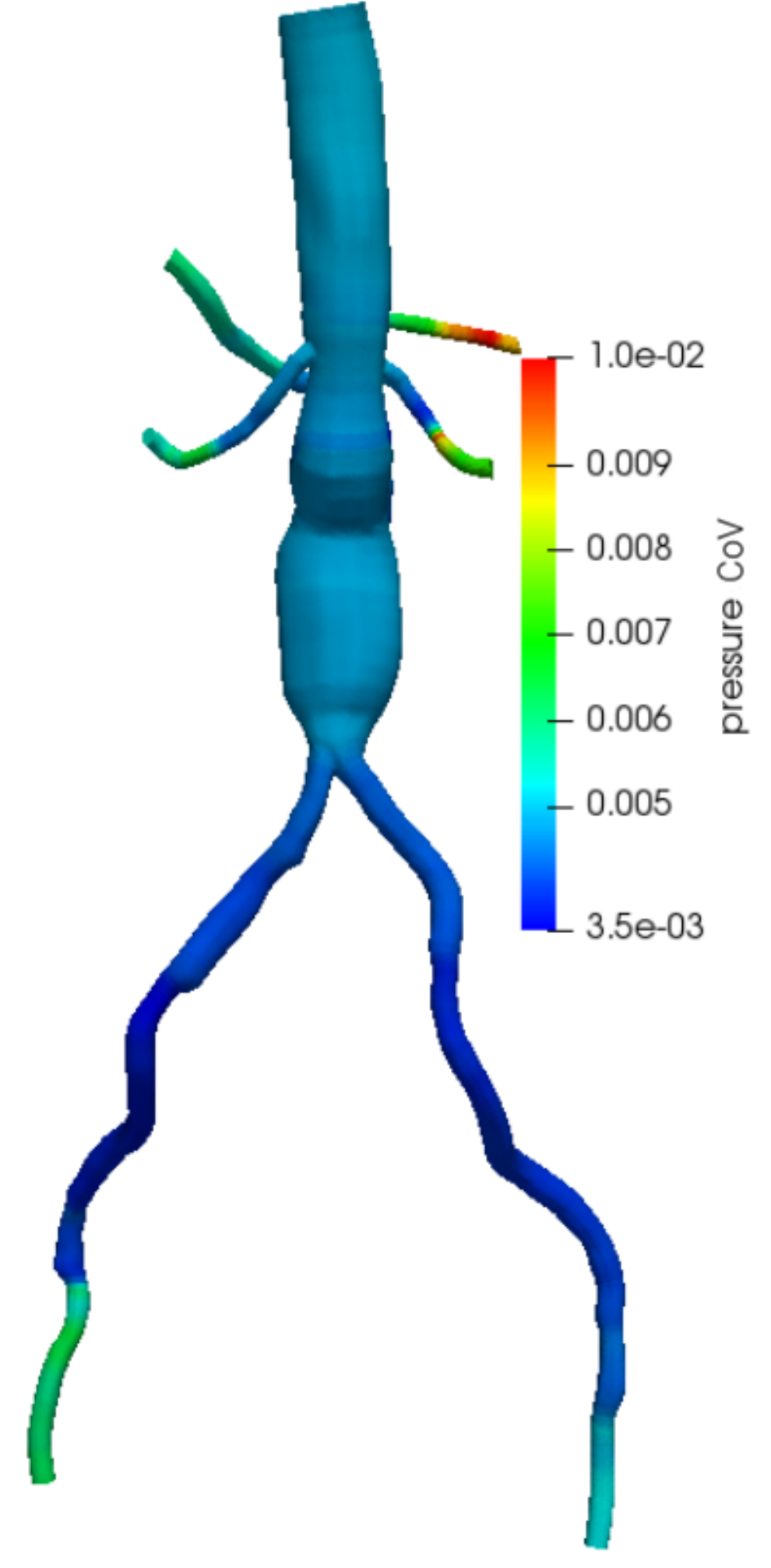}
		\caption{}
		\label{fig:results:01443dpressure}
	\end{subfigure}
	\begin{subfigure}[b]{0.235\textwidth}
		\centering\includegraphics[width=\textwidth]{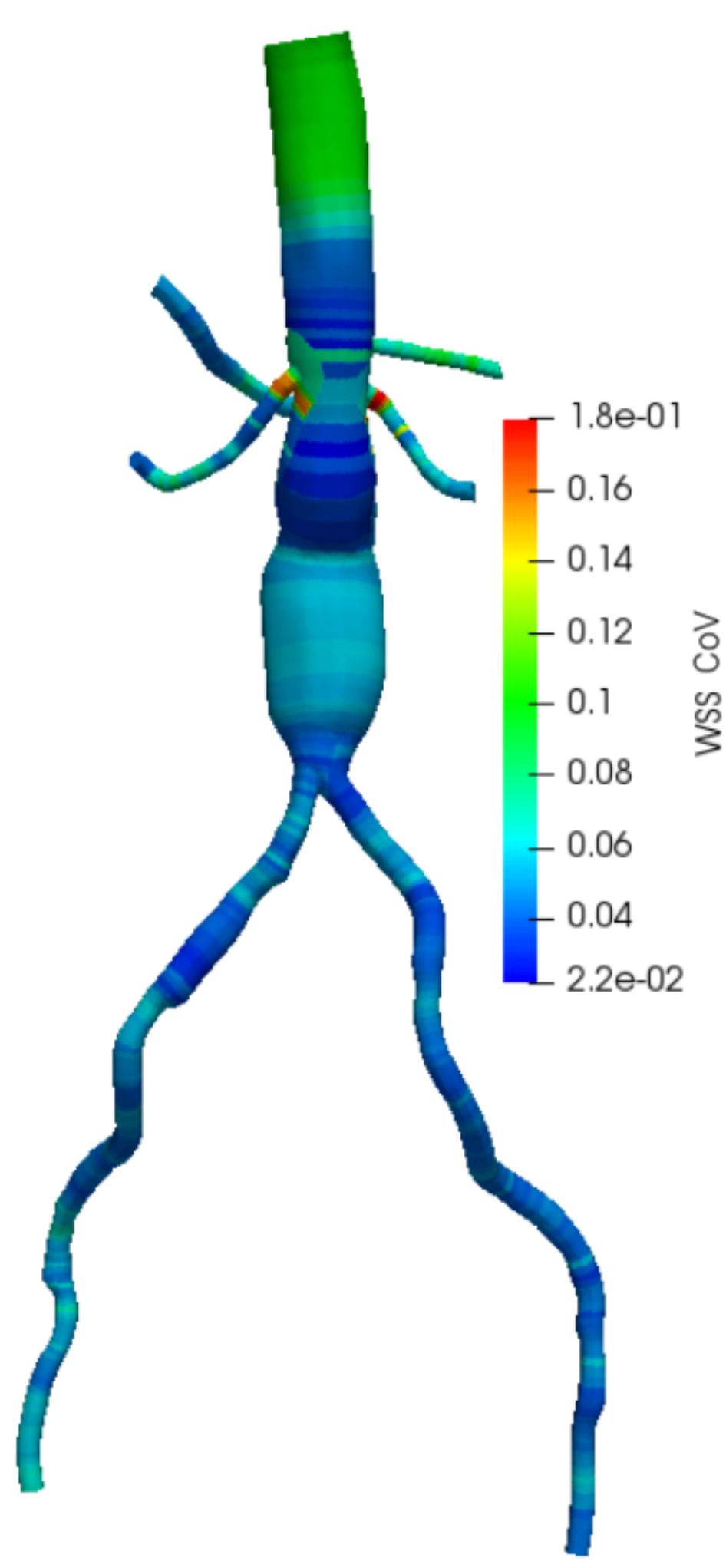}
		\caption{}
		\label{fig:results:01443dwss}
	\end{subfigure}
	\begin{subfigure}[b]{0.24\textwidth}
		\centering\includegraphics[width=\textwidth]{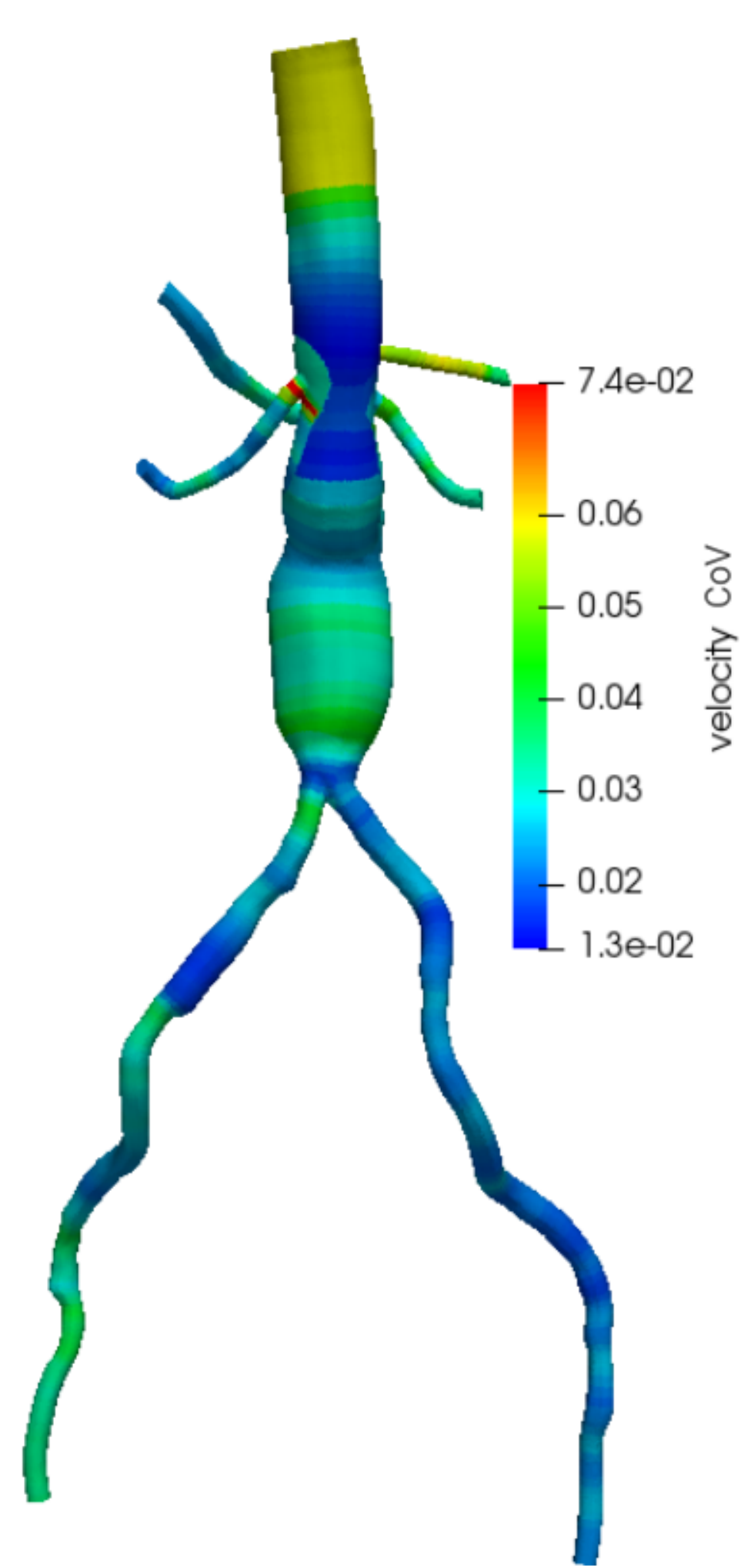}
		\caption{}
		\label{fig:results:01443dvelocity}
	\end{subfigure}
	\caption{Nearest neighbor interpolation of cross-sectional time-averaged CoVs for abdominal aortic aneurysm model.}
	\label{fig:01443d}
\end{figure}

As expected, radius CoV is inversely proportional to the vessel size (see Figure~\ref{fig:results:01443dradius}).
Relative pressure variability is approximately uniform along the path of each vessel, and particularly elevated in the celiac hepatic branch, due to the fact that it branches off of the celiac splenic which itself branches off of the aorta (see Figure~\ref{fig:results:01443dpressure}).
The TAWSS CoV appears to be significant for small branches (see Figure~\ref{fig:results:01443dwss}), and is equal to 0.06 in the aneurysm region and 0.02 in the proximal regions of the aorta. Similarly, the velocity magnitude CoVs is found to be 0.04 within the aneurysm and 0.02 in the upstream aorta. This illustrates how diseased regions like aneurysms can lead to increased geometric uncertainty, most likely due to increased ambiguity of the vessel lumen shape and increased surrounding noise sources in the input image volume, leading to higher neural network output variability. We note again that this is directly learned from the image volume by the neural network.

\begin{table}[!ht]
\centering
\caption{Monte Carlo sample mean, coefficient of variation (CoV) and 95\% relative confidence interval for all QoIs in abdominal aortic aneurysm model. $n$ indicates the number of cross-sectional slices for the associated vessel.}
\resizebox{\textwidth}{!}{
\begin{tabular}{lccccccc}
\toprule
{\bf Path} & {\bf Aorta} & {\bf Celiac hepatic} & {\bf Celiac splenic} & {\bf Ext. iliac left} & {\bf Renal left} & {\bf Renal right} & {\bf SMA}\\
& $(n=160)$ & $(n=30)$ & $(n=69)$ & $(n=169)$ & $(n=51)$ & $(n=35)$ & $(n=61)$ \\
\midrule
{\bf Radius mean} [cm] & 0.63 & 0.26 & 0.36 & 0.38 & 0.32 & 0.29 & 0.38\\
{\bf Radius CoV} & 0.007 & 0.018 & 0.016 & 0.019 & 0.018 & 0.021 & 0.015\\
{\bf Radius conf.} & 0.0012 & 0.0032 & 0.0028 & 0.0035 & 0.0033 & 0.0038 & 0.0027\\
\midrule
{\bf Pressure mean} [mmHg] & 96.45 & 89.11 & 94.51 & 94.35 & 95.29 & 92.35 & 99.26\\
{\bf Pressure CoV} & 0.004 & 0.004 & 0.004 & 0.004 & 0.004 & 0.005 & 0.004\\
{\bf Pressure conf.} & 0.0007 & 0.0008 & 0.0008 & 0.0006 & 0.0007 & 0.0008 & 0.0008\\
\midrule
{\bf TAWSS mean} [dyne/cm$^2$] & 47.66 & 125.70 & 85.11 & 58.47 & 87.28 & 117.50 & 35.35\\
{\bf TAWSS CoV} & 0.024 & 0.040 & 0.034 & 0.035 & 0.036 & 0.043 & 0.034\\
{\bf TAWSS conf.} & 0.0044 & 0.0073 & 0.0062 & 0.0064 & 0.0065 & 0.0079 & 0.0061\\
\midrule
{\bf Velocity mean} [cm/s] & 38.05 & 79.84 & 52.21 & 43.88 & 51.43 & 66.67 & 31.98\\
{\bf Velocity CoV} & 0.015 & 0.036 & 0.020 & 0.027 & 0.019 & 0.028 & 0.024\\
{\bf Velocity conf.} & 0.0028 & 0.0065 & 0.0036 & 0.0048 & 0.0034 & 0.0052 & 0.0043\\
\bottomrule
\end{tabular}}
\label{tab:results:0144converge}
\end{table}

%% file: res_LCA.tex
\subsection{Left Coronary Artery Model}\label{section:paper3:coronary}

\begin{figure}[!ht]
\centering
\begin{subfigure}[b]{0.4\textwidth}
\centering
\includegraphics[height=1.2\textwidth]{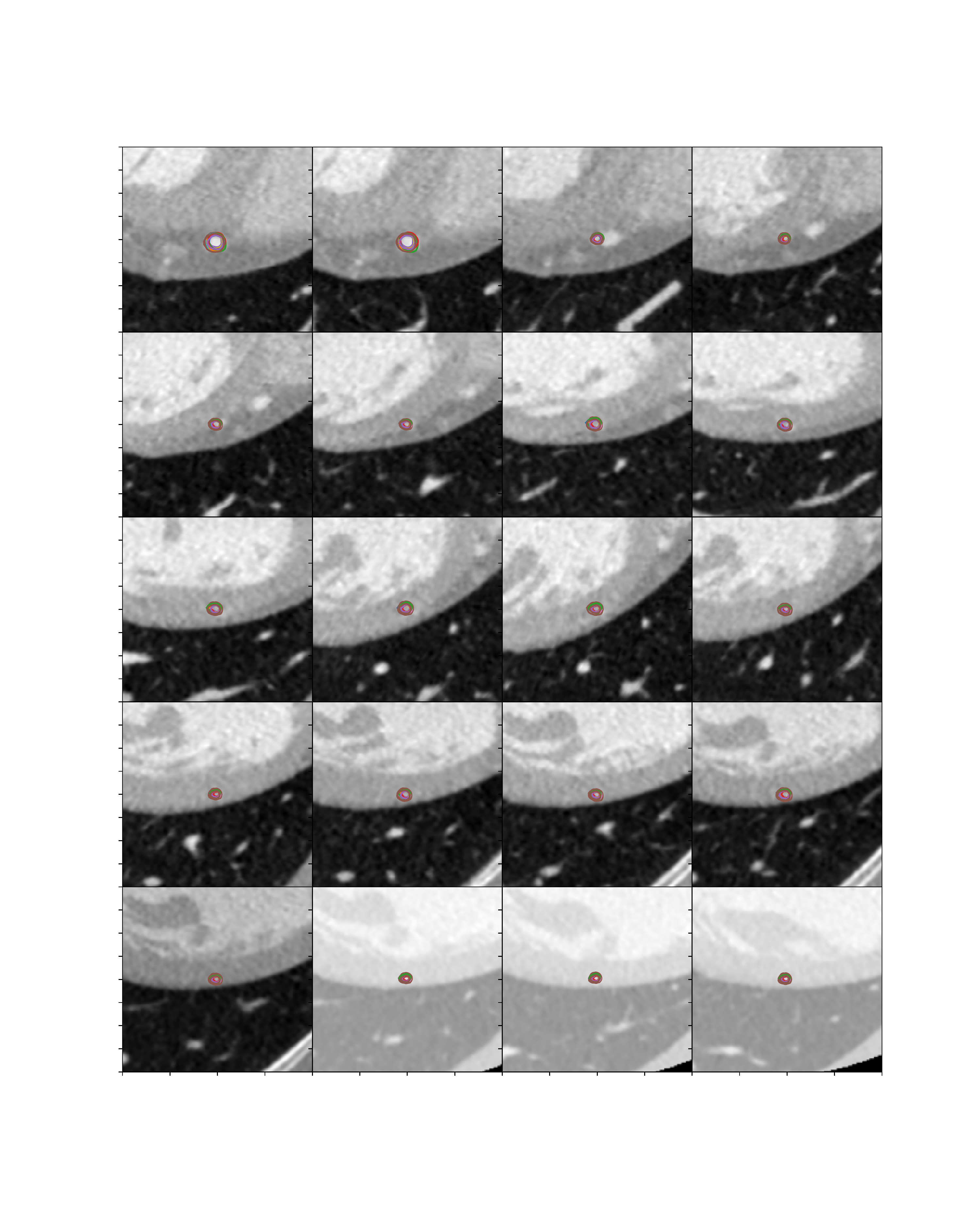}
\caption{Left Coronary Artery case - $LAD-D_1$}\label{fig:results:segcoronarylc2sub1}
\end{subfigure}
\begin{subfigure}[b]{0.29\textwidth}
\centering\includegraphics[scale=0.5]{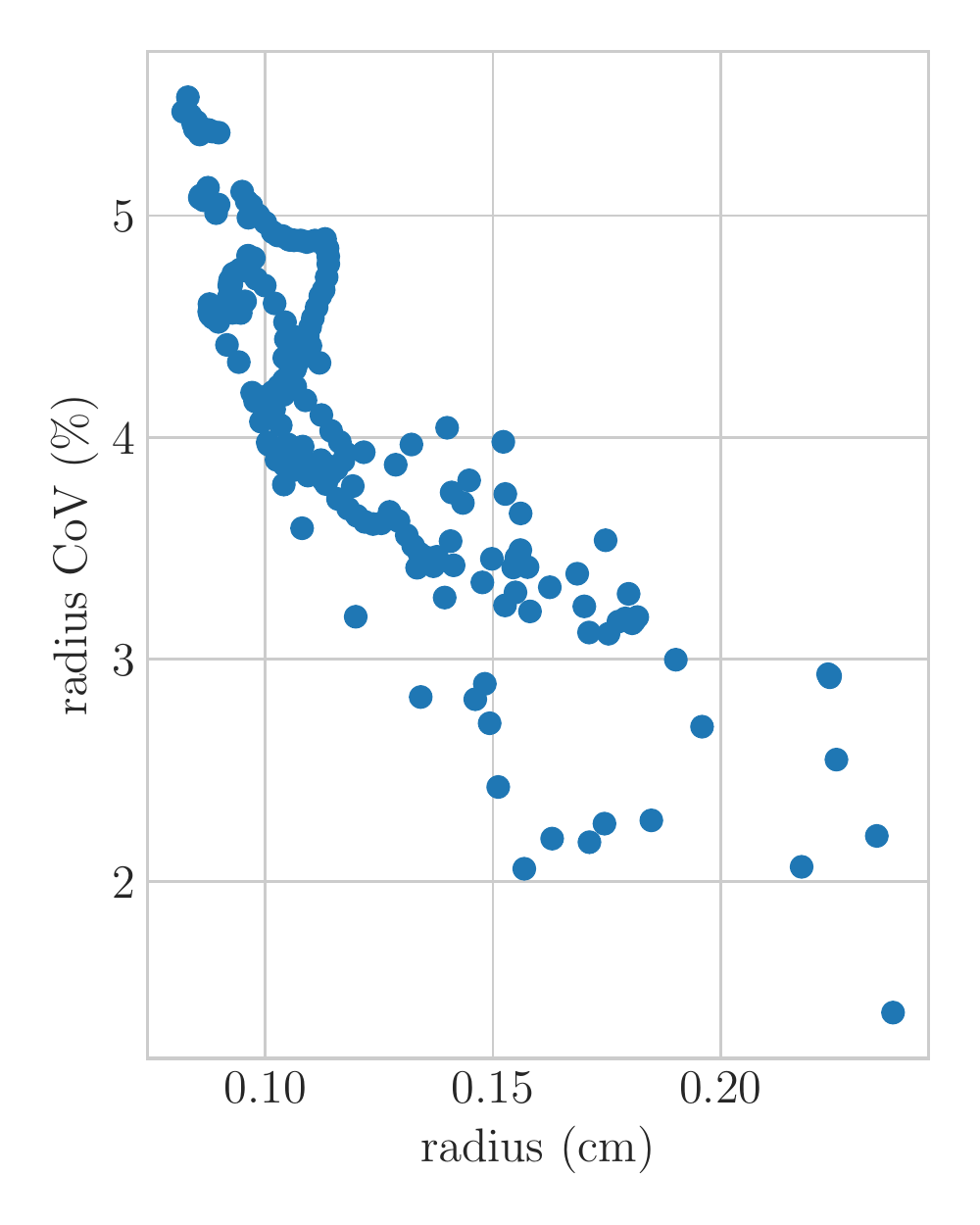}
\caption{Radius CoV}
\label{fig:results:rvrcvcoronary}
\end{subfigure}
\begin{subfigure}[b]{0.29\textwidth}
\centering\includegraphics[scale=0.5]{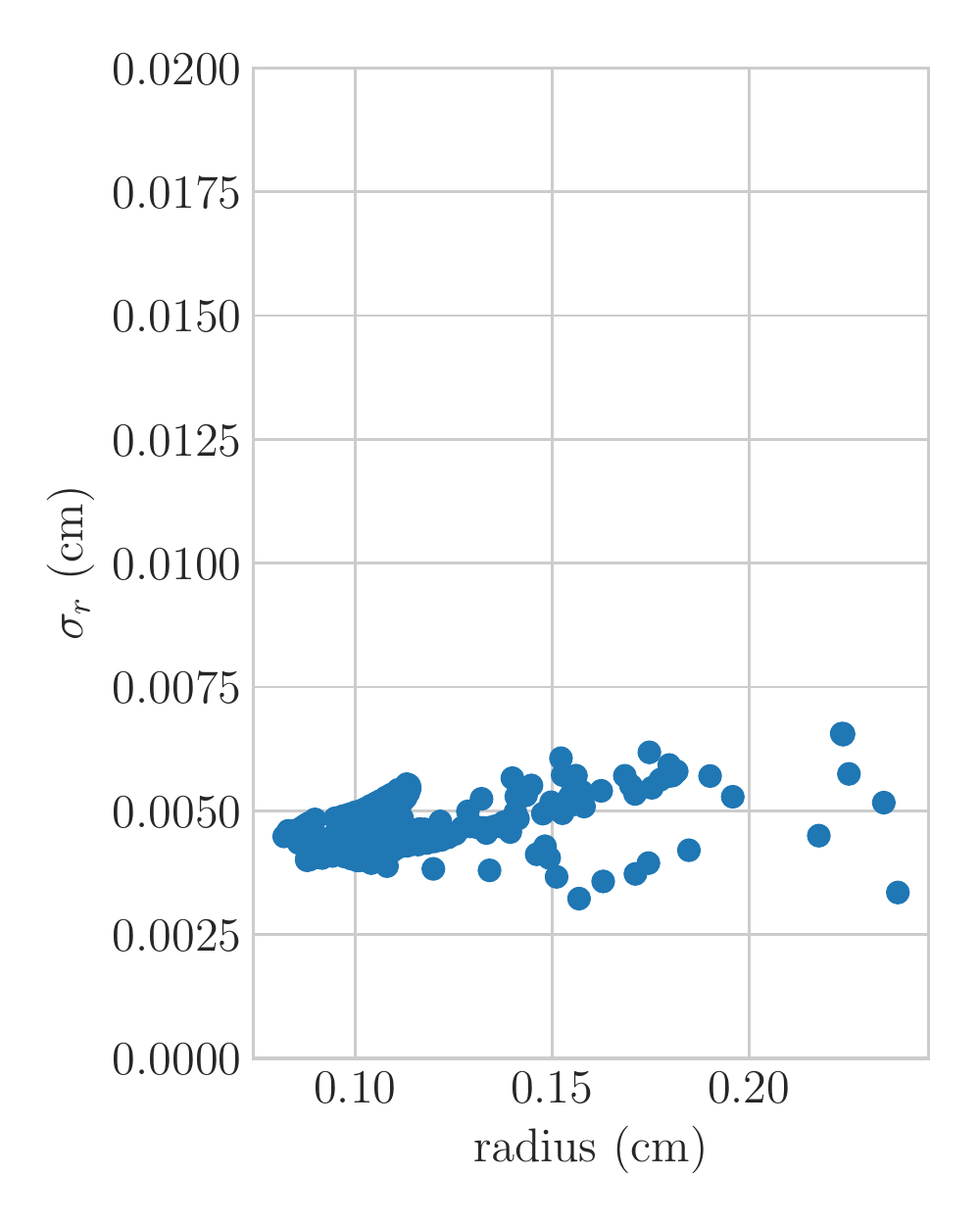}
\caption{Radius standard deviation}
\label{fig:results:rvrsigcoronary}
\end{subfigure}
\caption{Lumen segmentation samples and radius CoV/standard deviation for left coronary artery test case, computed over cross-sectional slices $\mathbf{x}_i,\,i=1,\dots,222$.}
\label{fig:results:rgraphcoronary}
\end{figure}

\begin{figure}[!ht]
\centering
\begin{subfigure}[b]{0.14\textwidth}
\centering\includegraphics[width=\textwidth]{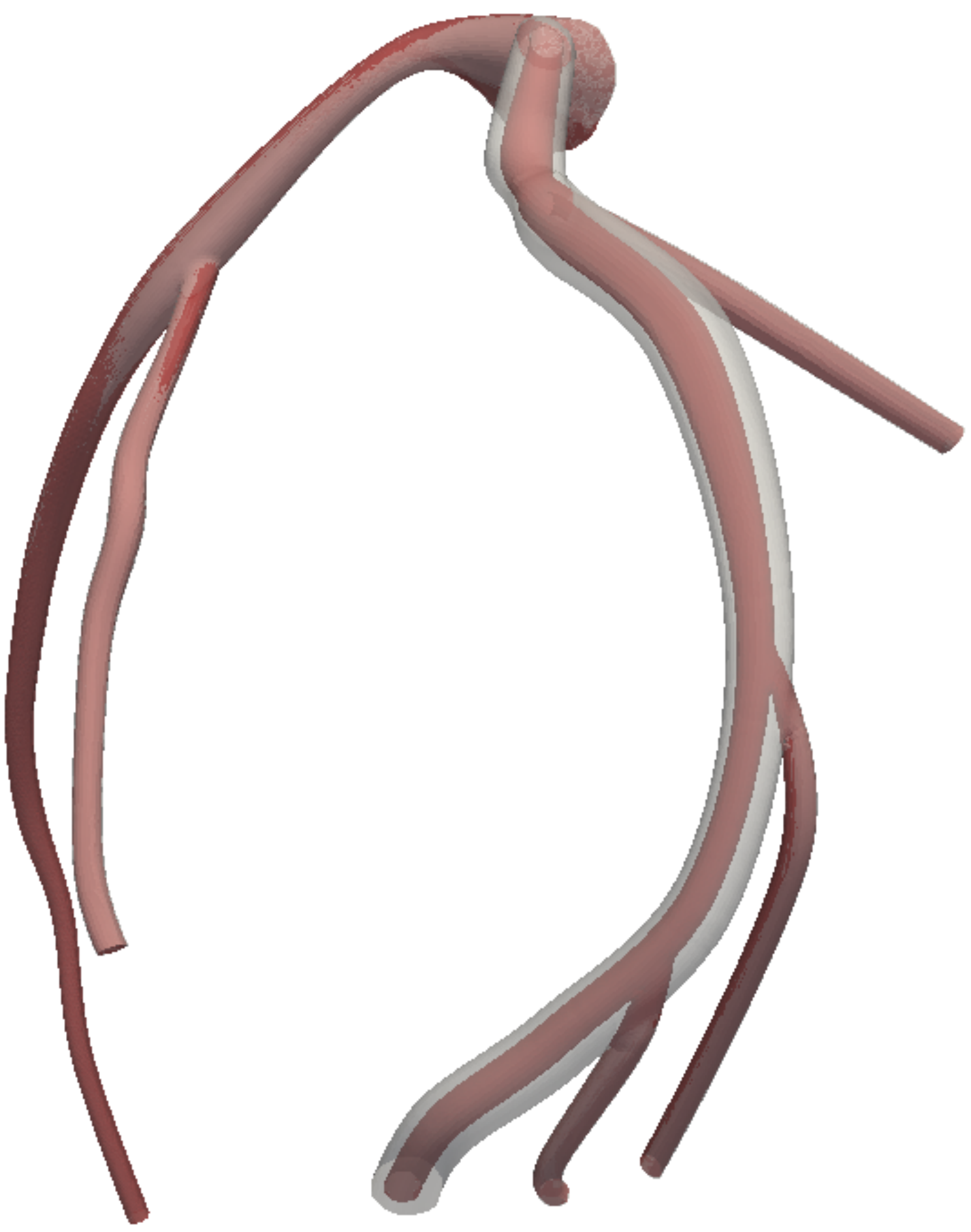}
\caption{$\lambda_1=8.09$}
\label{fig:results:coronarypca1}
\end{subfigure}
$\quad\quad$
\begin{subfigure}[b]{0.14\textwidth}
\centering\includegraphics[width=\textwidth]{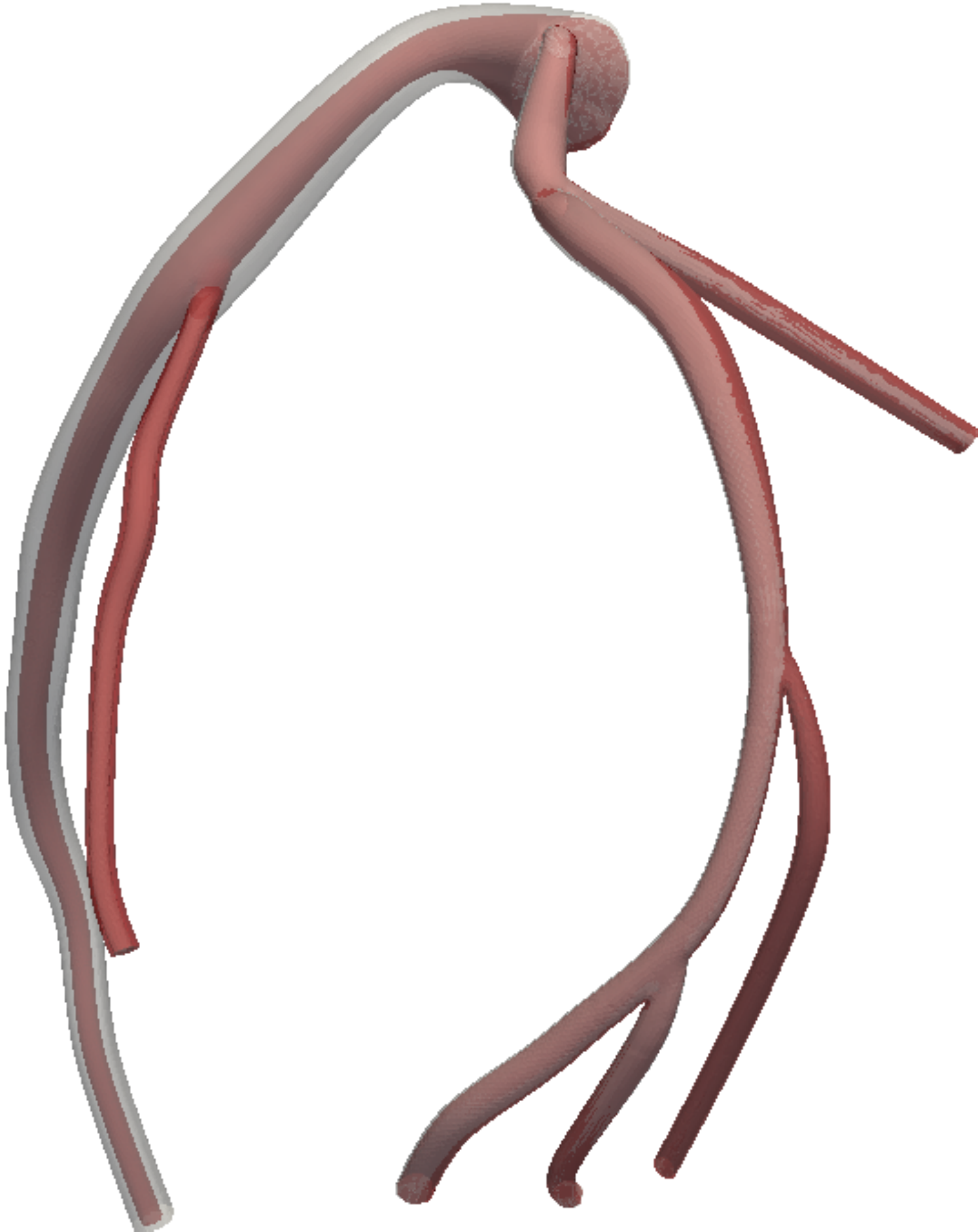}
\caption{$\lambda_2=6.72$}
\label{fig:results:coronarypca2}
\end{subfigure}
$\quad\quad$
\begin{subfigure}[b]{0.14\textwidth}
\centering\includegraphics[width=\textwidth]{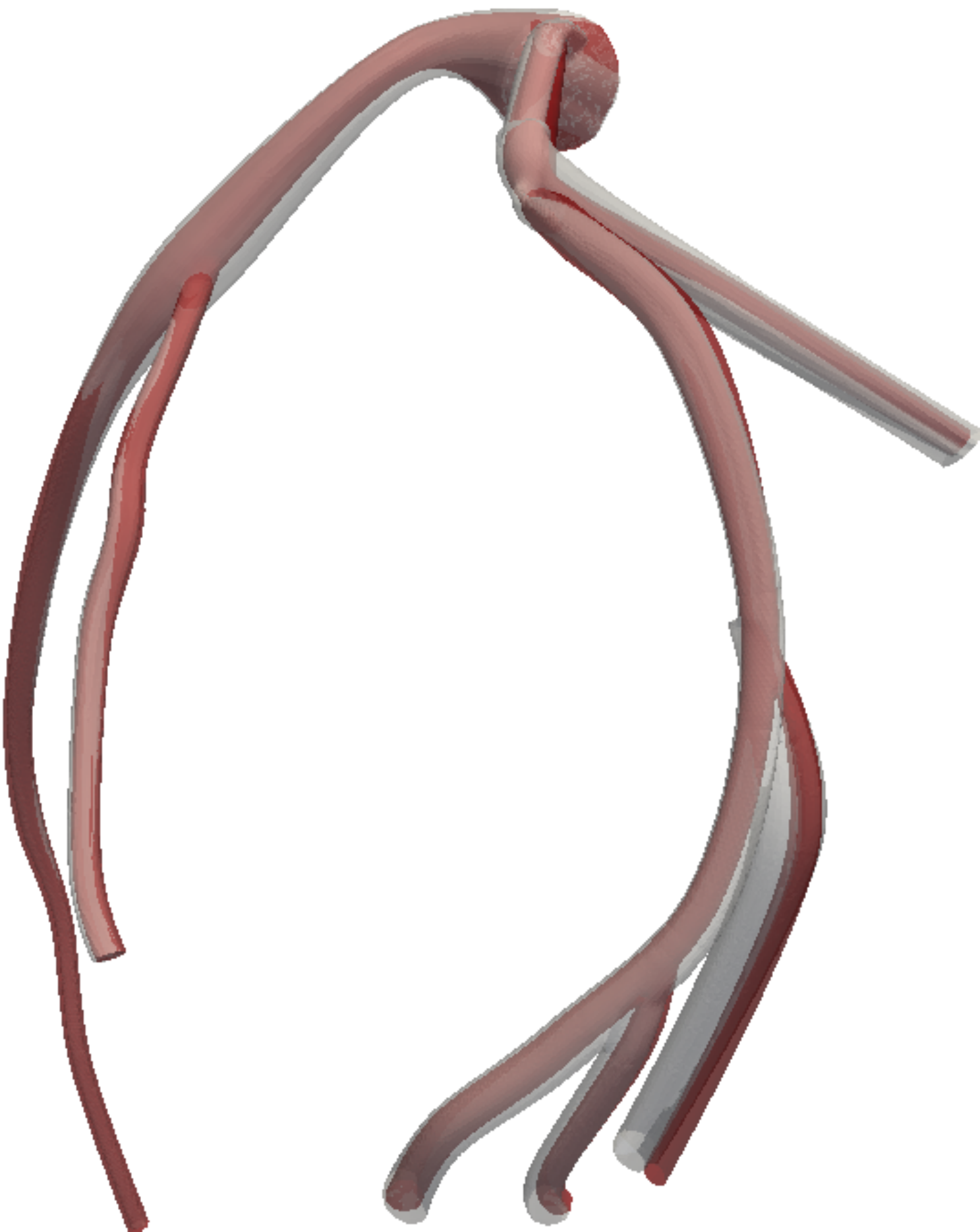}
\caption{$\lambda_{19}=1.91$}
\label{fig:results:coronarypca3}
\end{subfigure}
$\quad\quad$
\begin{subfigure}[b]{0.14\textwidth}
\centering\includegraphics[width=\textwidth]{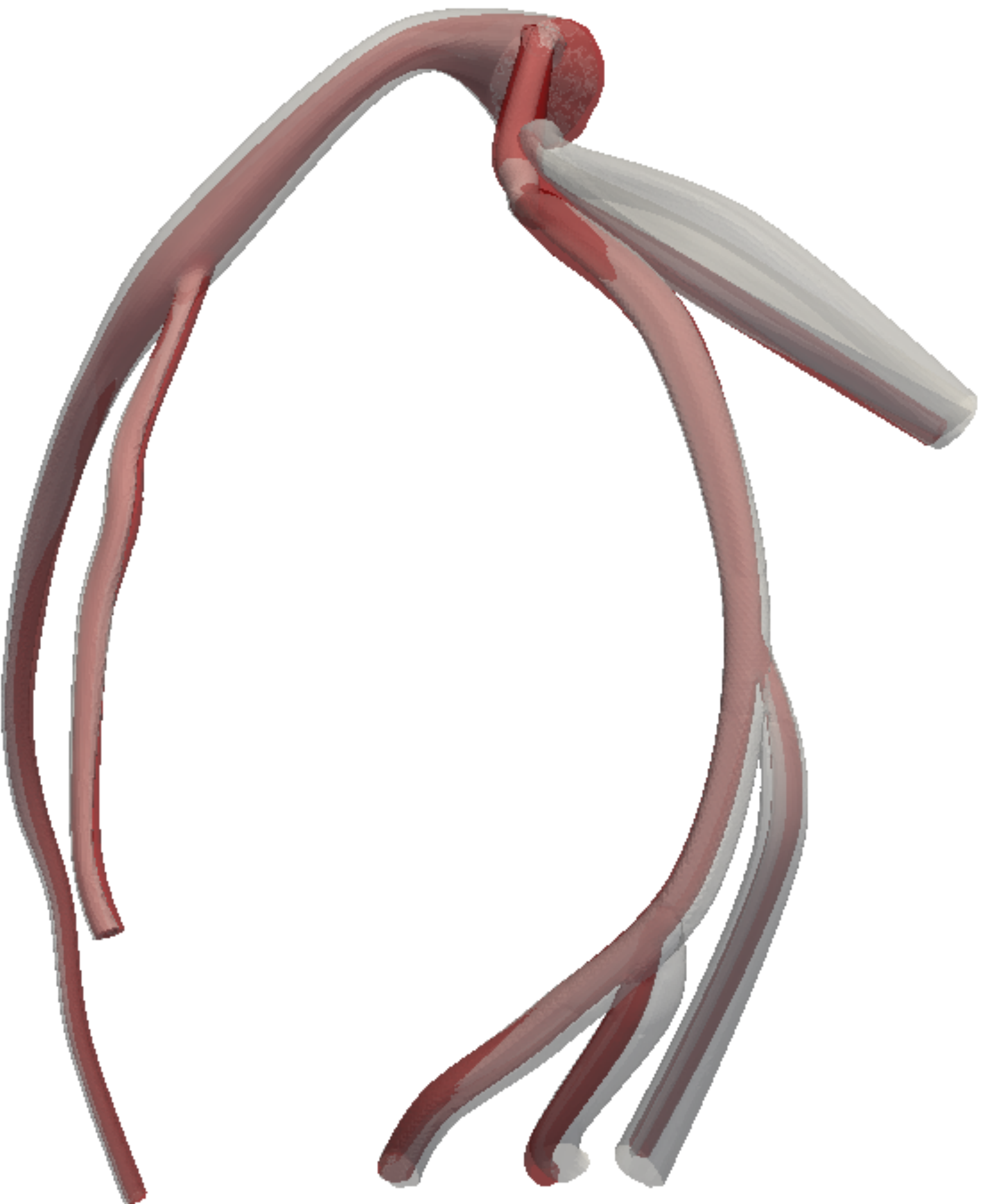}
\caption{$\lambda_{20}=1.89$}
\label{fig:results:coronarypca4}
\end{subfigure}
\begin{subfigure}[b]{0.25\textwidth}
\centering\includegraphics[width=\textwidth]{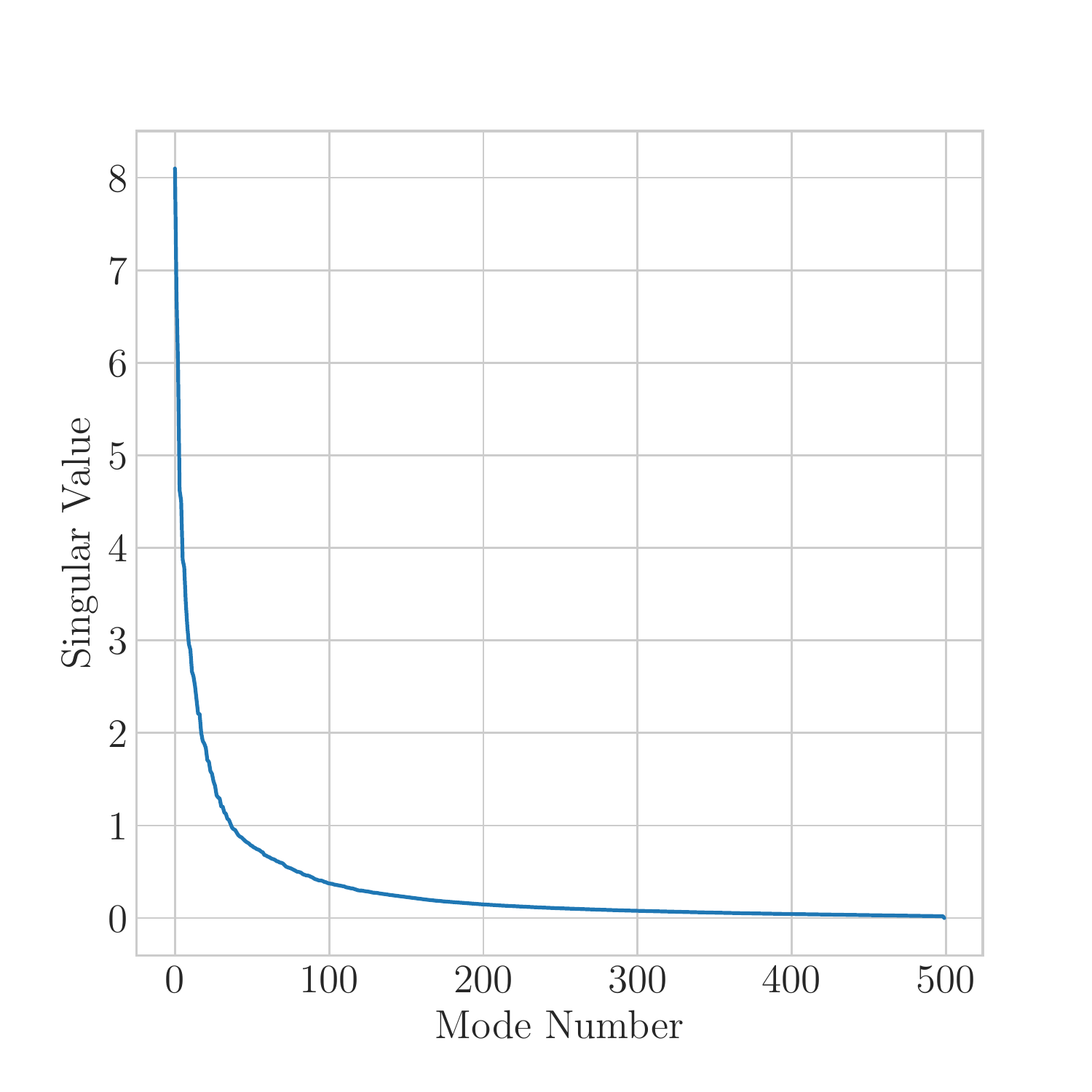}
\caption{Singular Values}
\label{fig:results:coronarypca4}
\end{subfigure}
\caption{PCA modes overlayed on mean left coronary artery model geometry.}\label{fig:coronarypca}
\end{figure}

\noindent Even for the smallest coronary arteries and in the presence of significant surrounding heart tissue, the vessel lumen shape was qualitatively well captured by the proposed dropout network (see Figure~\ref{fig:results:segcoronarylc2sub1}).
A PCA quantification of segmentation uncertainty shows dominant modes distributed along major arteries, and higher modes inducing local changes to smaller branches (see Figure~\ref{fig:results:coronarypca1}).
While the relative variance is higher in the small vessel branches when compared to the larger branches, PCA determines modes based on absolute variance.
The absolute variance is larger in the large vessels and explains their presence in the dominant PCA modes.

\begin{figure}[!ht]
\centering
\includegraphics[scale=0.475]{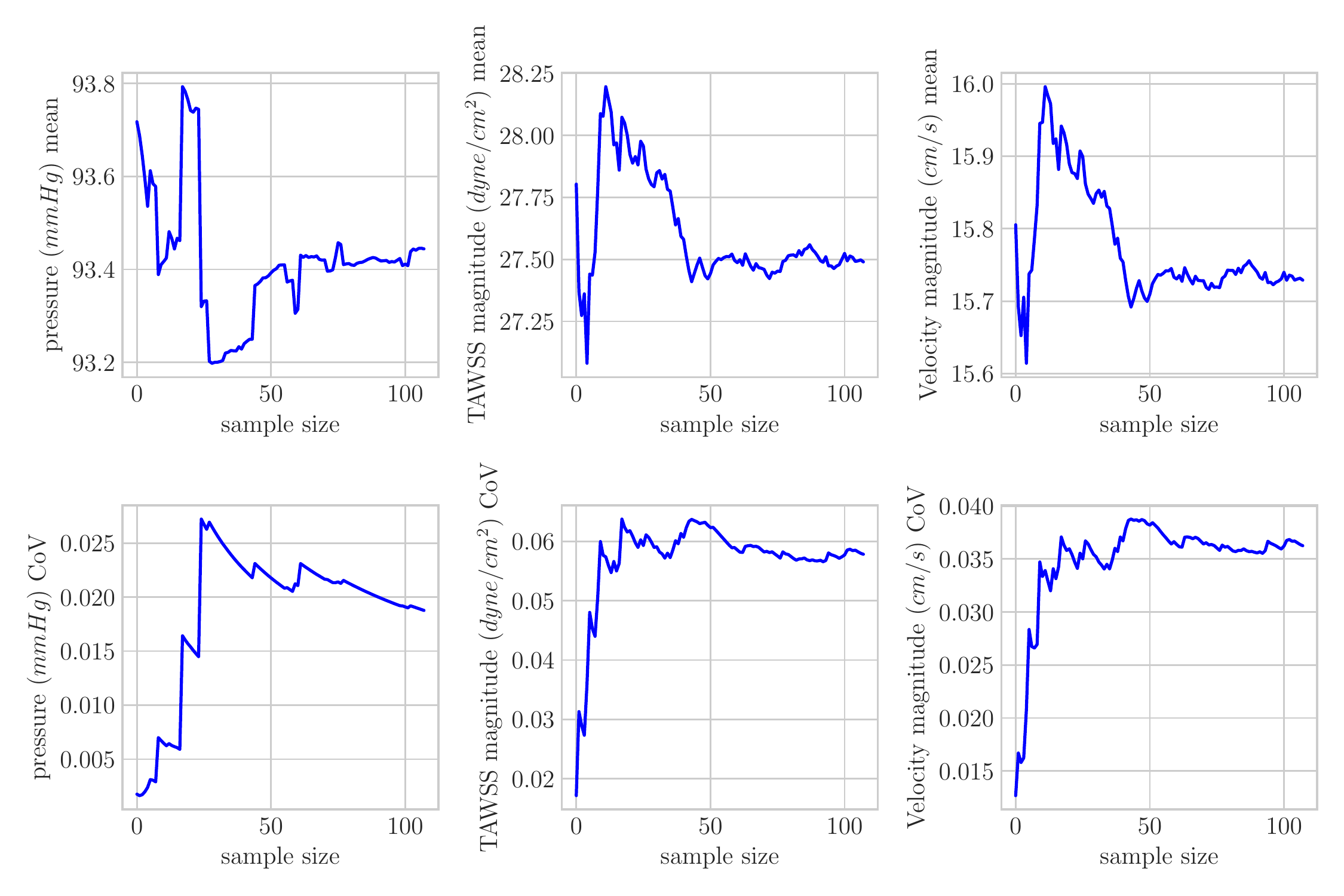}
\caption{Monte Carlo moment traces for left coronary artery model QoIs.}
\label{fig:results:mcconvergelc2}
\end{figure}

A satisfactory convergence is observed for the Monte Carlo statistical moments after 110 model evaluations, with asymptotic traces for more than 50 samples (see Figure~\ref{fig:results:mcconvergelc2}).
The 95\% relative confidence intervals for the Monte Carlo estimates of the mean are approximately equal to 0.4\%, 3\% and 1.5\% for pressure, TAWSS and velocity magnitude, respectively, and are well below the CoV computed for the same QoIs (see Tables~\ref{tab:results:coronaryconf}).

\begin{figure}[!ht]
	\centering
	\begin{subfigure}[b]{0.48\textwidth}
		\centering\includegraphics[width=\textwidth]{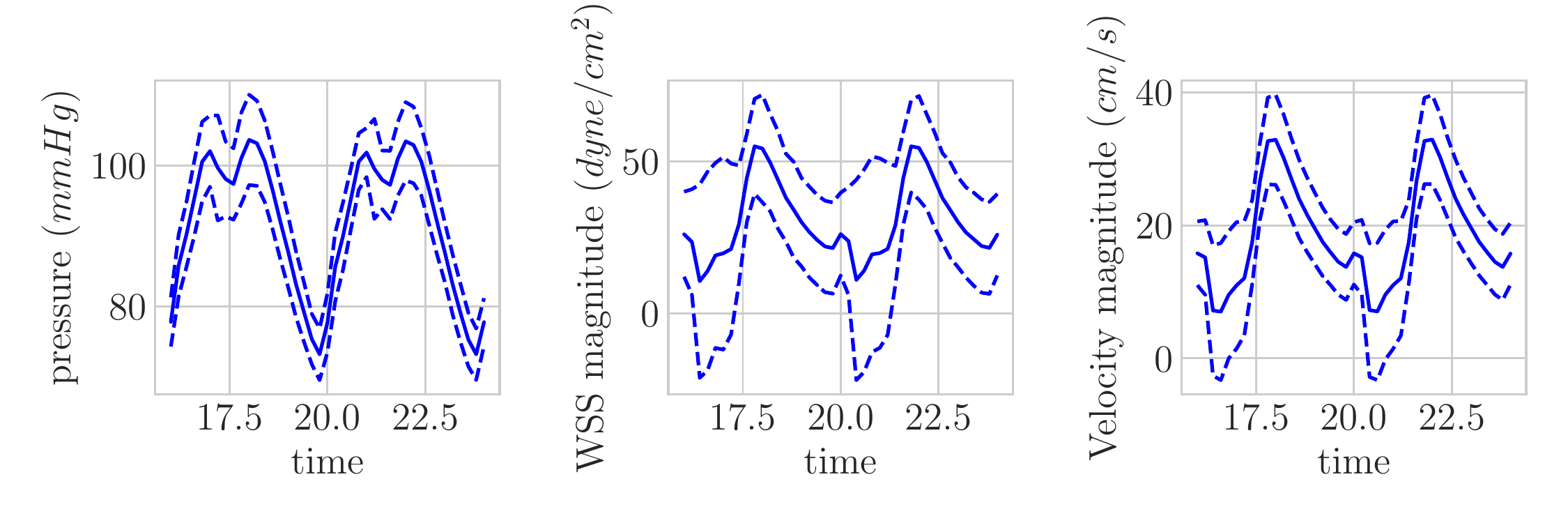}
		\caption{$LCx$}
		\label{fig:results:outletcoronarylc1}
	\end{subfigure}
	\begin{subfigure}[b]{0.48\textwidth}
		\centering\includegraphics[width=\textwidth]{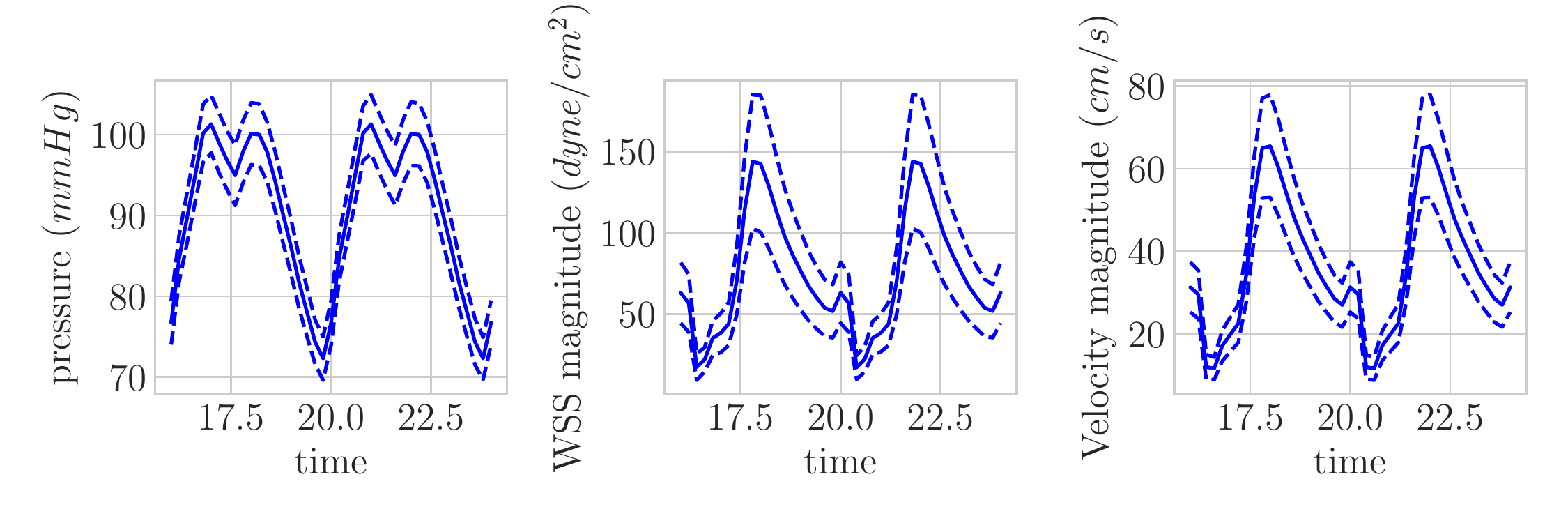}
		\caption{$LCx-OM_1$}
		\label{fig:results:outletcoronarylc1_sub1}
	\end{subfigure}
	
	\begin{subfigure}[b]{0.48\textwidth}
		\centering\includegraphics[width=\textwidth]{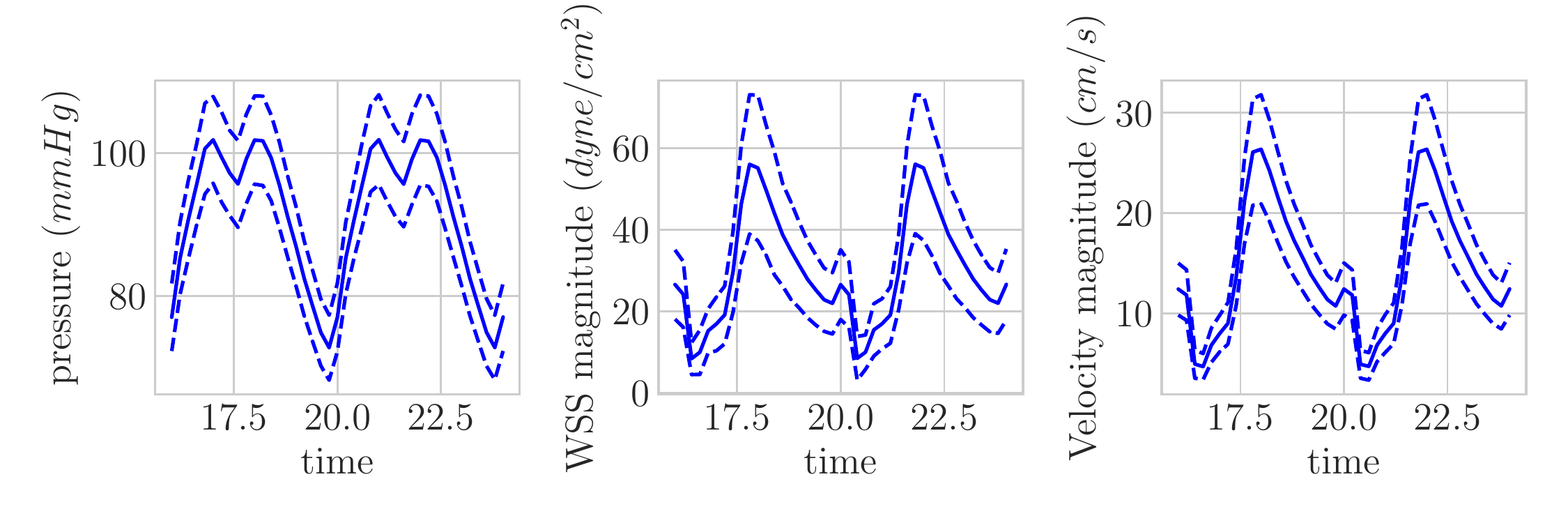}
		\caption{$LCx-OM_2$}
		\label{fig:results:outletcoronarylc1sub2}
	\end{subfigure}
	\begin{subfigure}[b]{0.48\textwidth}
		\centering\includegraphics[width=\textwidth]{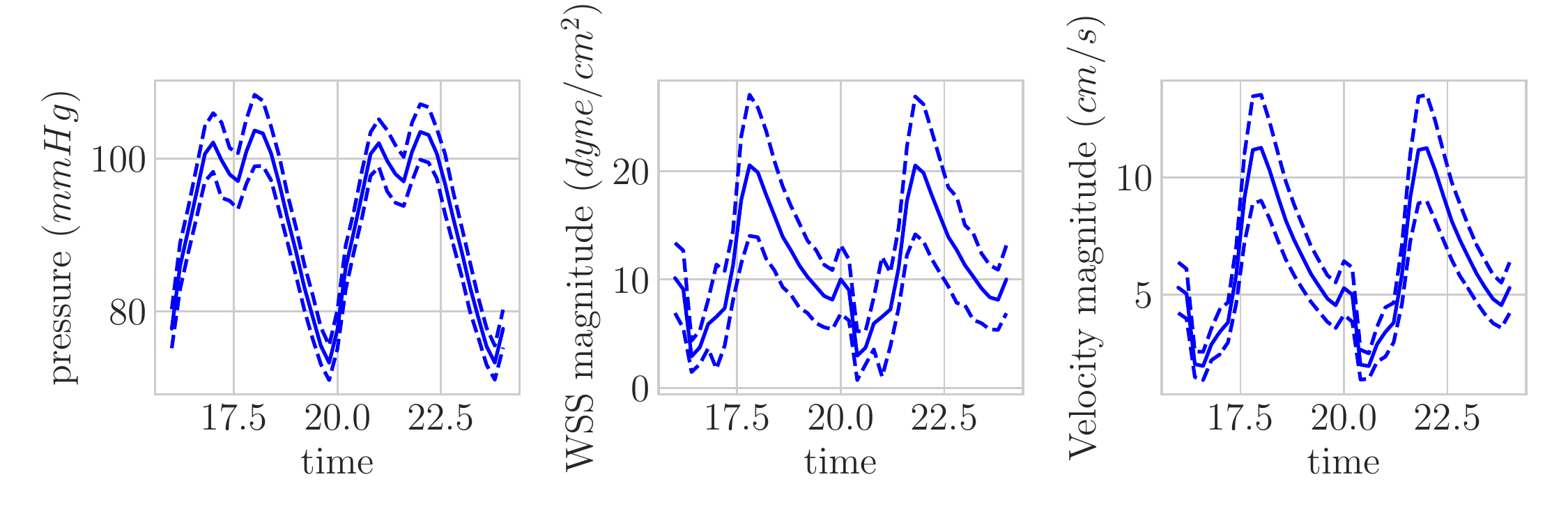}
		\caption{$LCx-OM_3$}
		\label{fig:results:outletcoronarylc1sub3}
	\end{subfigure}
	
	\begin{subfigure}[b]{0.48\textwidth}
		\centering\includegraphics[width=\textwidth]{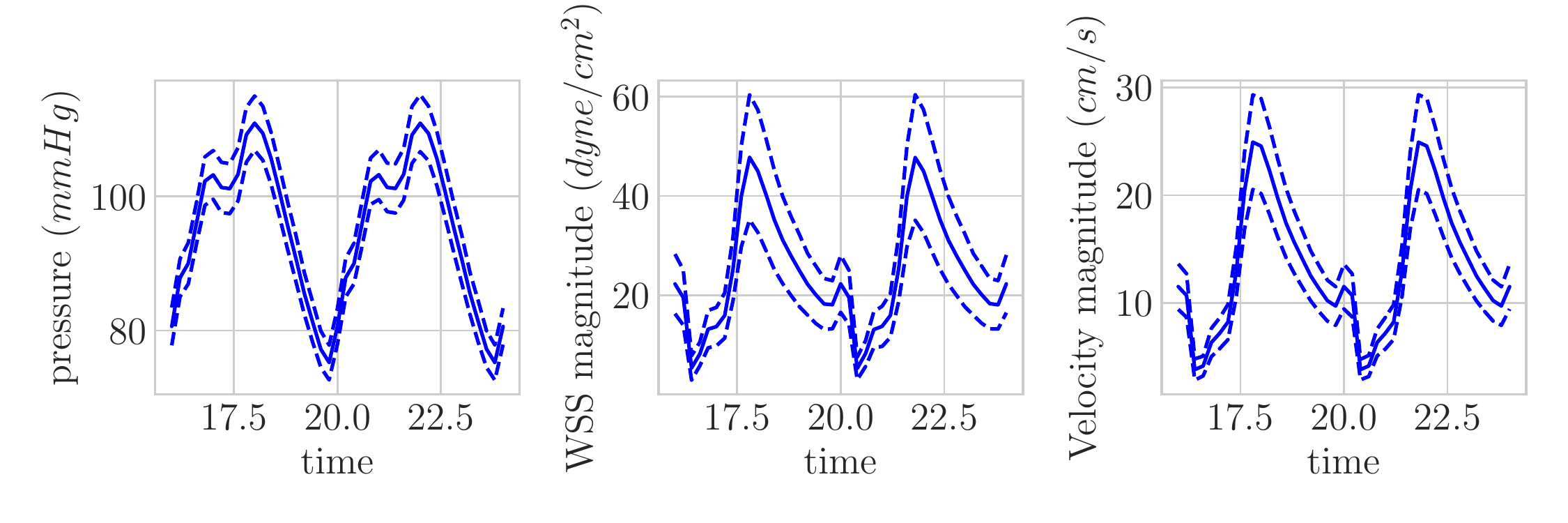}
		\caption{$LAD$}
		\label{fig:results:outletcoronarylc2}
	\end{subfigure}
	\begin{subfigure}[b]{0.48\textwidth}
		\centering\includegraphics[width=\textwidth]{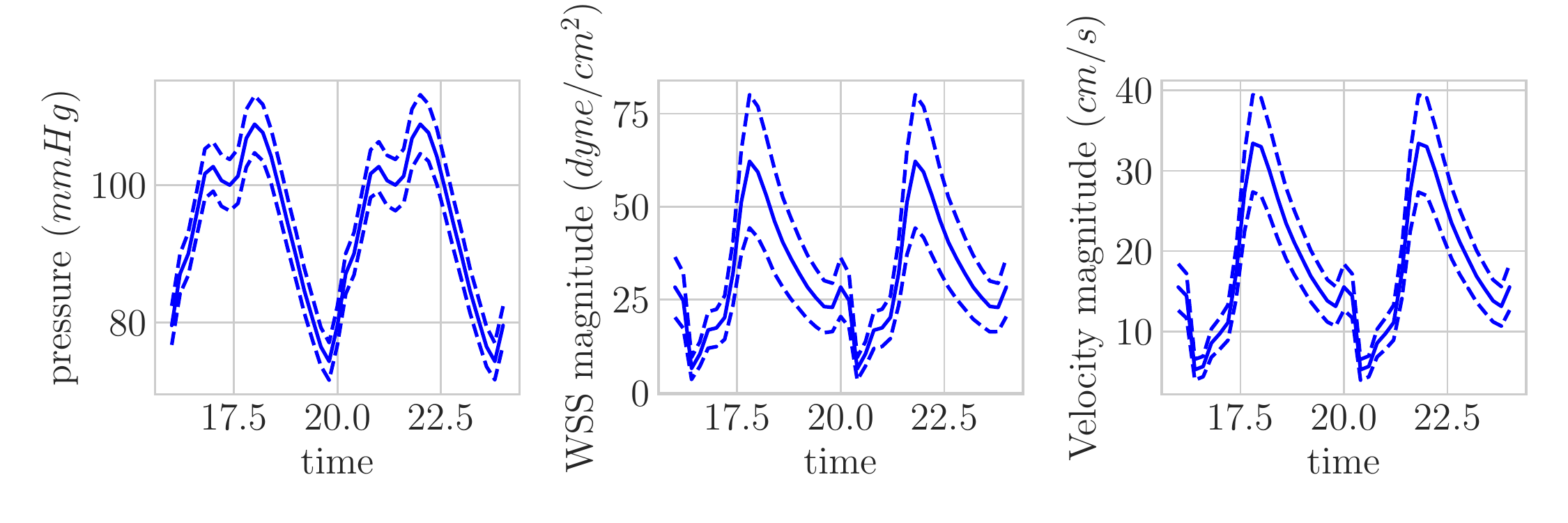}
		\caption{$LAD-D_1$}
		\label{fig:results:outletcoronarylc2sub1}
	\end{subfigure}
	\caption{Outlet QoIs and $\pm 2 \sigma$ interval for left coronary artery model.}
\end{figure}

\begin{figure}[!ht]
	\centering
	\begin{subfigure}[b]{0.48\textwidth}
		\centering\includegraphics[width=\textwidth]{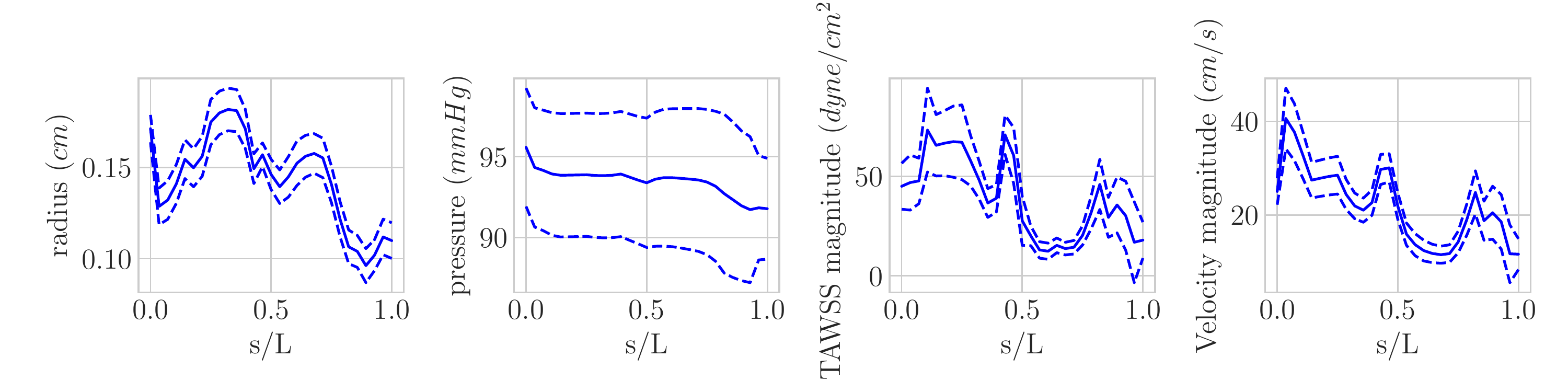}
		\caption{$LCx$}
		\label{fig:results:timeavgcoronarylc1}
	\end{subfigure}
	\begin{subfigure}[b]{0.48\textwidth}
		\centering\includegraphics[width=\textwidth]{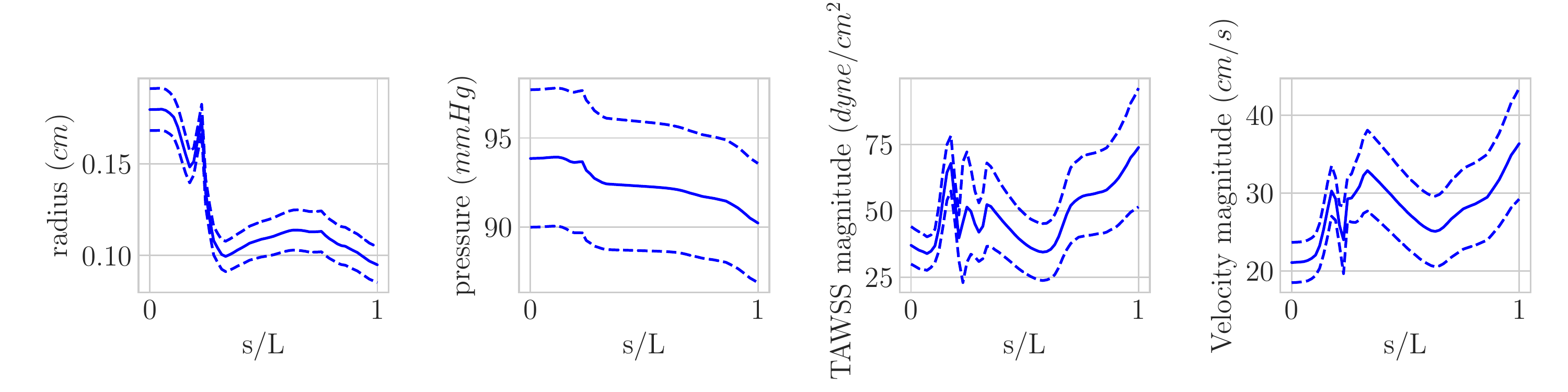}
		\caption{$LCx-OM_1$}
		\label{fig:results:timeavgcoronarylc1_sub1}
	\end{subfigure}
	
	\begin{subfigure}[b]{0.48\textwidth}
		\centering\includegraphics[width=\textwidth]{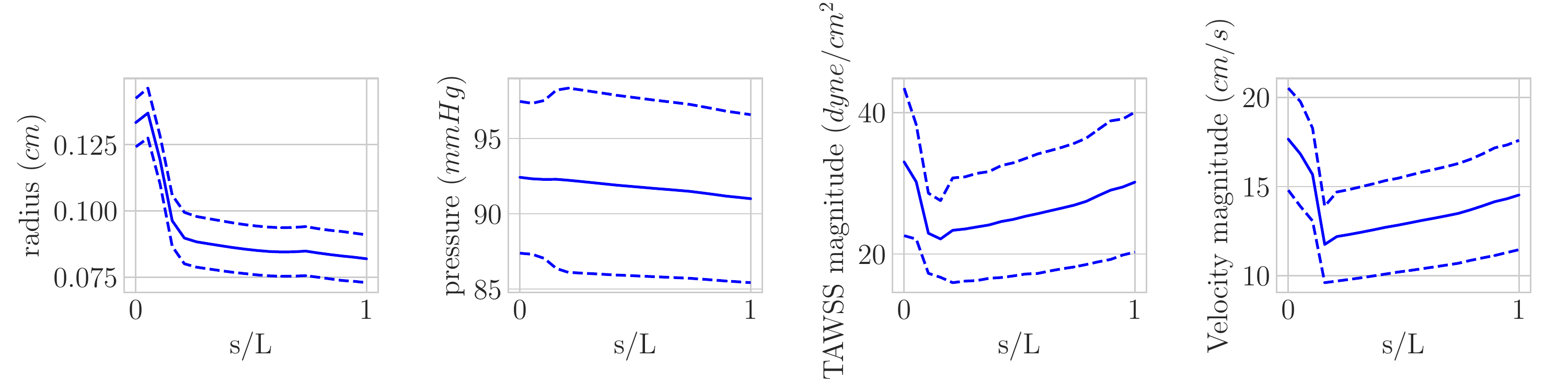}
		\caption{$LCx-OM_2$}
		\label{fig:results:timeavgcoronarylc1sub2}
	\end{subfigure}
	\begin{subfigure}[b]{0.48\textwidth}
		\centering\includegraphics[width=\textwidth]{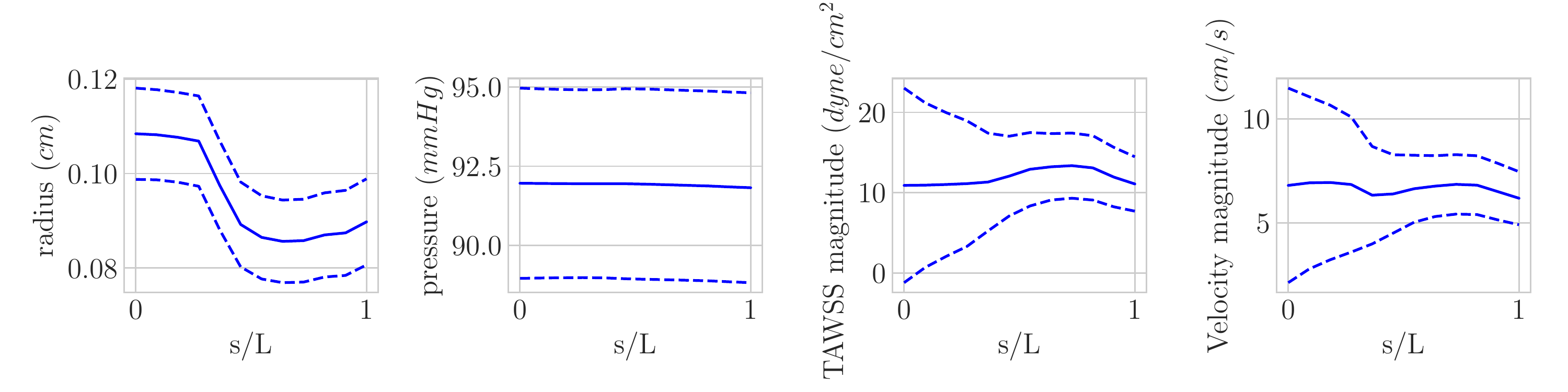}
		\caption{$LCx-OM_3$}
		\label{fig:results:timeavgcoronarylc1sub3}
	\end{subfigure}
	
	\begin{subfigure}[b]{0.48\textwidth}
		\centering\includegraphics[width=\textwidth]{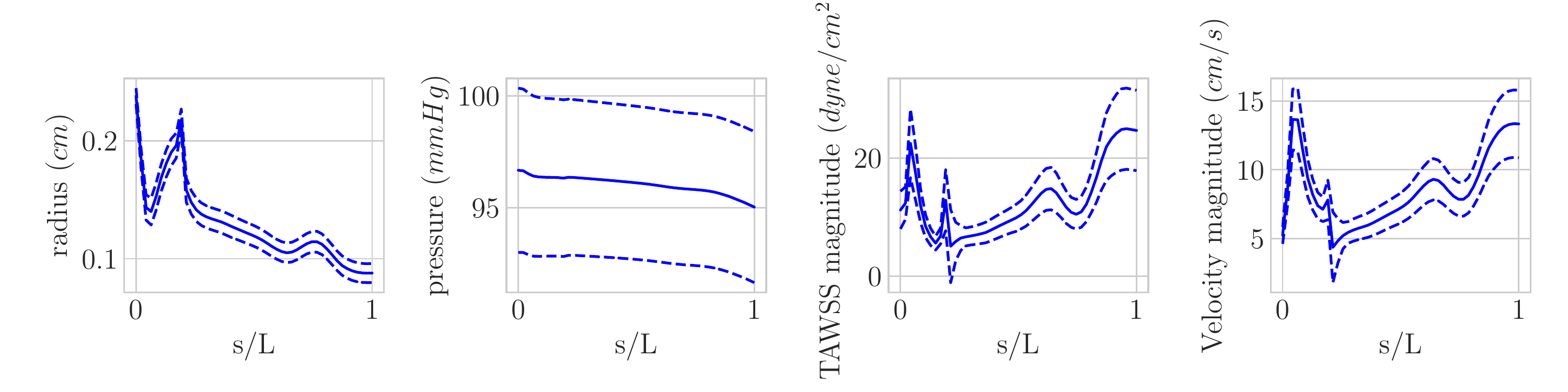}
		\caption{$LAD$}
		\label{fig:results:timeavgcoronarylc2}
	\end{subfigure}
	\begin{subfigure}[b]{0.48\textwidth}
		\centering\includegraphics[width=\textwidth]{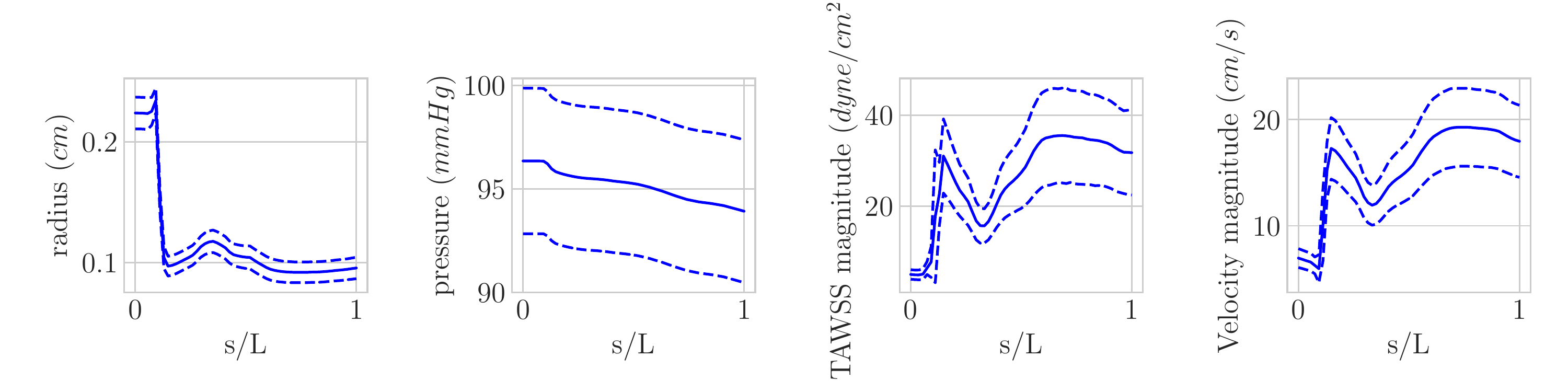}
		\caption{$LAD-D_1$}
		\label{fig:results:timeavgcoronarylc2sub1avg}
	\end{subfigure}
	\caption{Time averaged QoIs and $\pm 2 \sigma$ interval for left coronary artery model, plotted along the vessel centerline.}
\end{figure}

The outlet time histories reflect the diastolic nature of the coronary flow (see Figures~\ref{fig:results:outletcoronarylc1}-\ref{fig:results:outletcoronarylc2sub1}), where the WSS exhibits the largest uncertainty followed by velocity magnitude and pressure. Unlike the other two anatomies considered in the previous sections, geometrical uncertainty significantly affects hemodynamic model outputs due to smaller vessel sizes, as previously suggested in the literature in the context of coronary artery disease~\cite{sankaran16}.
Time averaged quantities over the vessel length show a similar pattern (see Figure~\ref{fig:results:timeavgcoronarylc1}-\ref{fig:results:timeavgcoronarylc2sub1avg}).
In particular, TAWSS and velocity uncertainty appear to be very similar and correlated with the vessel radius. Specifically, smaller radii (with larger radius variability) produce larger TAWSS and velocity uncertainty. A different behavior is instead observed for the $LCx-OM_3$ branch, where large TAWSS and velocity uncertainty are associated with a larger radius. However, this phenomenon is localized at the proximal end of the vessel and probably triggered by the bifurcation nearby.

\begin{figure}[!ht]
	\centering
	\begin{subfigure}[b]{0.23\textwidth}
		\centering\includegraphics[width=\textwidth]{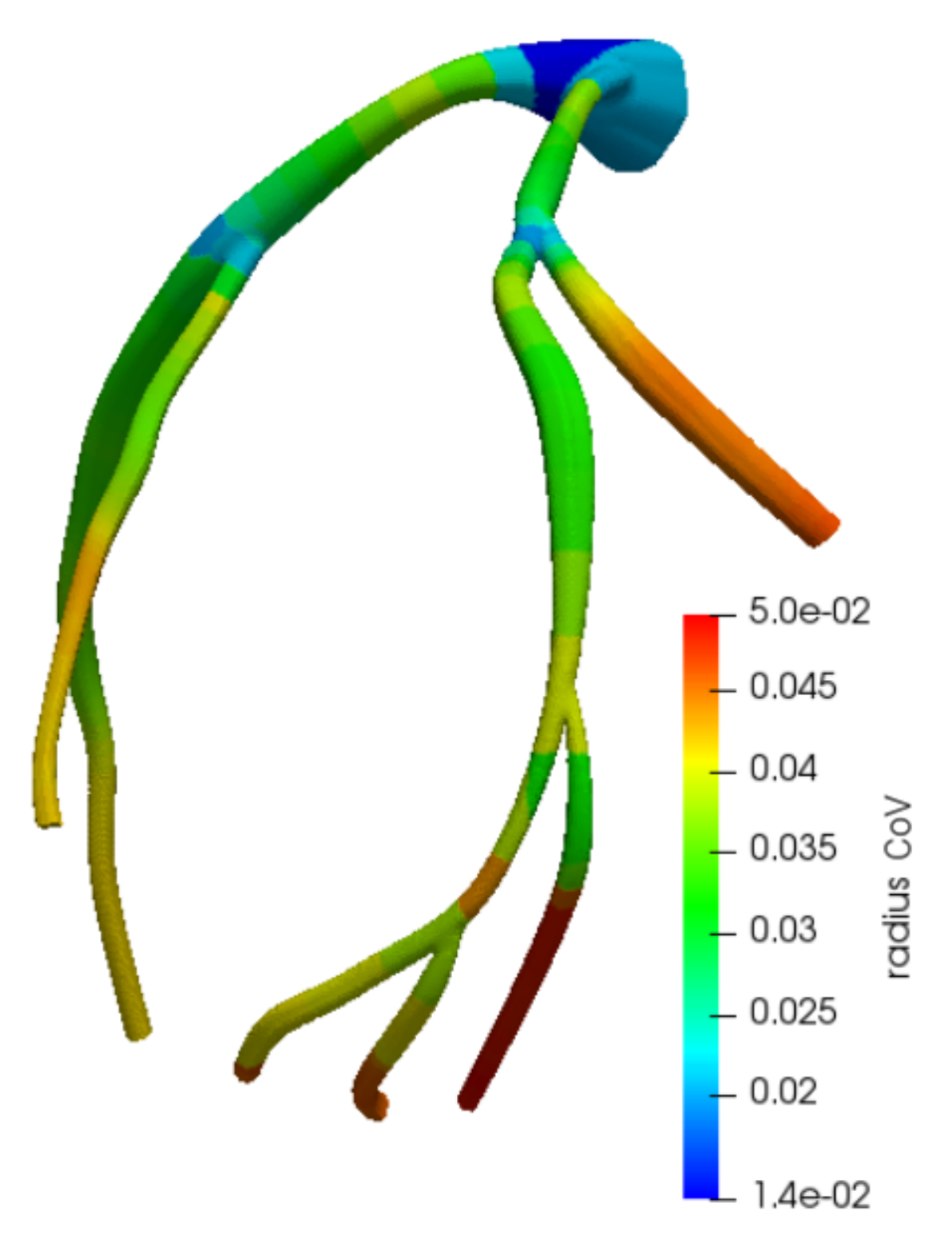}
		\caption{}
		\label{fig:results:coronary3dradius}
	\end{subfigure}
	\begin{subfigure}[b]{0.24\textwidth}
		\centering\includegraphics[width=\textwidth]{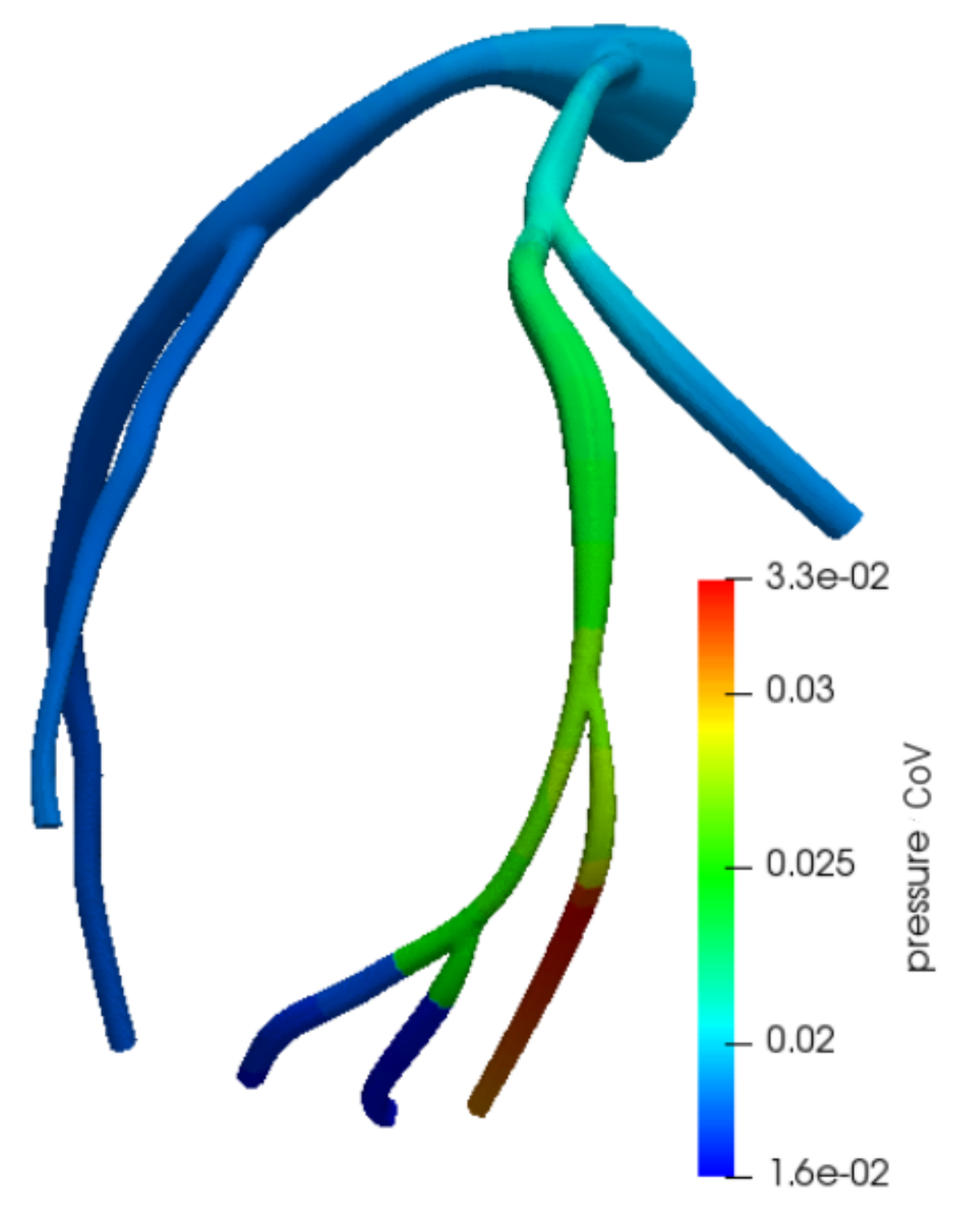}
		\caption{}
		\label{fig:results:coronary3dpressure}
	\end{subfigure}
	\begin{subfigure}[b]{0.22\textwidth}
		\centering\includegraphics[width=\textwidth]{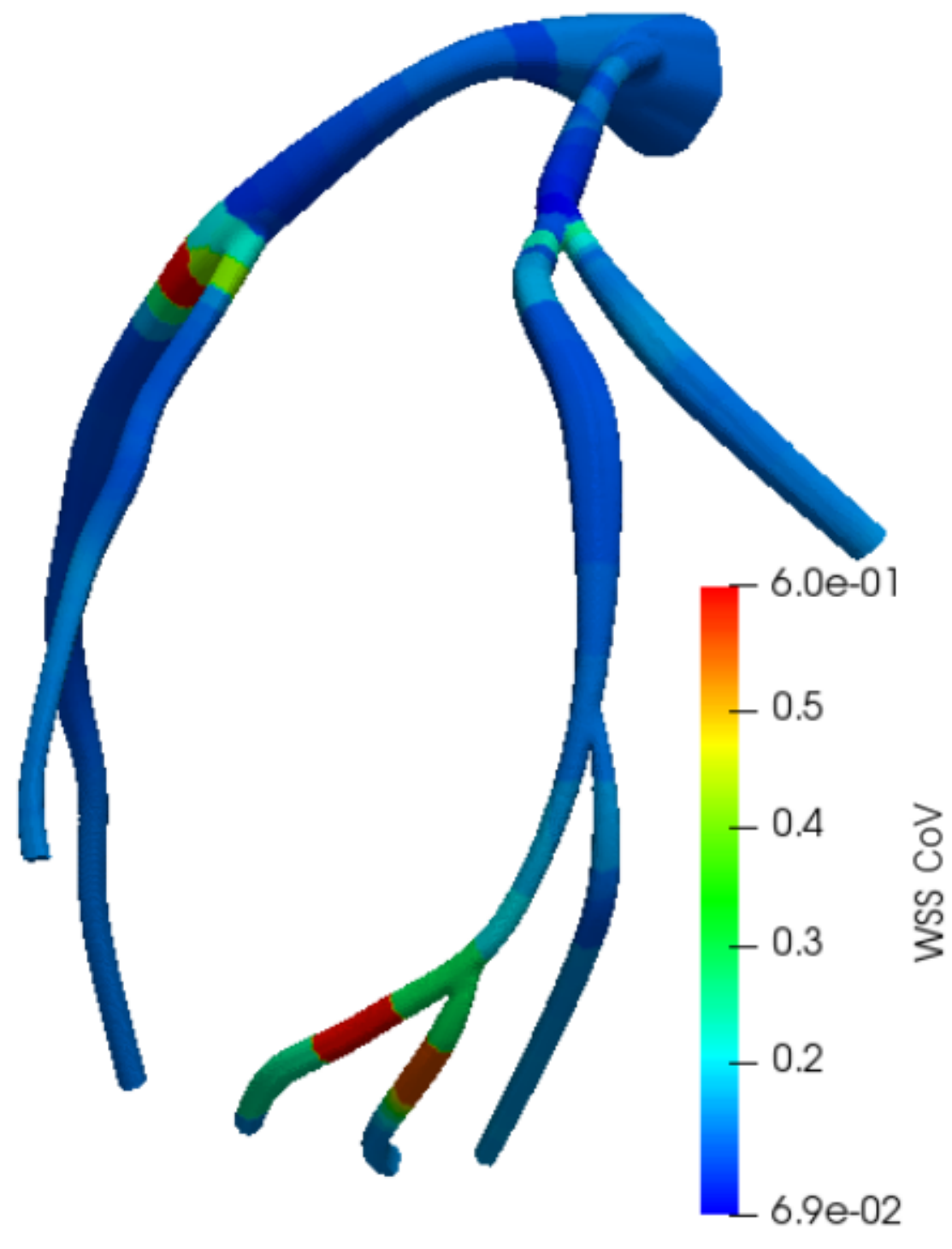}
		\caption{}
		\label{fig:results:coronary3dwss}
	\end{subfigure}
	\begin{subfigure}[b]{0.225\textwidth}
		\centering\includegraphics[width=\textwidth]{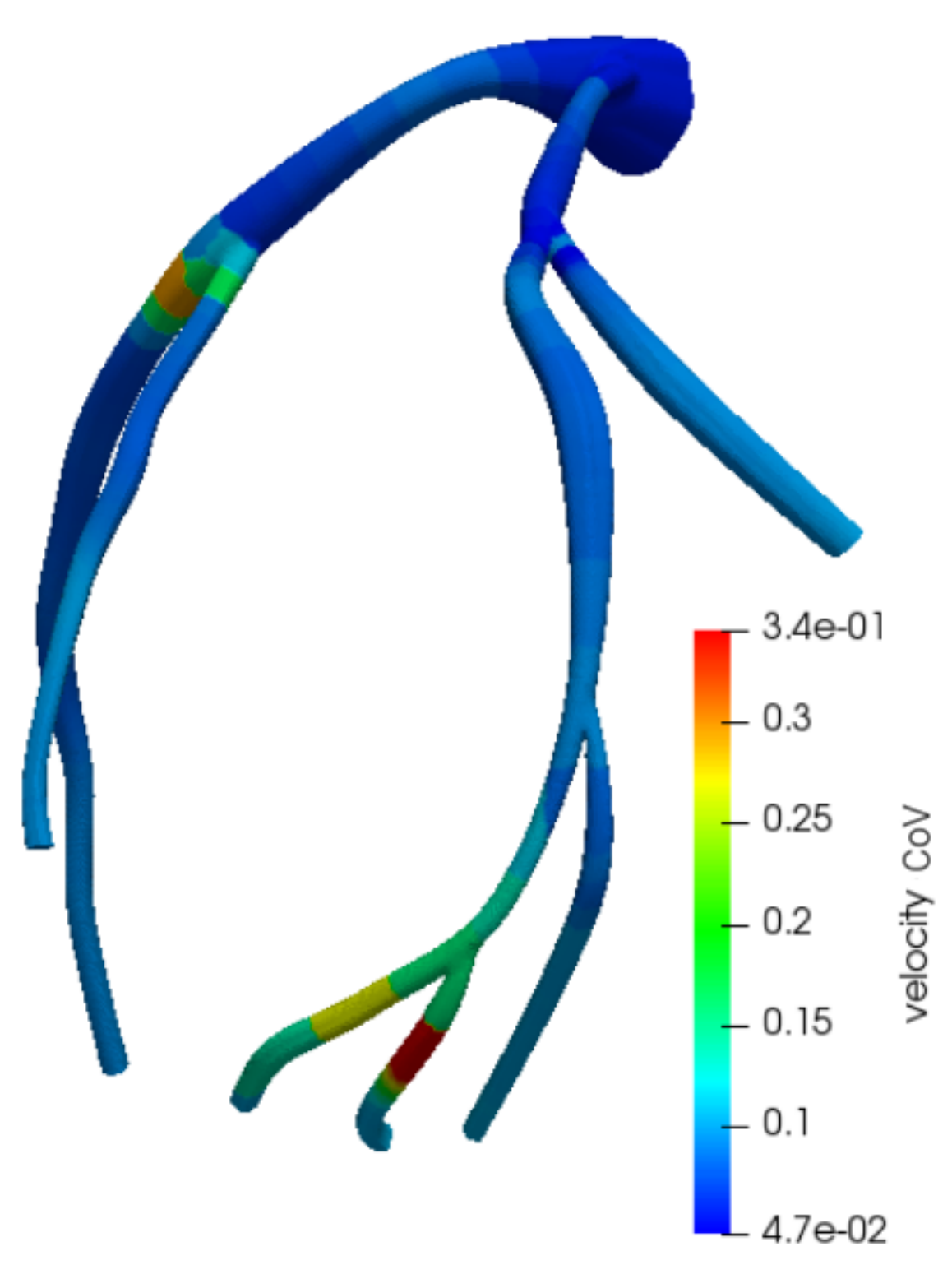}
		\caption{}
		\label{fig:results:coronary3dvelocity}
	\end{subfigure}
	\caption{Nearest neighbor interpolation of cross-sectional time-averaged CoVs for left coronary artery model.}
	\label{fig:coronary3d}
\end{figure}

Pressure, TAWSS and velocity magnitude CoVs were approximately equal to 2\%, 10-20\% and 6-15\%, respectively (see Table~\ref{tab:results:coronaryconf}). Notably, the $LAD-D_1$, $LCx-OM_2$, and $LCx-OM_3$ branches showed larger TAWSS and velocity CoVs, equal to 16\%, 24.5\% and 92.9\%, still explained by the small range of vessel sizes present in the coronary anatomy, which amplifies the segmentation uncertainty, particularly towards the distal end of each vessel (see Table~\ref{tab:results:coronaryconf} and Figure~\ref{fig:results:coronary3dradius}).
The pressure CoV appears to be larger in the $LCx$ and its branches $LCx-OM_1$ and particularly $LCx-OM_2$. This is explained by the relatively small radius of such vessels, which increases resistance and amplifies geometric uncertainty and by the variations in flow split caused by the relatively smaller $LCx$ branching off the $LAD$ trunk (see Figure~\ref{fig:results:coronary3dpressure}).
Finally, the largest WSS and velocity CoVs are located near bifurcations, due to higher local flow variability and more ambiguity in the definition of the vessel lumen (see Figure~\ref{fig:results:coronary3dwss} and Figure~\ref{fig:results:coronary3dvelocity}).

\begin{table}[!ht]
\centering
\caption{Monte Carlo sample mean, coefficient of variation (CoV) and 95\% relative confidence interval for all QoIs in left coronary artery model. $n$ indicates the number of cross-sectional slices for the associated vessel.}
\small{
\begin{tabular}{lcccccc}
\toprule
{\bf Path} & {\bf LCx} & {\bf LCx-OM$_1$} & {\bf LCx-OM$_2$} & {\bf LCx-OM$_3$} & {\bf LAD} & {\bf LAD-D$_1$}\\
& $(n=29$ & $(n=58)$ & $(n=20)$ & $(n=12)$ & $(n=48)$ & $(n=55)$ \\
\midrule
{\bf Radius mean} [cm] & 0.14 & 0.12 & 0.09 & 0.10 & 0.13 & 0.11\\
{\bf Radius CoV} & 0.032 & 0.031 & 0.047 & 0.046 & 0.034 & 0.034\\
{\bf Radius conf.} & 0.0061 & 0.0059 & 0.0090 & 0.0088 & 0.0064 & 0.0064\\
\midrule
{\bf Pressure mean} [mmHg] & 93.46 & 92.37 & 91.78 & 91.92 & 96.02 & 95.14\\
{\bf Pressure CoV} & 0.021 & 0.020 & 0.031 & 0.016 & 0.018 & 0.018\\
{\bf Pressure conf.} & 0.0040 & 0.0037 & 0.0059 & 0.0031 & 0.0034 & 0.0034\\
\midrule
{\bf TAWSS mean} [dyne/cm$^2$]& 39.27 & 48.43 & 26.37 & 11.91 & 12.78 & 26.17\\
{\bf TAWSS CoV} & 0.106 & 0.108 & 0.142 & 0.220 & 0.114 & 0.133\\
{\bf TAWSS conf.} & 0.0201 & 0.0206 & 0.0269 & 0.0418 & 0.0217 & 0.0252\\
\midrule
{\bf Velocity mean} [cm/s]& 21.91 & 28.04 & 13.69 & 6.66 & 8.56 & 15.50\\
{\bf Velocity CoV} & 0.069 & 0.067 & 0.091 & 0.158 & 0.075 & 0.084\\
{\bf Velocity conf.} & 0.0131 & 0.0126 & 0.0172 & 0.0299 & 0.0143 & 0.0160\\
\bottomrule
\end{tabular}}
\label{tab:results:coronaryconf}
\end{table}

%% file: discussion.tex
\section{Discussion}\label{section:paper3:discussion}

Our experiments in the previous sections illustrate how the proposed Bayesian dropout network generates vessel lumen segmentations characterized by a $\sigma_{r}$ between 0.005 cm and 0.01 cm. This translates in radius CoVs in the range 1-5\%, where smaller vessel sizes are associated with larger radius variability, as expected.
Furthermore, our cardiovascular model sampling process resulted in realistic whole model variation as observed from the PCA modes across model realizations.

Amongst the output QoIs considered in this study, wall shear stress was the most impacted by geometry uncertainty, followed by velocity magnitude and then pressure. Wall shear stress CoV ranged from 3\% to 20\%, whereas velocity magnitude and pressure resulted in CoV ranges equal to 1.4-15\% and 0.2-3\%, respectively. Larger variability is observed in the left coronary artery model, due to the prevalence of vessels with small radius.

To quantify the relative importance of geometric uncertainty with respect to other sources relevant in hemodynamic simulations, we compare the output variability found in our study to those found in studies investigating other sources of uncertainty.
Uncertainty due to coronary pressure waveform, intramyocardial pressure, morphometry exponent and vascular wall Young's modulus was recently investigated in~\cite{seo19}, in the context of coronary artery modeling. Coronary pressure waveform uncertainty resulted in a 7\% CoV for the average pressure and $<$7\% for TAWSS and velocity magnitude. Intramyocardial pressure uncertainty produced CoV of roughly 25\% for TAWSS and velocity magnitude. Morphometry exponent uncertainty resulted in a 2\% CoV for velocity magnitude and negligible impact on TAWSS magnitude and pressure. Finally, vascular wall Young's modulus uncertainty had negligible impact on the hemodynamics, as suggested in the literature~\cite{Tran2019}. The CoV of 1.4-15\% and 3-20\% for velocity magintude and TAWSS produced by geometric uncertainty in this work is thus comparable to the CoV due to coronary pressure and intramyocardial pressure uncertainty, but larger than the CoV due to morphometry exponent and vascular wall Young's modulus.
The pressure CoV of 0.2-3\% due to geometry uncertainty was smaller than that produced by intramyocardial pressure uncertainty.

A multifidelity uncertainty quantification approach was used in~\cite{Fleeter2019} to investigate simulation uncertainty due to material property and boundary condition uncertainty, for both healthy and diseased aortic and coronary anatomies. The uncertain parameters were uniformly distributed with $\pm30\%$ variation around their means. Resulting CoV were $\sim$3\% for pressure, $<$1\% for velocity magnitude and $\sim$10-20\% for TAWSS, regardless of model and disease condition.
These pressure and TAWSS CoV values were comparable to the CoVs produced by geometric uncertainty.
However the velocity magnitude CoV was smaller than the 1.4-15\% CoV produced by geometric uncertainty.

The effect of variability in closed-loop boundary conditions assimilated from uncertain clinical data including aortic pressure, cardiac output, pulmonary pressure, peak and total flow volumes is investigated in the context of coronary artery disease in~\cite{Tran2017}.
Uncertainties in a number of independent clinical measurements were assumed to be normally distributed with standard deviation in the range 10-40\% of the corresponding (measured) mean value, and the boundary conditions parameters learned using a Bayesian parameter tuning framework, based on adaptive Markov chain Monte Carlo sampling. The results show CoVs for TAWSS in the range of 5\% to 10\%, which is comparable or smaller than the 3-20\% CoV observed for TAWSS in this work.

The results of the three test cases discussed in the previous sections seem to suggest, on the one hand, that geometrical uncertainty has a generally limited impact on hemodynamics compared to other sources of uncertainty, and, in practice, could be disregarded if pressure is the sole QoI. On the other hand, the velocity magnitude and TAWSS variability for the left coronary artery model were found to be approximately 3 to 5 times higher compared to the other two anatomies.
This suggests that geometric uncertainty might play a dominant role for anatomies characterized by small vessel sizes (the coronary circulation is a particularly relevant case), especially for stenotic lesions, associated with substantial radius uncertainty.

\begin{table}[!ht]
\centering
\caption{Comparison between the hemodynamic effect geometry uncertainty and other sources of uncertainty.}
\resizebox{\textwidth}{!}{
\begin{tabular}{l c c c c c}
\toprule
{\bf Model type} & \multicolumn{4}{c}{\bf Coronary}\\
\midrule
{\bf Uncertainty source} & Pressure waveform & Intramyocardial pressure & Morphometry Exponent & Wall Young's modulus\\
{\bf Input distribution} & Uniform & Uniform & Uniform & Uniform\\
{\bf Reference} & \cite{seo19} & \cite{seo19} & \cite{seo19} & \cite{seo19} \\
{\bf Pressure CoV} & 7\% & negligible & negligible & negligible\\
{\bf TAWSS CoV} & $<$7\% & 25\% & negligible & negligible\\
{\bf Velocity CoV} & $<$7\% & 25\% & 2\% & negligible\\
\midrule
{\bf Model Type} & {\bf Aorta/Coronary} & {\bf Coronary bypass graft} & {\bf Aorta/AAA/Coronary}\\
\midrule
{\bf Uncertainty source} & Boundary conditions & Multiple clinical targets &  Full model geometry\\
{\bf Input distribution} & Uniform $\pm30\%$ & Assimilated from clinical data &  Dropout sampling\\
{\bf Reference} & \cite{Fleeter2019} & \cite{Tran2017} & \emph{This paper}\\
{\bf Pressure CoV} & $\sim$3\% & - & 0.2-3\%\\
{\bf TAWSS CoV} & $\sim$10-20\% & 5-10\% & 3-20\%\\
{\bf Velocity CoV} & $<$1\% & - & 1.4-15\%\\
\bottomrule
\end{tabular}}
\label{tab:discussion:uq}
\end{table}

Even though this is the first systematic study in the literature combining machine learning and high-fidelity cardioascular models to study the effect of geometric uncertainty, we recognize several limitations. 
First, uncertainty propagation is performed using standard Monte Carlo sampling. Even though a number of approaches in the literature have shown promise to accelerate convergence to the true statistical moments (such as, e.g., stochastic collocation~\cite{Sankaran2011}, or generalized multiresolution expansions~\cite{Schiavazzi2017} in the context of cardiovascular flow), the Bernoulli random vectors used in the dropout layer in this study have a dimension of around 10,000, which is extremely challenging for approaches based on stochastic spectral expansion. 
Second, our study includes only one diseased anatomy, and showed that geometrical uncertainty is amplified in the aneurysm region of the abdominal aorta. We also found such amplification in a healthy left coronary artery model due to the typically smaller vessel radii. This suggests how cases of stenosed or calcified coronary arteries and vascular lesions, that are not often seen during network training, may offer new insights on the role of geometric uncertainty.
Third, we assume that dropout networks are able to learn output uncertainty from their training data~\cite{gal16}, while, in practice, they provide only an approximate representation of the true distribution of vessel lumen for a given image.
Additionally, our segmentations are generated at discrete slices along the centerline path, and thus the training data might act as a filter on the whole geometric variability, leading the proposed algorithm to underestimate the true underlying geometric uncertainty. Removing these limitations would require new three-dimensional vessel segmentation paradigms, which is an active area of research.
Fourth, the path planning approach intrinsically limits the uncertainty at the bifurcations, as it requires users to adjust pathlines so they originate within the parent vessel. This introduces user bias into the cardiovascular model samples and may constrain the underlying geometric uncertainty. Future work will be devoted to produce improved estimates for bifurcation uncertainty. 

%% file: conclusion.tex
\section{Conclusions}\label{section:paper3:conclusion}

We have developed a Bayesian dropout network to generate families of two-dimensional lumen segmentations from slices of a clinically acquired image volume. Of particular note is the fact that our neural network learns lumen segmentation uncertainty directly from the image training data and is thus able to generate lumen samples with a realistic uncertainty distribution. This was combined with vessel centerlines and a path-planning model building workflow to create realizations of high-fidelity cardiovascular models with uncertain lumen surface. Finally, we characterized simulation output variability due to geometric uncertainty using Monte Carlo sampling.

We have also analyzed the principal components of the lumen surfaces we generated, showing how, despite segmenting slices independently and analyzing geometries characterized by a wide range of vessel radii, dominant modes appear to be equally distributed on the entire model, without amplifying any particular local feature.
Additionally, our network generated vessel lumens with relatively constant radius standard deviation that were found to be independent of vessel size.  This resulted in increasing relative uncertainty for smaller vessels, similar to manual segmentations generated by expert users.

Experiments on an aortic bifurcation model, an abdominal aortic aneurysm model and a left coronary artery model showed that geometry uncertainty primarily resulted in wall shear stress and velocity magnitude uncertainty. This was true in particular for the coronary anatomy, characterized by smaller vessel sizes. 
Moreover, while TAWSS and velocity magnitude were impacted by geometrical uncertainty, especially near the distal ends of small vessels and near bifurcations, pressure was only marginally affected.
Compared to other sources of uncertainty, for example, the boundary conditions or material properties, the relative importance of geometry uncertainty was found to be determined by the particular patient-specific geometry being investigated and the vessel radius.

Our method still requires one to manually create the vessel centerlines, which may be laborious, time consuming and introduce additional uncertainty. Automated methods to predict vessel centerlines and corresponding uncertainty could therefore be combined with the proposed dropout network to further improve performance and increase model building efficiency.
Finally, we have explored geometry uncertainty independently from other sources, disregarding their interaction. Future work will be devoted to combine all three sources of simulation uncertainty, i.e., boundary conditions, material properties and geometry.